\documentclass{aa}  
\usepackage{academicons}
\usepackage[usenames,dvipsnames]{xcolor}
\usepackage{hyperref}
\hypersetup{colorlinks=true,allcolors=blue}

\usepackage{pdflscape}
\usepackage{comment}
\usepackage{longtable}
\usepackage{lscape}
\usepackage{orcidlink}
\usepackage{bm}

\newcommand{\simgt}{\lower.5ex\hbox{$\; \buildrel > \over \sim \;$}}
\newcommand{\simlt}{\lower.5ex\hbox{$\; \buildrel < \over \sim \;$}}

\def\btheta{\mbox{\boldmath $\theta$}}

\def\hubbleMpc{\mathrel{h_{70}^{-1}{\rm Mpc}}}

\def\hubbleMsol{\mathrel{h_{70}^{-1}M_\odot}}

\usepackage{graphicx}
\usepackage{txfonts}

\begin{document} 

   \title{The SRG/eROSITA all-sky survey}
     \subtitle{Subaru/HSC-SSP weak-lensing mass measurements for eRASS1 galaxy clusters}
      \titlerunning{HSC-SSP WL  analysis of the eRASS1 clusters}

\author{Nobuhiro~Okabe\orcidlink{0000-0003-2898-0728}\inst{1,2,3}  \and
 Thomas H. Reiprich\inst{4} \and
 Sebastian Grandis\inst{5} \and
 I-Non Chiu\inst{6} \and
 Masamune Oguri\inst{7,8} \and
 Keiichi Umetsu\inst{6}  \and
 Esra Bulbul\inst{9} \and
 Emre Bahar\inst{9} \and
 Fabian Balzer\inst{9} \and
 Nicolas Clerc\inst{10} \and
 Johan Comparat\inst{9} \and
 Vittorio Ghirardini\inst{9,11} \and
 Florian Kleinebreil\inst{5} \and
 Matthias Kluge\inst{9} \and
 Ang Liu\inst{9,12} \and
 Rog{\'e}rio Monteiro-Oliveira\inst{6} \and
 Florian Pacaud\inst{4} \and
 Miriam Ramos Ceja\inst{9} \and
 Jeremy Sanders\inst{9} \and
 Tim Schrabback\inst{5} \and
 Riccardo Seppi\inst{9} \and
 Martin Sommer\inst{4} \and
Xiaoyuan Zhang\inst{9}}

   \institute{Affiliations at the end of the paper,   \email{okabe@hiroshima-u.ac.jp}}

\date{}

  \abstract
 { We performed individual weak-lensing (WL) mass measurements for 78 eROSITA's first All-Sky Survey (eRASS1) clusters in the footprint of Hyper Suprime-Cam Subaru Strategic Program (HSC-SSP) S19A. We did not adopt priors on the eRASS1 X-ray quantities or assumption of the mass and concentration relation. In the sample, we found three clusters are misassociated with optical counterparts and 12 clusters are poorly fitted with an NFW profile. The average mass for the 12 poor-fit clusters changes from $\sim 10^{14}h_{70}^{-1}M_\odot$ to $\sim 2\times 10^{13}h_{70}^{-1}M_\odot$ when lensing contamination from surrounding mass structures is taken into account.
  The scaling relations between the true mass and cluster richness and X-ray count-rate agree well with the results of the eRASS1 western Galactic hemisphere region based on count-rate-inferred masses, which were calibrated with the HSC-SSP, DES, and KiDS surveys. We developed a Bayesian framework for inferring the mass-concentration relation of the cluster sample, explicitly incorporating the effects of weak-lensing mass calibration in the mass-concentration parameter space. The redshift-dependent mass and concentration relation is in excellent agreement with predictions of dark-matter-only numerical simulations and previous studies using X-ray-selected clusters. 
  Based on the two-dimensional (2D) WL analysis, the offsets between the WL-determined centers and the X-ray centroids for 36 eRASS1 clusters with high WL S/N can be described by two Gaussian components. We find that the miscentering effect with X-ray centroids is smaller than that involving peaks in the galaxy maps. Stacked mass maps support a small miscentering effect, even for clusters with a low WL S/N.
  The projected halo ellipticity is $\langle \varepsilon \rangle=0.45$ at $M_{200}\sim 4\times10^{14}h_{70}^{-1}M_\odot$, which is in agreement with the results of numerical simulations and previous studies of clusters characterized by masses greater than twice the mass treated here.}


   \maketitle

\section{Introduction}

The evolution of the cluster mass function provides powerful constraints on cosmological parameters, in particular, the energy density of the
total matter,  normalization of density fluctuation, and  dark energy \citep[e.g.,][]{2023MNRAS.522.1601C,2024A&A...689A.298G,2024A&A...691A.301A,2025A&A...696A...5A}. Since the X-ray emissivity of thermal bremsstrahlung is proportional to the square of the electron number density, X-ray observations, which are less sensitive to projection effects, serve as a powerful tool for searching for galaxy clusters.
The eROSITA \citep[extended ROentgen Survey with an Imaging Telescope Array;][]{2012arXiv1209.3114M,2021A&A...647A...1P} instrument on board the Spectrum Roentgen Gamma (SRG) mission \citep{2021A&A...656A.132S} is a unique tool that enables the discovery of galaxy
clusters on an order of $10^5$.
Indeed, the eROSITA Final Equatorial Depth Survey (eFEDS) in a footprint
with an area of $\sim140\,{\rm deg}^2$, significantly overlapping the Subaru/HSC-SSP footprint \citep{HSC1styrOverview,2022PASJ...74..247A}, found 542 cluster candidates \citep{2022A&A...661A...2L} and confirmed an optical counterpart of 477 clusters and groups \citep{2022A&A...661A...4K}. 
Recently, the Western Galactic Hemisphere of the eROSITA’s first All-Sky Survey \citep[the eRASS1;][]{2024A&A...685A.106B,2024A&A...688A.210K} discovered 12,247 optically confirmed galaxy groups and clusters detected in the 0.2-2.3 keV over 13,116 deg$^2$ and constructed a sample for cosmology \citep{2024A&A...685A.106B}, with a high purity level of $\sim95\%$, comprising 5259 optically confirmed clusters over an area of 12,791 deg$^2$.

Accurate mass measurements of galaxy clusters are vitally important for cluster cosmology. Thanks to deep and wide optical surveys of ground-based telescopes of Subaru/HSC-SSP, DES, and KiDS \citep{2022A&A...661A..11C,Chiu25, 2024A&A...687A.178G,2025A&A...695A.216K}, the weak gravitational lensing effect can measure mass structures without assumptions of dynamical states. 
The weak-lensing (WL) mass measurements for eROSITA utilize a hierarchical Bayesian approach to simultaneously fit the tangential shear profiles, accounting for the mass and count-rate relation in individual clusters. \citep{2022A&A...661A..11C,2024A&A...687A.178G,Chiu25}. The resulting WL mass is, therefore, a kind of count-rate-related mass. The adopted mass model assumes the mass and concentration relation for a Navarro-Frenk-White (NFW) model \citep{NFW96,NFW97} statistically treats  miscentering effects by convolving a probability function,  using the relationship between the true and WL masses based on numerical simulations.

In contrast, the individual cluster WL analysis allows for the concurrent determination of the two parameters of mass and concentration; however, it does necessitate a large number of background galaxies \citep[e.g.,][]{Okabe16b,2020ApJ...890..148U}.
Furthermore, two-dimensional WL analyses \citep{2010MNRAS.405.2215O} enable us to measure central positions for the assumed mass models to assess the miscentering effect utilized in the statistical WL mass measurement. The lensing information also independently provides us with the purity of the detected clusters. Therefore, individual cluster lensing measurement is complementary to statistical cluster-lensing analysis. 

Thanks to the deep imaging of the Subaru/HSC-SSP, we were able to perform cluster mass measurements for the eRASS1 clusters to understand and control WL systematics.
The paper is organized as follows. We describe the WL mass measurement techniques in Sect. \ref{sec:WL}, followed by our results in Sect. \ref{sec:result}, a discussion in Sect. \ref{sec:dis}, and conclusions in Sect. \ref{sec:con}, respectively.
Throughout the paper, we use $\Omega_{m,0}=0.3$, $\Omega_{\Lambda,0}=0.7$, and $H_0=70h_{70}\,{\rm km s^{-1}Mpc^{-1}}=100 h \,{\rm km s^{-1}Mpc^{-1}}$, with $h_{70}=1$ and $h=0.7$.

\section{Weak-lensing analysis} \label{sec:WL}

\begin{figure*}[!ht]
   \centering 
   \includegraphics[width=9cm]{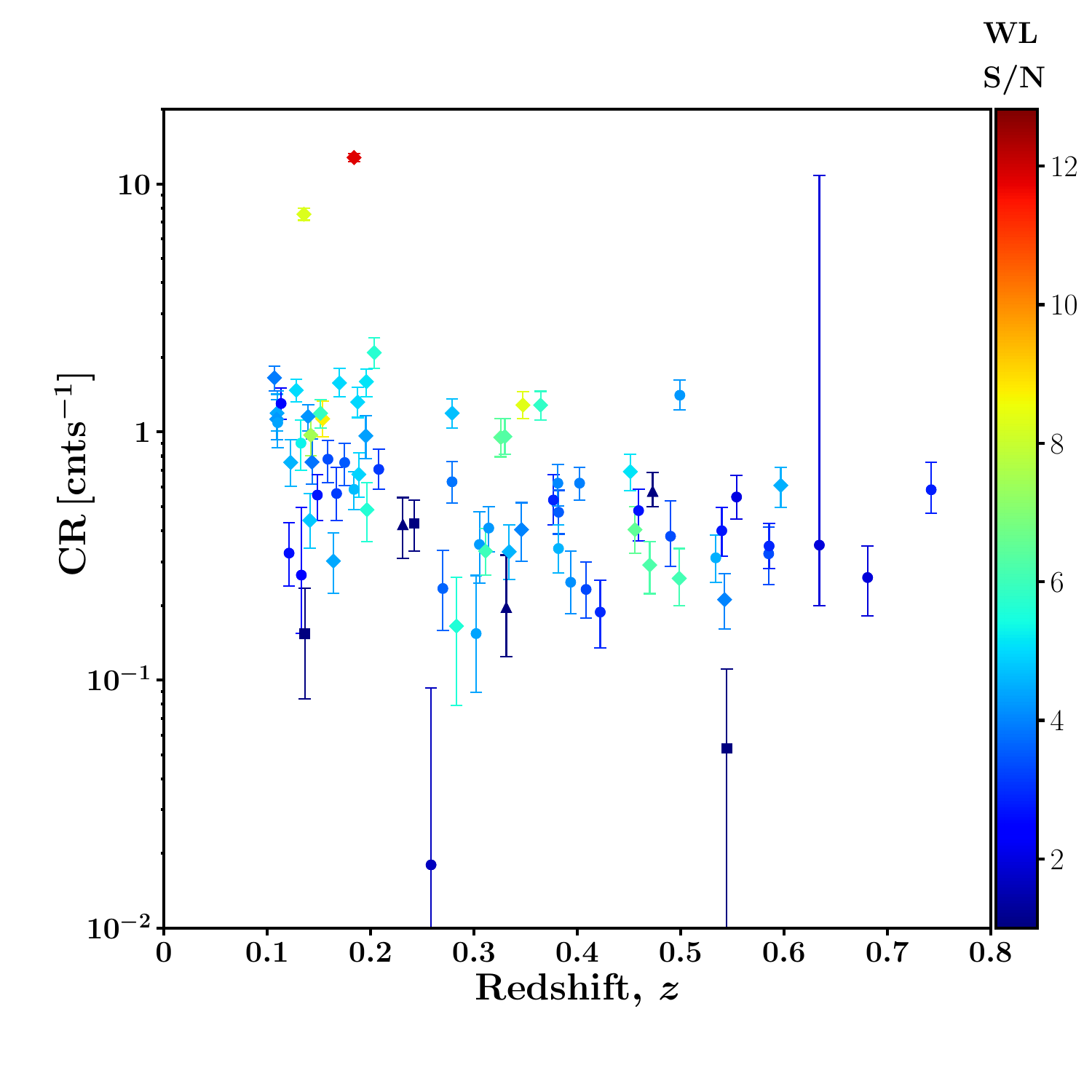} 
   \includegraphics[width=9cm]{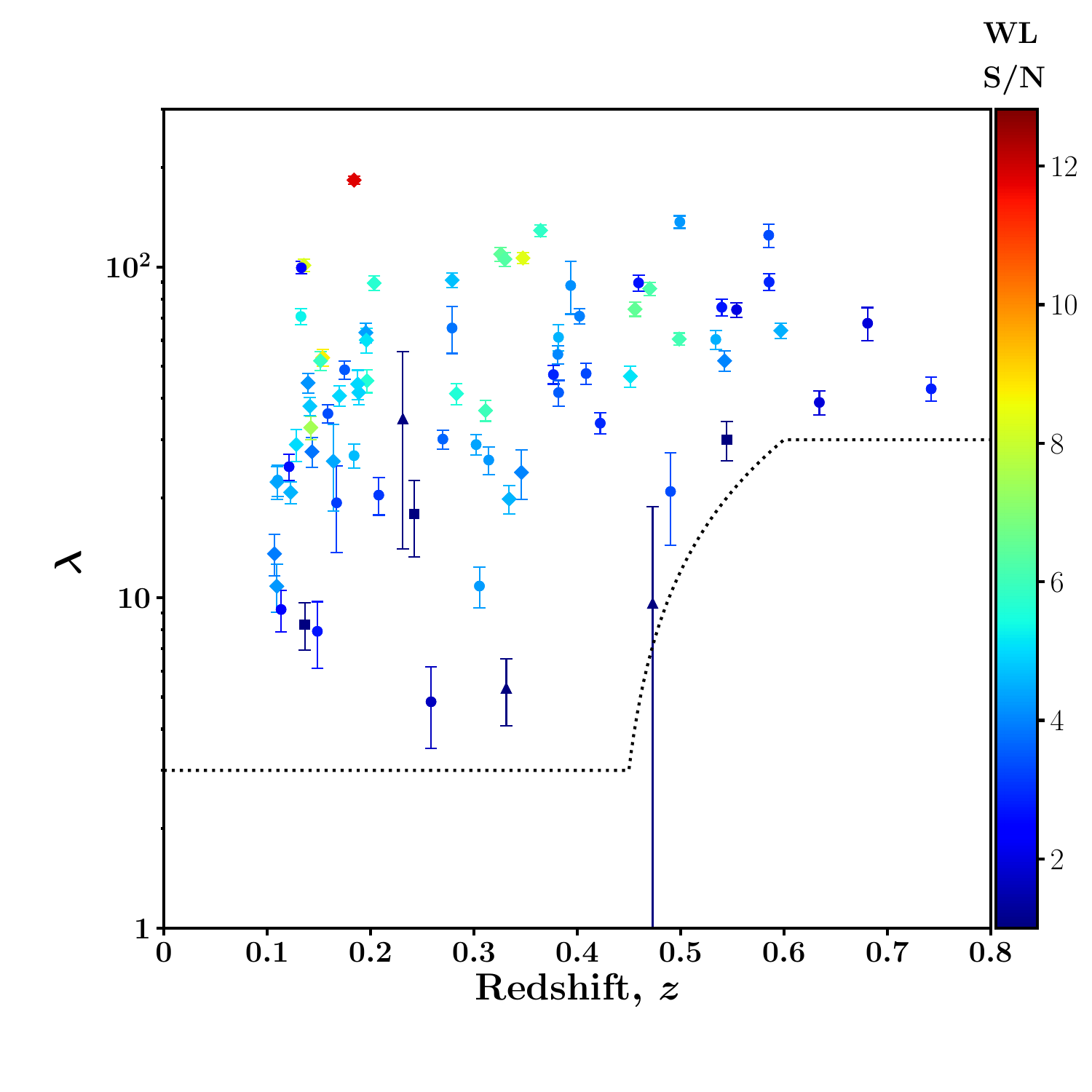}
      \caption{Left: Count-rate (CR) versus redshift ($z$).  Right: Richness ($\lambda$) versus redshift ($z$). The dotted line is the threshold described by a function connecting two constants with a linear function. Colors in all the panels denote the S/N of the reduced tangential shear profile. The circles, diamonds, squares, and up-triangles denote the clusters with 1D WL analysis only, 1D and 2D WL analysis, non-WL analysis, and misassociation, respectively.
              }
         \label{fig:cluster_sample}
   \end{figure*}

 We selected 78 clusters (Table \ref{table:cog1} and Fig. \ref{fig:cluster_sample}) for weak-lensing mass measurements from the 103 eRASS1 clusters in the HSC-SSP S19A footprint, according to the following criteria; the area fraction within $r<3\hubbleMpc$ centering the eRASS1 primary clusters \citep{2024A&A...689A.298G}, where an HSC-Y3 shape catalog \citep[see details in][]{HSC-3Y-Shape} is available, is greater than $70\%$ and the innermost radius is less than $0.7\hubbleMpc$. For instance, 23 clusters were removed because they are located around the edge of the HSC-Y3 footprint, so the galaxies in the shape catalog are not distributed in all azimuthal directions but only on one side or a small patch region. The central areas of the other two clusters are unavailable due to star masks, rendering them inaccessible.

\subsection{Background selection}

\begin{figure}[!ht]
   \centering 
   \includegraphics[width=9cm]{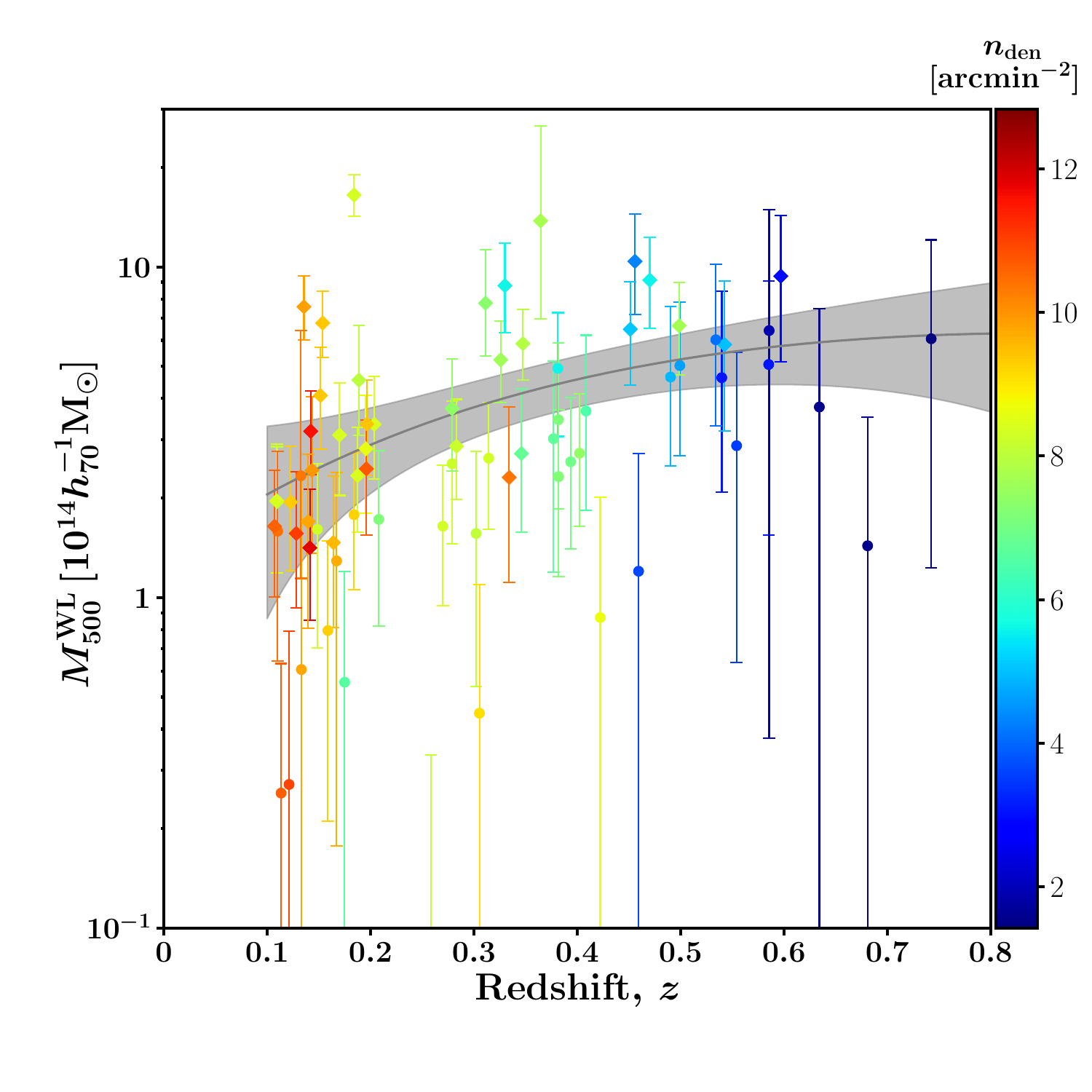}
      \caption{Weak-lensing masses ($M^{\rm WL}_{500}$ [$10^{14}\hubbleMsol$]) versus redshift. The symbols are the same as in Fig. \ref{fig:cluster_sample}. Colors denote the background number density, $n_{\rm den}\,{\rm [arcmin^{-2}]}$. The gray solid line and regions are the mean and $1\sigma$ region for the WL mass distribution in the parent sample obtained by the multivariate scaling relation analysis (Sec. \ref{subsec:scaling}), respectively.
              }
         \label{fig:z_vs_M500}
   \end{figure}

The HSC-Y3 shape catalog uses a method based on point spread function (PSF) correction
known as re-Gaussianization \citep{Hirata03}, which is implemented in the HSC pipeline \citep[see details in][]{HSCWL1styr,HSC-3Y-Shape}. 
In weak-lensing mass measurements, we utilize galaxies from the HSC galaxy catalog that meet the full-color and full-depth criteria, ensuring accurate shape measurement and reliable photometric redshift estimation.
We selected background galaxies behind each cluster based on the condition \citep{Medezinski18}:
\begin{eqnarray}
    \int_{z_l+0.1} P(z) dz>0.98, \label{eq:Pz_sel}
\end{eqnarray}
where $P(z)$ is the probability of the photometric redshift from the machine learning method \citep[MLZ;][]{MLZ14} calibrated with spectroscopic data \citep{2018PASJ...70S...9T,2020arXiv200301511N} and $z_l$ is a cluster redshift.  After the background selection, the number density of background galaxies has a strong redshift dependence, varying from $\sim 2$ to $\sim 12$ $[{\rm arcmin}^{-2}]$, with an average of $7.0\pm 2.5\,[{\rm arcmin}^{-2}]$ (Figs. \ref{fig:z_vs_M500} and \ref{fig:z_vs_M_WLcal}).

\subsection{Mass estimation}
The reduced tangential shear $\Delta \Sigma_{+}$ was computed by the azimuthal averaging the measured tangential ellipticity, $e_+$, for $i$-th galaxy in a given $k$-th annulus,
\begin{eqnarray}
\Delta \Sigma_{+}(r_k) = 
\frac{\sum_{i} e_{+,i} w_{i} \langle \Sigma_{{\rm cr}}(z_{l}, z_{s,i})^{-1}\rangle ^{-1}}{2 {\mathcal R}_k (1+K_k) \sum_{i} w_{i}}, \label{eq:g+}
\end{eqnarray}
where $e_+$ is the tangential ellipticity ($e_{+}=-(e_{1}\cos2\varphi+e_{2}\sin2\varphi)$, $w_i$ is 
the weighting function, $\langle \Sigma_{{\rm cr}}(z_{l},z_{s,i})^{-1}\rangle$ is the inverse of the mean critical surface mass density, $\mathcal{R}$ is the shear responsivity, and $K$ is the calibration factor. 
The weighting function is given by
\begin{eqnarray}
w_{i}=\frac{1}{e_{\rm rms,i}^2+\sigma_{e,i}^2}\langle \Sigma_{{\rm cr}}(z_{l,j},z_{s,i})^{-1}\rangle^2\label{eq:weight}.
\end{eqnarray}
where $e_{\rm rms}$ and $\sigma_e$ are the root mean square of intrinsic ellipticity and the measurement error per component
($\alpha=1$ or $2$), respectively. The inverse of the mean critical surface mass density is computed by the probability function, $P(z)$, following 
\begin{eqnarray}
 \langle \Sigma_{{\rm cr}}(z_{l},z_{s,i})^{-1}\rangle =
  \frac{\int^\infty_{z_{l}}\Sigma_{{\rm cr}}^{-1}(z_{l},z_{s,i})P(z_{s,i})dz_{s,i}}{\int^\infty_{0}P(z_{s,i})dz_{s,i}}.
\end{eqnarray}
Here, $z_{l}$ and $z_{s}$ are the cluster and source redshift, respectively.
The critical surface mass density is $\Sigma_{{\rm cr}}=
c^2D_{s}/4\pi G D_{l} D_{ls}$, where $D_s$ and $D_{ls}$ are the
angular diameter distances from the observer to the sources and from the
lens to the sources, respectively.  
The shear responsivity is expressed as
\citep[see ][]{Mandelbaum05,Reyes12}:\begin{eqnarray}
{\mathcal R}= 1-\frac{\sum_i e_{\rm rms,i}^2 w_i}{\sum_i w_i}, \label{eq:R}
\end{eqnarray}

The measured values are corrected by the shear calibration factor
$(m,c)$ for individual objects \citep{HSC-3Y-Shape}, where
$m$ is a multiplicative calibration bias and $c$ an additive residual shear
offset in the relation between the input and output shear component, $\gamma_{\rm output,\alpha} =(1+m_\alpha) \gamma_{\rm input,\alpha}+c_\alpha$, as defined by STEP (Shear TEsting Programme) simulations
\citep{Heymans06,Massey07}.
The calibration factor, $K$, is computed by
\begin{eqnarray}
 K=\frac{\sum_i m_i w_i}{\sum_i w_i}. \label{eq:K}
\end{eqnarray}
where $m_i$ for individual galaxies are estimated based on GREAT3-like simulations
\citep{Mandelbaum18, Mandelbaum14,Mandelbaum15}  as a part of the
GREAT (GRavitational lEnsing Accuracy Testing) project. 
We then subtract  
\begin{eqnarray}
\tilde{c}=\frac{\sum_i c_{+,i} w_{i} \langle \Sigma_{{\rm
 cr},i}^{-1}\rangle^{-1}}{(1+K) \sum_n w_{i}}
\end{eqnarray}
from $\langle \Delta \Sigma_{+}\rangle(r_i)$  (Eq.
\ref{eq:g+}). The additional offset term is negligible ${\mathcal
O}(<10^{-4})$ compared to $\langle \Delta
\Sigma_{+}\rangle\sim{\mathcal O}(10^{-1})$.
The effective radius utilized to characterize the tangential shear in each annulus is shifted from the intermediate radius since parts of the galaxy distributions are partially missing due to the presence of a star mask. We thus adopted the weighted harmonic mean as the radius positions, 
\begin{eqnarray}
    r_k=\frac{\sum_i w_i }{\sum_i r_i^{-1} w_i}, \label{eq:whm_radius}
\end{eqnarray}
which aptly explains a power-law tangential shear profile \citep{Okabe16b}.

We employed an NFW profile \citep{NFW96} for model fitting.
The NFW mass density profile is described as
\begin{equation}
\rho_{\rm NFW}(r)=\frac{\rho_s}{(r/r_s)(1+r/r_s)^2},\label{eq:rho_nfw}
\end{equation}
where $\rho_s$ is the central density parameter and $r_s$ is the scale
radius. We used two parameters of the spherical mass, $M_\Delta$, and the halo concentration, $c_\Delta$, instead of $\rho_s$ and $r_s$.
The spherical mass and the halo concentration are defined by
\begin{eqnarray}
    M_{\Delta}=\frac{4}{3}\pi \Delta \rho_{\rm cr} r_{\Delta}^3,  \quad\quad    c_{\Delta}=\frac{r_{\Delta}}{r_s},  
\end{eqnarray}
respectively. Here, $\rho_{\rm cr}(z)$ is the critical mass density at the cluster redshift and $\Delta$ is the overdensity.  
The reduced shear model, $f_{\rm model}$, is described using the differential surface mass density
$\Delta \tilde{\Sigma}_+$ and the local surface mass density $\Sigma$ as follows,
\begin{eqnarray}
f_{\rm model}=\frac{\Delta \tilde{\Sigma}_+}{1-\mathcal{L}_z \Sigma},
\end{eqnarray}
where $\mathcal{L}_z=\sum_{i} \langle \Sigma_{{\rm cr},i}^{-1}\rangle w_{i}/\sum_{i} w_{i}$.
The log-likelihood is expressed as
\begin{eqnarray}
-2\ln {\mathcal L}&=&\ln(\det(C_{nm})) +  \label{eq:likelihood} \\
 &&\sum_{n,m}(\Delta \Sigma_{+,n} - f_{{\rm model}}(r_n))C_{nm}^{-1} (\Delta
 \Sigma_{+,m} - f_{{\rm model}}(r_m)), \nonumber
\end{eqnarray}
where the covariance matrix, $C$, is composed of the uncorrelated large-scale structure (LSS), $C_{\rm LSS}$, along the line of sight \citep{Schneider98}, the shape noise, $C_g$, and the error of photometric redshift, $C_s$. We also adopted the maximum likelihood estimation (MLE).

When we compute the reduced tangential shear from a small number of background galaxies, any internal substructures or surrounding clusters might accidentally affect mass measurements. In such cases, we can adopt the adaptive choice of binning scheme \citep[e.g.,][]{Okabe16b} via the following procedure. We first fit the NFW model to 270 radial combinations of the measured $\Delta \Sigma_{+}$ profile; the innermost radii $r_{\rm in}=[0.1-0.2]\hubbleMpc$ stepped by 0.05$\hubbleMpc$, the outermost radii $r_{\rm out}=[2-3.6] \hubbleMpc$ stepped by 0.2 $\hubbleMpc$, and the number of bins $N_{\rm bin}=[4-8]$ stepped by 1. We then estimated the mean masses of the suite of $M_{\Delta}$ within the overdensities of $\Delta=2500,100,500,200,$ and vir and chose a radial bin set closest to the mean values. If the innermost galaxy is farther away than $r_{\rm in}$, the first radius of $r_{\rm in}$ is taken as its position without changing the annulus width. To minimize lensing contamination from the nearest clusters in the reduced tangential shear profile, we computed the distance between the target cluster and the nearest one from the eRASS1 cluster catalog. If the distance between two clusters is smaller than the maximum radius $r_{\rm out}$, the maximum radius is set to half this distance. It is important to note that we refrain from utilizing any external catalogs to identify neighboring clusters when aiming to achieve mass measurements in a self-consistent way.

\subsection{Weak-lensing mass reconstruction} \label{sub:massreconst}

After background selection, we pixelize the shear distortion data in a regular grid of pixels using a Gaussian kernel, $G\propto \exp[-\btheta^2/(2\sigma_g^2)]$ with ${\rm FWHM}=2\sqrt{2\ln 2}\sigma_g$. In the mass reconstruction, we assumed the WL limit as $\Delta \Sigma_\alpha\simeq \Delta \tilde{\Sigma}_\alpha$ where $\alpha=1,2$. The dimensional shear field at angular position $\btheta$ is obtained by 
\begin{eqnarray}
\Delta \tilde{\Sigma}_{\alpha}(\btheta) = 
\frac{\sum_{i} e_{\alpha ,i}G(\btheta_i-\btheta) w_{i} \langle \Sigma_{{\rm cr}}(z_{l}, z_{s,i})^{-1}\rangle ^{-1}}{2 \tilde{\mathcal R} (1+\tilde{K}) \sum_{i}  G(\btheta_i-\btheta) w_{i}}, \label{eq:massmap}
\end{eqnarray}
where $\tilde{\mathcal R}$ and $\tilde{K}$ are computed with a lensing weight added by the Gaussian kernel, that is, $w_i$ is replaced by $w_i G(\btheta_i-\btheta)$ in  Eqs. (\ref{eq:R}) and (\ref{eq:K}). Since the above Gaussian smoothing is the convolution integral, we performed it in the Fourier space at the same time as the Kaiser \& Squires (KS) inversion method \citep{1993ApJ...404..441K} to invert the pixelized shear field to the mass map. If there were blank areas due to the bright star mask in a map-making field, we filled in dummy data in the region and mask it after the mass reconstruction. 
The spatial positions of the dummy are generated from a random uniform distribution. The shape and photo-z data of the dummy are taken randomly from the real data and the ellipticities were then randomly rotated. We estimated the mass reconstruction errors by the bootstrapping method. We fixed the positions of background galaxies and randomly rotated an ellipticity component taken from other galaxies. This was repeated for 500 realizations and the error is estimated as the standard error of the mock mass maps. The S/N for the mass map is obtained by dividing the reconstructed mass maps by the noise maps. As for the stacked mass maps (Sects. \ref{subsec:massmap_result} and \ref{sec:2DhaloE}), we first combined the background shape catalogs, converting the celestial positions and ellipticities to those measured at the reference coordinates \citep{2013ApJ...769L..35O,2019PASJ...71...79O}, and then we ran the KS inversion method. We adopted FWHM=$400\,h_{70}^{-1}{\rm kpc}$ and $1'$ for the individual and stacked map making, respectively.

\subsection{2D WL analysis}

The mass modeling using the 2D shear pattern enables us to parameterize both the central positions and the mass parameters of clusters \citep{2010MNRAS.405.2215O}. 
The miscentering effect is one of the important parameters for the simultaneous analysis of ensemble samples of clusters \citep{2022A&A...661A..11C,Chiu25, 2024A&A...687A.178G}. For this purpose, we computed the shear grids every 1.5 arcmin in a square box of half-size 3 $h_{70}^{-1}$Mpc, centering the eRASS1 positions. The ensemble averages of the ellipticity components and positions are calculated similarly to  Eqs. \ref{eq:g+} and \ref{eq:whm_radius}, respectively. 
The log-likelihood is expressed as
\begin{eqnarray}
-2\ln {\mathcal L}&=&\ln(\det(C_{\alpha\beta,nm})) +  \label{eq:likelihood2} \\
 &&\hspace{-3em}\sum_{\alpha,\beta=1}^{2}\sum_{n,m}(\Delta \Sigma_{\alpha,n} - f_{{\rm model},\alpha}(r_n))C_{\alpha\beta,nm}^{-1} (\Delta
 \Sigma_{\alpha,m} - f_{{\rm model},\beta}(r_m)), \nonumber
\end{eqnarray}
where the indices $\alpha$ and $\beta$ are the two components of shear distortion. We considered shape noise only in the covariance matrix to reduce the computation time. 
We employed three distinct mass models: the spherical NFW model, an elliptical NFW mass model, and a multi-component model of the spherical NFW model \citep{2011ApJ...741..116O}. The spherical NFW model is the same as that used in the 1D WL analysis. 

The elliptical NFW model \citep[see also][for fast computation]{2021PASP..133g4504O} takes into account the ellipticity of the mass distribution on the sky plane \citep{2010MNRAS.405.2215O,2002A&A...390..821G}. The 2D halo ellipticity, $\varepsilon$, is defined by $1-b/a$, where $a$ and $b$ are the major and minor axes of the mass distribution on the sky plane, respectively. The distances from the centers to the iso-contours are calculated by
\begin{eqnarray}
    r&=&(x'^2/(1-\varepsilon)+(1-\varepsilon)y'^2)^{1/2}, \nonumber \\
     x'&=&x\cos \phi_e + y \sin \phi_e, \label{eq:elliptical}\\
     y'&=&-x\sin \phi_e +y\cos \phi_e. \nonumber 
\end{eqnarray}
Here, $\phi_e$ is an orientation angle of the major axis measured from the north to the east. 

If there are multiple clusters in the data field, the center of a single NFW model might be affected. 
In this case, we consider the multi-component spherical NFW model. 
If the mass map has a peak of more than $4\sigma$, which is the FWHM away from the eRASS1 centroids, we would add mass components other than the target clusters at its peak-finding position to investigate how much the WL-determined central positions are changed.
We assumed the mass-concentration relation \citep{2015ApJ...799..108D} for clusters that are not being specifically targeted.

We used the Markov chain Monte Carlo (MCMC) method and treated the logarithmic quantities for $M_{200}$ and $c_{200}$ as parameters because they are positive. We used flat priors for all the parameters; $|{\rm R.A.}-{\rm R.A.}_{\rm eRASS1}|<3\,[h_{70}^{-1}{\rm Mpc}]/D_l$, $|{\rm DEC}-{\rm DEC}_{\rm eRASS1}|<3\,[h_{70}^{-1}{\rm Mpc}]/D_l$,
$\ln(0.1) < \ln ( M_{200} /10^{14} h_{70}^{-1}M_\odot) <  \ln 50$, and $\ln 1 < \ln c_{200} < \ln 30$. 
Regarding halo ellipticity and orientation angle, we used $-1<e<1$ and $0<\phi_e < 2\pi$ to avoid an artificial boundary around $e=0$ and $\phi_e=0,\pi$ and then used absolute values for $e$ and superimposed a double-peak posterior distribution for $\phi_e$ in the estimation. All parameters were estimated using a central bi-weight estimation from marginalized posterior distributions to down-weight outliers in skewed distributions.

\section{Results} \label{sec:result}

\subsection{Cluster sample} \label{subsec:clustersample}

The X-ray count-rate (CR) was estimated in the soft-band (0.2-2.3 keV) with correction of Galactic absorption \citep{2024A&A...685A.106B}. The CR of the eRASS1-HSC clusters has a weak dependence on redshift (the left panel of Fig. \ref{fig:cluster_sample}), as expected by the CR selection function. The average and median CRs are $0.91_{-0.14}^{+0.19}$ and $0.56$, respectively. When we remove the first two with the highest CR, the average drops to $0.66$. The richness is measured by an adapted version of the \texttt{redMaPPer} \citep{2014ApJ...785..104R} algorithm \citep{2024A&A...688A.210K}. The lower boundary of the cluster richness $\lambda$ (the right panel of Fig. \ref{fig:cluster_sample}) at high redshift ($z\simgt 0.6$) is higher than at low redshift ($z\simlt 0.4$). This indicates that less massive clusters are harder to detect at higher redshifts \citep{2024A&A...687A.238C}. We effectively adopted the detection threshold $\lambda=3$ at $z<0.45$ and $30$ at $z>0.6$ and a linear function between the two redshifts to consider the selection function in the mass-richness scaling relation.

\begin{figure}[!ht]
    \centering
    \includegraphics[width=\hsize]{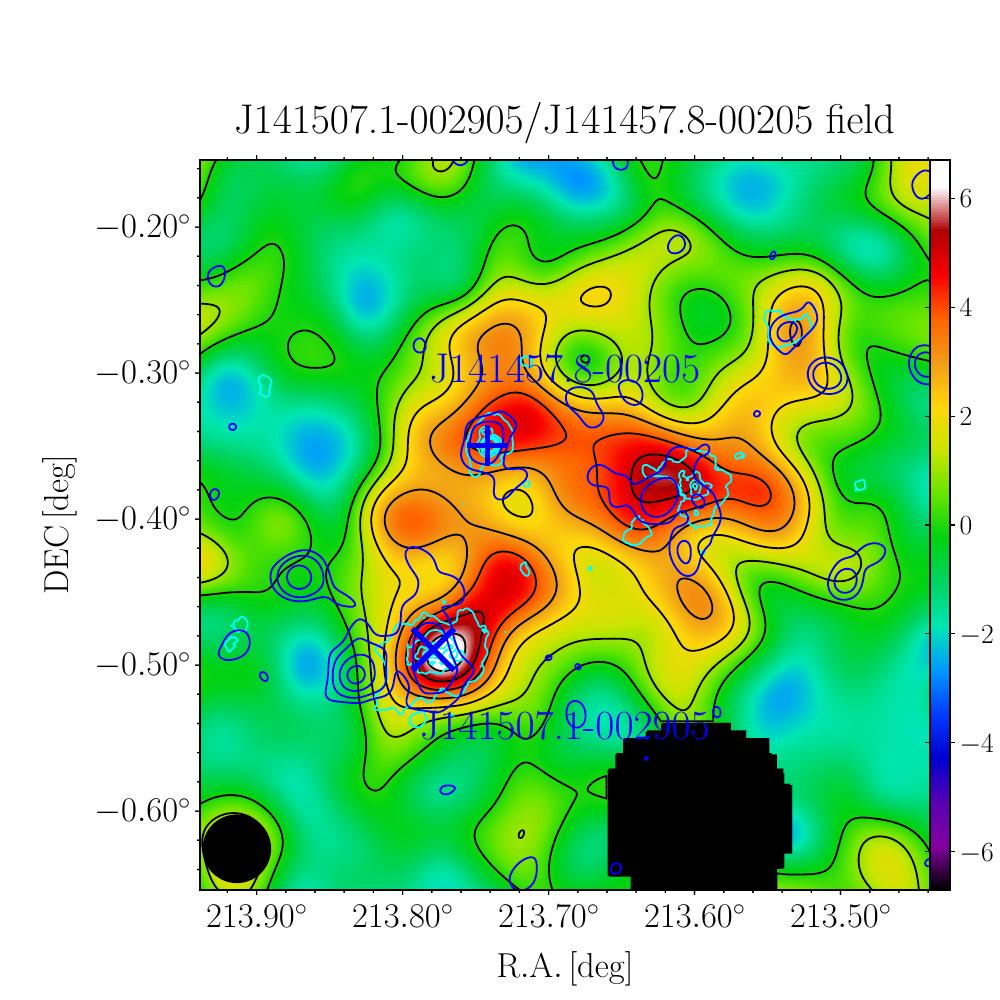}
    \caption{Mass map for the J141507.1-002905 and J141457.8-00205 field ($30'\times30'$). The black contours represent the reconstructed WL mass map spaced in units of 1$\sigma$ bootstrapping error.
    The diameter of the black circle in the lower left corner represents the Gaussian smoothing FWHM=$400$ kpc in the mass reconstruction. Blacked-out areas are the masked regions of bright stars.  The blue $+$ and $\times$ are positions of J141507.1-002905 and J141457.8-00205, respectively.
    The system is MCXC J1415.2-0030 \citep{2011A&A...534A.109P}.
    The blue and cyan contours are the eRASS1 and XMM-Newton X-ray contours \citep{2018PASJ...70S..22M}, respectively. An X-ray bright point source in the western component is removed in the XMM-Newton contours. It was properly identified when the eRASS1 cluster catalog was constructed. At the same time, diffuse X-ray emission in the XMM-Newton image appeared even after removing the X-ray point source thanks to a higher angular resolution.
  The system is composed of three components with ${\rm S/N}\sim 6.7\sigma$ for J141457.8-00205, $\sim 4.7\sigma$ for J141507.1-002905, and $\sim 5.3\sigma$ for the western component.  }
    \label{fig:massmap_J141457.8-002050}
\end{figure}

We matched the eRASS1 clusters in the HSC footprint to existing cluster catalogs \citep{1958ApJS....3..211A,1961cgcg.book.....Z,2009ApJS..183..197W,2018PASJ...70S..20O,2007ApJ...660..239K,2009AJ....137.2981G,2002AJ....123.1807G,2007MNRAS.379..867V,2007A&A...470..821D,2004A&A...423..449P,2012MNRAS.423.1024M,2021ApJS..253....3H}, as summarized in Table \ref{table:cog1}. The tolerance of cross-matching is $3$ arcmin for the spatial separation and $|z-z_{\rm eRASS1}|<0.05$. If there are multiple associations of optically selected clusters, the priority is given to putting WHL clusters \citep{2009ApJS..183..197W} or HSC CAMIRA clusters \citep{2018PASJ...70S..20O}. From the well-known cluster catalogs, we found 12 Abell clusters \citep{1958ApJS....3..211A}, 2 Zwicky clusters \citep{1961cgcg.book.....Z}, 2 X-ray clusters, and 5 SZ clusters \citep{2021ApJS..253....3H} in the sample. 
As one of the samples, Fig. \ref{fig:massmap_J141457.8-002050} shows the mass map overlaid with X-ray contours from eRASS1 and XMM-Newton data for J141507.1-002905 and J141457.8-00205 field, which is known as  MCXC J1415.2-0030 \citep{2011A&A...534A.109P}.

   \begin{figure*}[!ht]
   \centering
   \includegraphics[width=19cm]{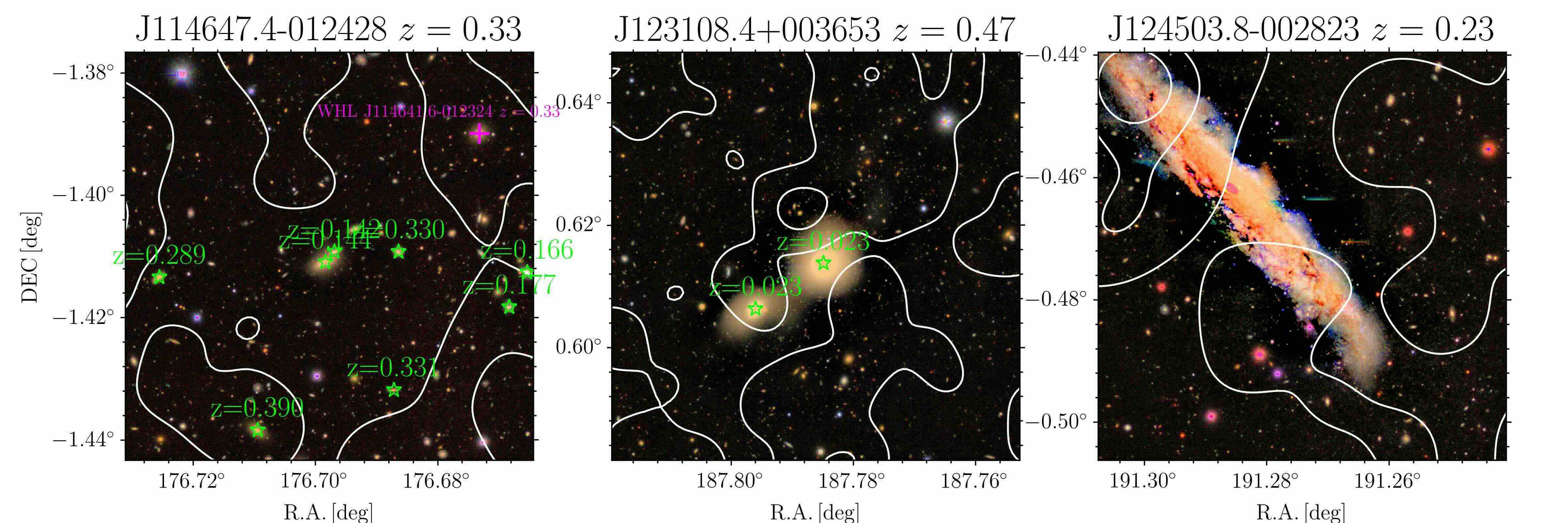}
   \caption{HSC images for misassociation clusters. Overlaid with white contours representing red galaxy distribution, stepped by two galaxies per each pixel above two galaxies. }
              \label{fig:junk}
    \end{figure*}

We carried out a visual inspection of the purity using HSC images, publicly available spectroscopic redshifts, and the spatial distribution of red-sequence galaxies.
It would be difficult to do this with each of the over 5,000 clusters of the primary sample \citep{2024A&A...689A.298G}, but it is practically possible in the current HSC sample. The HSC images were extracted with $4'\times 4'$ centered on the centroid of the eRASS1 cluster. We overlaid spectroscopic redshifts from the publicly available catalogs \citep{2018MNRAS.474.3875B,2019ApJS..240...23A,2009ApJS..184..218L,2003astro.ph..6581C} and monitored the difference with the eRASS1 redshifts.  We constructed Gaussian smoothing maps (FWHM$=0.2\hubbleMpc$) of the number distribution of red sequence galaxies selected in the color-magnitude plane following \cite{Nishizawa18,2019PASJ...71...79O}. The apparent z-band magnitudes of the red-sequence galaxies are brighter than the observer-frame magnitude with the constant z-band absolute magnitude of $M_z=-18$ ABmag and the limiting apparent magnitude $m_z=25.5$. We employed K-correction to convert between apparent and absolute magnitudes, considering passive evolution. Henceforth, clusters possessing and lacking optical counterparts are referred to as association clusters and misassociation clusters, respectively.

Based on the spectroscopic redshifts, we found three misassociation clusters: J114647.4-012428, J123108.4+003653, and J124503.8-002823 (Fig. \ref{fig:junk}). 
Although we examined the galaxy overdensities sliced by each redshift, we could not find any significant excess, as shown by the white contours.
The X-ray centroid of J114647.4-012428 (left panel, $z=0.33$) is associated with two ellipticals at $z\sim 0.14$. 
WHL J1146416-012324, with a close redshift of $z=0.3318,$ is 1.9 arcmin away from J114647.4-012428.
The X-ray emission of J123108.4+003653 (middle panel, $z=0.47$) is associated with two galaxies identified as MCXC J1231.0+0037 at $z\sim 0.023$. WHL J123122.9+003718 at $z=0.4354$ is $3.6$ arcmin away from J123108.4+003653 outside the HSC image (Fig. \ref{fig:junk}).
The X-ray emission from J124503.8-002823 (right panel, $z=0.23$)  apparently comes from a nearby spiral galaxy, NGC 4666.
WHL J124454.5-002640 at $z=0.231$ is 3.3 arcmin away from J124503.8-002823 and its galaxy concentration is found.

We note that the fraction of the misassociation clusters in the HSC-SSP field, $3/78\sim4\%$, is consistent with the $\sim95\%$ purity of the primary clusters \citep{2024A&A...687A.238C,2024A&A...688A.210K}.

\subsection{Weak-lensing mass measurements}\label{subsec:M_WL}

We first calculated the signal-to-noise ratio (S/N) of the reduced tangential shear profile by $(\sum_{ij} \Delta \Sigma_{+,i} C_{ij}^{-1}\Delta\Sigma_{+,j})^{1/2}$, where the indexes $i$ and $j$ denote the $i$-th and $j$-th radial bins, respectively. The average and median S/N are $4.5$ and $4.3$, respectively. In the CR and $z$ plane (left panel in \ref{fig:cluster_sample}), the S/N is higher for clusters with higher CR around $z\sim0.1-0.4$, because the clusters have good lensing efficiency and a large number of background galaxies. 
The  clearer feature is found in the richness and $z$ plane (right panel of Fig. \ref{fig:cluster_sample}). The S/N tends to be higher for the clusters in the upper left corner of the figure.

The individual WL masses for the 72 association clusters are shown in Table \ref{table:cog1}. Among all the 78 clusters, we cannot measure WL masses of the three association clusters and the three misassociation clusters. 
The number of non-measurable clusters in the 75 association clusters is consistent with the expectation computed by mock catalogs using a realistic number density of background galaxies (see Appendix \ref{app2} and Fig. \ref{fig:z_vs_M_WLcal}). The measurement uncertainties for nine association clusters are too large. Thus, only an upper limit on the WL masses can be constrained. We refer to the 12 association clusters above, whose WL masses at $\Delta=500$ are difficult or impossible to measure, as poor-fit clusters (Table \ref{table:cog1}). 
The WL masses as a function of the redshift are shown in Fig. \ref{fig:z_vs_M500}. With increasing redshifts, the WL masses tend to be larger and at lower redshifts ($z\simlt 0.2$), they span a broad range of values. The modeled distribution of the WL masses in the parent population obtained via a scaling relation analysis (see the details in Sec \ref{subsec:scaling}) follows this trend.

\subsection{Stacked tangential shear profiles}

\begin{figure}[!ht]
   \centering
   \includegraphics[width=9cm]{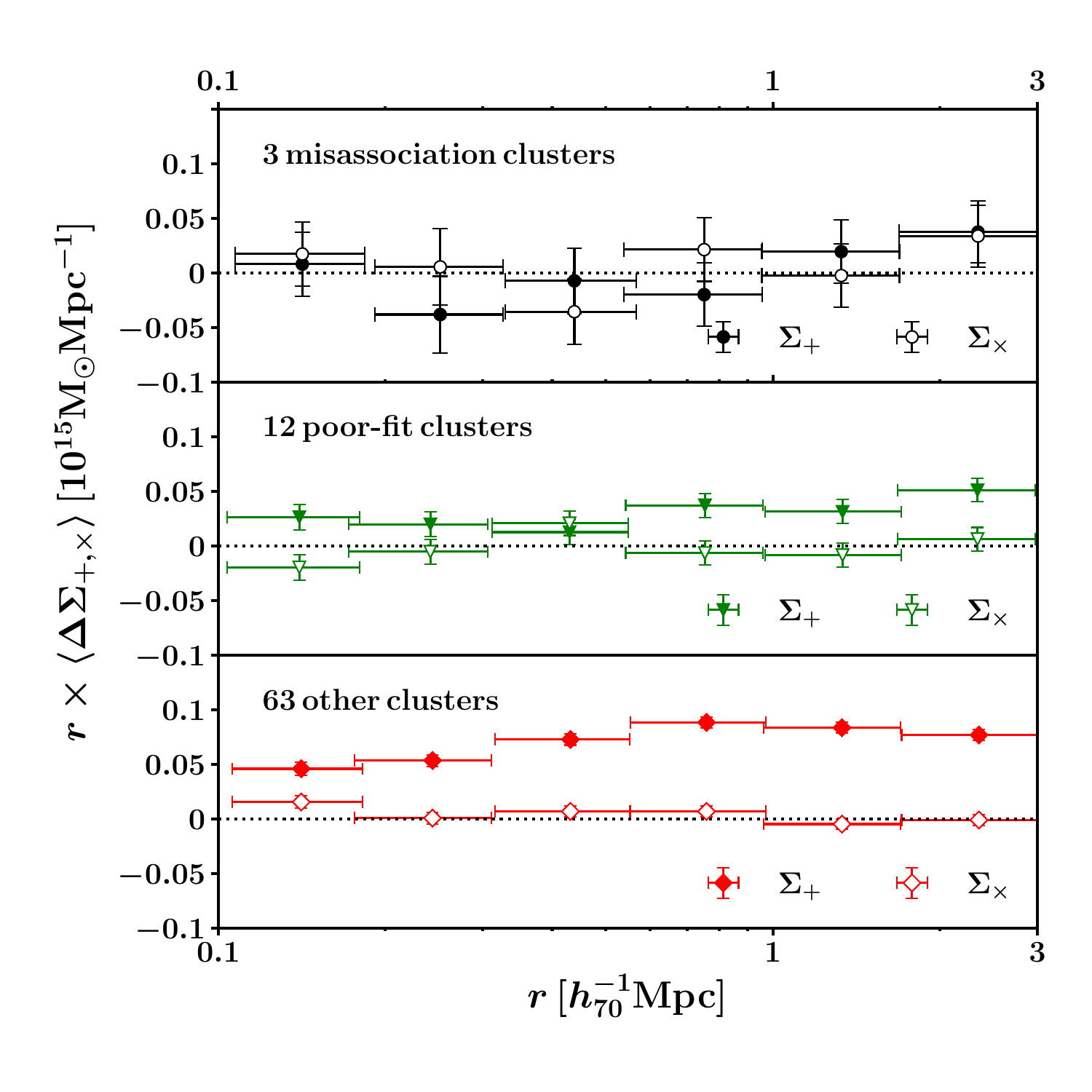}
      \caption{ Stacked profiles for the three misassociation clusters (top), the 12 poor-fit clusters (middle), and the 63 other clusters (bottom). The $y$-axis represents $r\times \langle \Sigma_+ \rangle$ (filled colors) or $r\times \langle \Sigma_\times \rangle$ (open colors).  The $+$ components for the 63 other clusters and the 12 poor-fit clusters are higher than the $\times$ components, while both the $+$ and $\times$ components for the 3 misassociation clusters are consistent with the null. }
         \label{fig:g+_stack}
   \end{figure}

We computed the stacked tangential shear profiles for the 3 misassociation clusters, the 12 poor-fit clusters, and 63 other association clusters (Fig. \ref{fig:g+_stack}). The number of bins was set to six because of the small number of background galaxies for the three misassociation clusters. To visualize the null signal for the misassociation clusters, we used $r$ times the tangential shear, $\langle \Sigma_+\rangle$, or the 45-degree rotated component, $\langle \Sigma_\times \rangle$, for the $y$ axis.
The tangential shear component for the misassociation clusters is comparable to the 45-degree rotated component and consistent with the null, supporting the visual inspection result (Sec. \ref{subsec:clustersample}). 
In contrast, the lensing signals for the poor-fit and remaining 63 clusters are significantly detected with S/N$=7.0$ and $34.0$, respectively.

\begin{figure}[!ht]
   \centering
   \includegraphics[width=9cm]{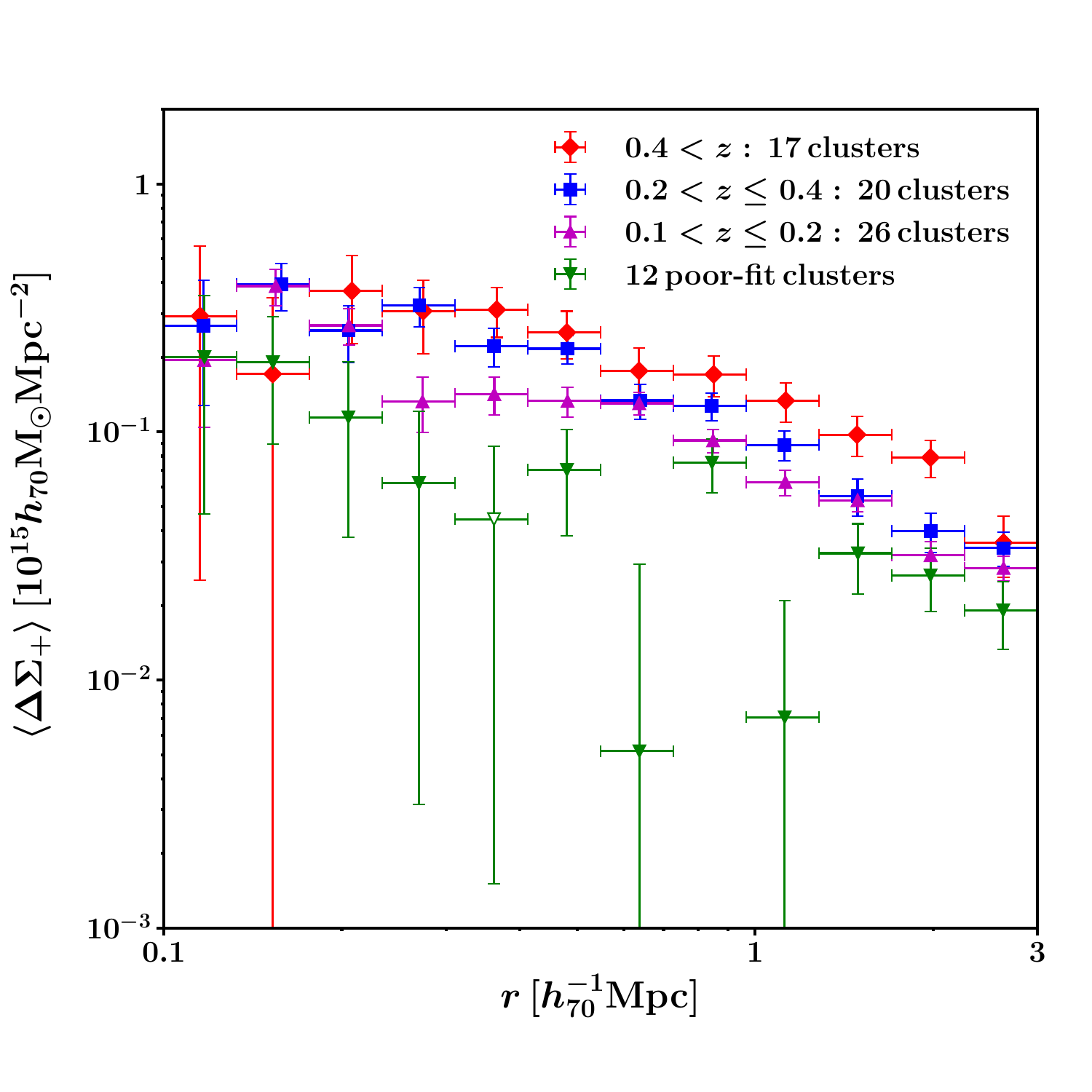}
      \caption{Stacked tangential profiles for 3 subsamples divided by redshift ranges of $0.4<z$ (red diamonds), $0.2<z\le 0.4$ (blue squares), and $0.1<z\le 0.2$ (magenta upward triangles), and the 12 poor-fit clusters (green downward triangles). The filled and open symbols denote positive and negative values, respectively. }
         \label{fig:g+_stack_log}
   \end{figure}

We divided the remaining 63 clusters into three samples by redshift: $0.4<z$, $0.2<z \leq 0.4$, and $0.1<z\leq 0.2$ (Fig. \ref{fig:g+_stack_log}).
The S/Ns are $12.9$, $19.1$, and $23.6$ in the order of higher redshift, respectively. 
The tangential shear profile distinctly exhibits curvature.
The tangential shear profile for the poor-fit clusters shows a depletion in an intermediate radius of $0.2\simlt r \simlt 1$ $\hubbleMpc$ (Sect. \ref{subsec:poorfit}).

\subsection{Mass-richness-CR relation} \label{subsec:scaling}

\begin{figure*}[!ht]
   \centering
   \includegraphics[width=8cm]{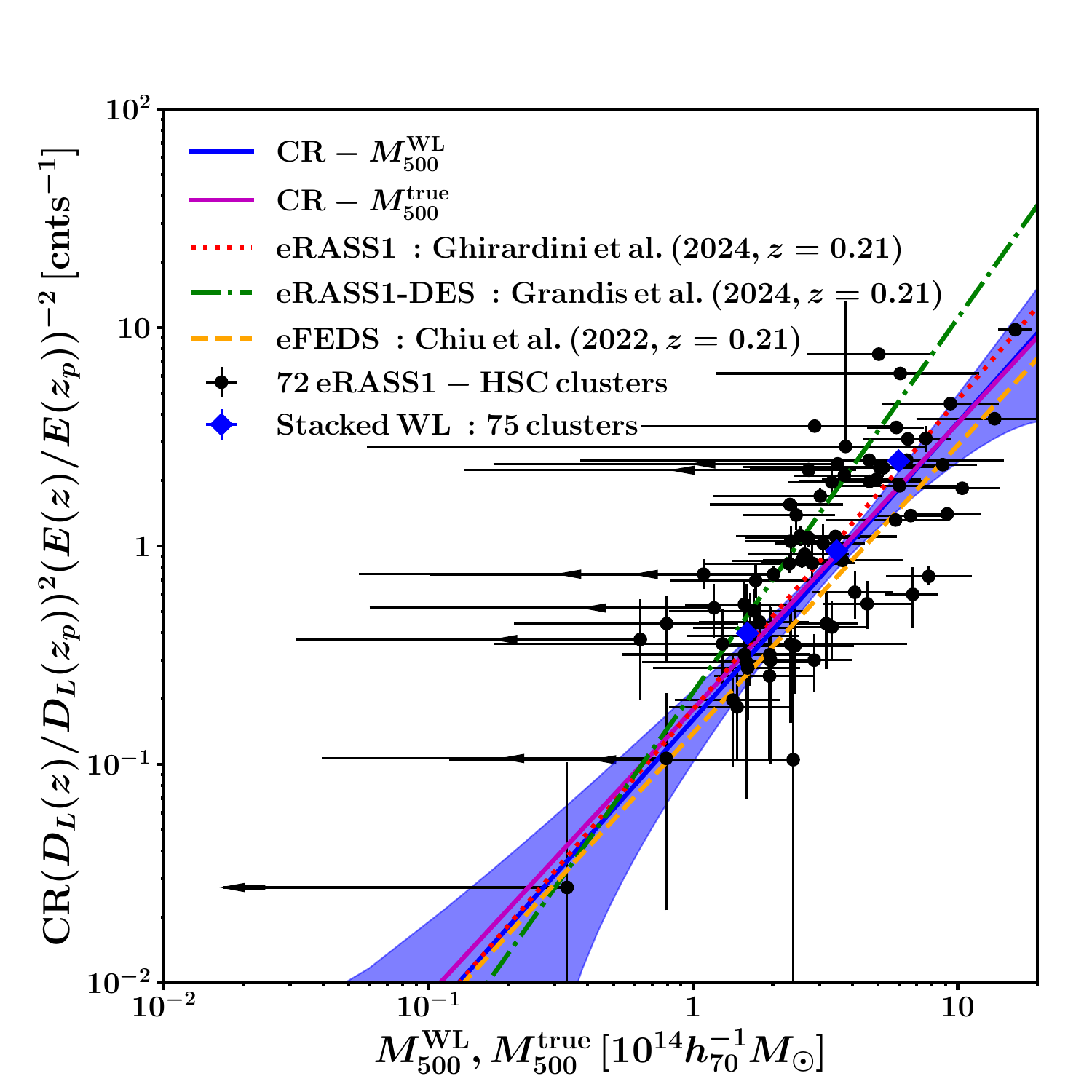}
   \includegraphics[width=8cm]{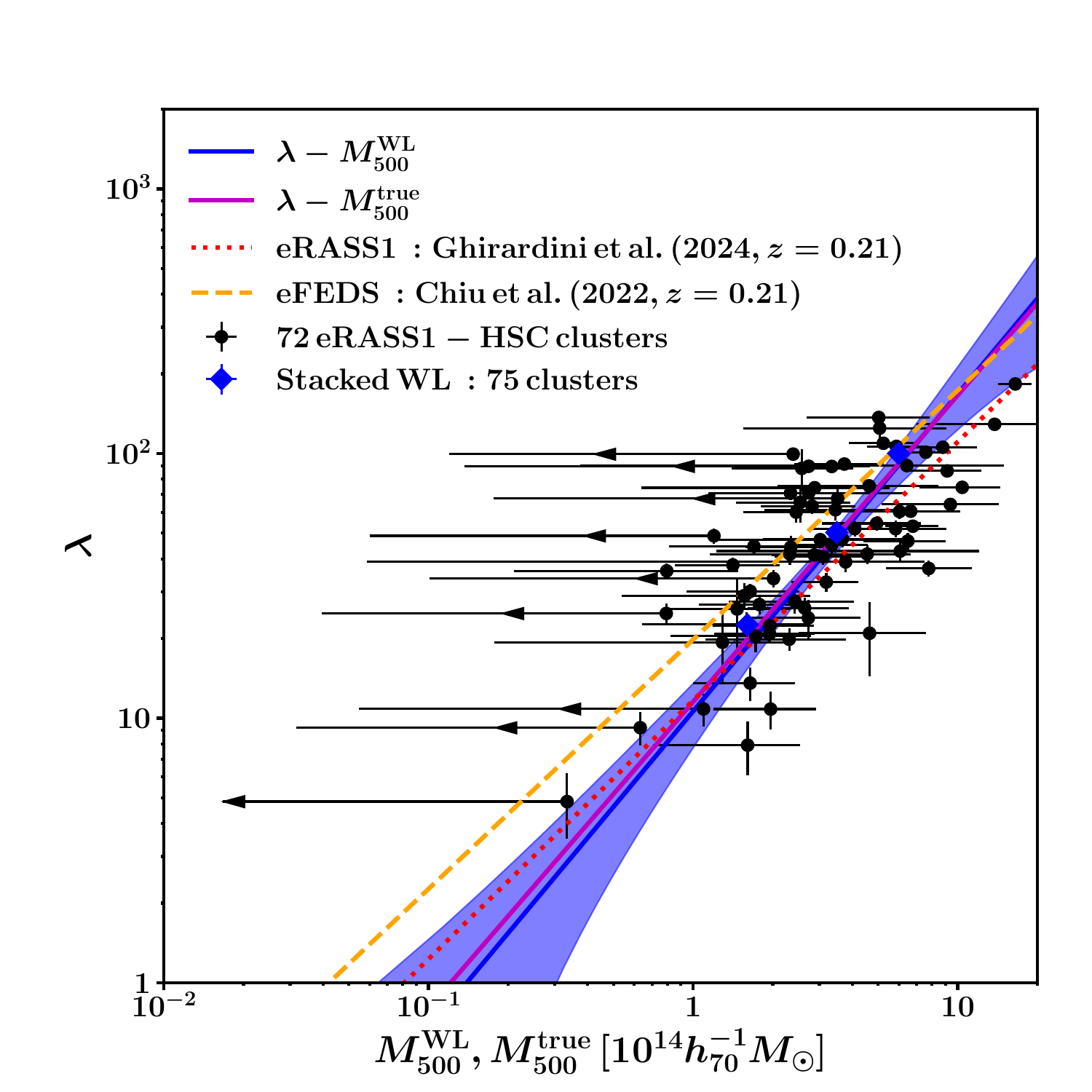}
      \caption{Scaling relations between the count-rate (left) the richness (right) and the WL ($M_{500}^{\rm WL}$) and true masses ($M_{500}$). 
      The black circles and blue diamonds denote the individual WL masses of the 72 association clusters and the stacked WL masses for the three sub-samples divided by richness from the 75 association clusters, respectively. The upper bounds of the WL masses and the associated arrows are used for the poor-fit clusters. The blue and magenta solid lines are the best-fit scaling relation with respect to the WL masses and the true masses ($\delta_{\rm CR}=\delta_{\lambda}=r_{\rm CR,WL}=r_{\lambda,{\rm WL}}=0$), respectively. The blue region is the $1\sigma$ uncertainty of the scaling relation with the WL masses. The red dotted, green dot-dashed, and orange dashed lines are the results of eRASS1 \citep{2024A&A...689A.298G}, eRASS-DES \citep{2024A&A...687A.178G}, and eFEDS \citep{2022A&A...661A..11C}, respectively.  }
         \label{fig:lambda-Mwl}
   \end{figure*}

We computed the scaling relations between the extinction corrected CR and the WL masses and between the richness of the cluster and the WL masses of the 72 individual clusters, excluding 3 clusters with no mass measurements (Fig. \ref{fig:lambda-Mwl}). The CR is multiplied by the square of the luminosity distance, $D_L(z)$,  and the inverse square of $E(z)=\sqrt{\Omega_{m,0}(1+z)^3+\Omega_{\Lambda,0}}$ to correct the redshift evolution \citep{2022A&A...661A..11C,2024A&A...689A.298G}, specified as
\begin{eqnarray}
    CR\left(\frac{D_L(z)}{D_L(z_{p})}\right)^2\left(\frac{E(z)}{E(z_{p})}\right)^{-2} \label{eq:cr_wgt},
\end{eqnarray}
where $z_{p}$ is a pivot redshift $0.21$ as a lensing weight average from stacked lensing analysis. 
For the logarithmic quantity, we express the correction term as 
\begin{eqnarray}
     E_x(z)&=&-2 \ln (D_L(z)/D_L(z_{p})) + 2 \ln (E(z)/E(z_{p}) ).\nonumber 
\end{eqnarray}

Both the corrected CR and the richness are highly correlated with the WL masses (Fig. 
\ref{fig:lambda-Mwl}). To quantify the relationships between mass, CR, and richness, we adopt the trivariate scaling relations, including the relation between the WL masses ($M_{500}^{\rm WL}$), and the true masses ($M_{500}^{\rm true}$), the so-called WL mass calibration. In general, the WL mass calibration consists of two effects: astrophysical (intrinsic) properties and observational conditions. The former is mainly due to the orientation of elliptical halos, the presence of subhalos, and their surroundings \citep{2010A&A...519A..90M,2011ApJ...740...25B,2012MNRAS.426.1558G,2024A&A...681A..67E}. 
The adaptive choice of radial bins could reduce these effects as much as possible \citep{Okabe16}. 
The latter is associated with the number of background galaxies.
When combining the gravitational lensing effect with the intrinsic ellipticity of a small number of background galaxies, the resulting S/N exhibits random variation and the measurable masses are restricted to those exhibiting a notably high bias.
Since the number density of our background galaxies strongly depends on cluster redshifts (Figs. \ref{fig:z_vs_M500} and \ref{fig:z_vs_M_WLcal}), we cannot rule out such a selection bias in successful individual WL mass measurements. 
To quantify selection bias, we performed mock simulations empirically representing observational conditions using a spherical NFW model (see details in Sec. \ref{app2}).
We made 9000 mock shape catalogs and repeated our analysis without lensing effect from large-scale structure. The measurable fraction decreases as the cluster redshift increases and the mass decreases, so the measured WL mass has a higher positive bias as the cluster redshift increases and the mass decreases. With a selection function, the WL mass calibration is well described by a redshift-dependent linear equation of the logarithmic quantities (Eq. \ref{eq:mwl-m}) and is used as a prior in a regression analysis. We note that our WL mass calibration based on two parameters is different from the one-parameter case \citep{2022A&A...661A..11C,2024A&A...687A.178G}.

We used a hierarchical Bayesian regression method \citep{2022PASJ...74..175A,2016MNRAS.455.2149S} to infer the true mass distribution of the parent population, taking into account the selection effect. We assume that the logarithm of the true mass follows a Gaussian distribution ${\mathcal N}(\mu(z),\sigma(z))$ of which mean ($\mu$) and standard error ($\sigma$) are a polynomial function of $F(z)=\ln((1+z)/(1+z_p))$. We used the second-order and first-order polynomial functions for the mean and standard error, respectively. As shown in \cite{2022PASJ...74..175A}, the model realizes well the input of multivariate scaling relations in samples of $\sim{\mathcal O}(100)$ clusters, regardless of the shape of the parent population, such as the Gaussian distribution and the cluster mass function.
Its advantage is that it is independent of the cluster mass function or cosmological parameters. 
We adopted the lower limits for the corrected CR and the richness as $0.01$ and the line in Fig. \ref{fig:cluster_sample}, respectively. In the regression analysis, we use the following equations \citep[see also][]{2024A&A...689A.298G,2024A&A...687A.178G}:
\begin{eqnarray}
    \ln \left(\frac{M_{500}^{\rm WL}}{M_p}\right)&=&\alpha_{\rm WL}+(\beta_{\rm WL}+\delta_{\rm WL}F(z))\ln \left(\frac{M_{500}^{\rm true}}{M_p}\right) \nonumber \\
   & & +\gamma_{\rm WL}F(z), \label{eq:mwl-m}\\
   \ln \left(\frac{CR}{CR_p}\right) -E_x(z) &=& \alpha_{\rm CR} +(\beta_{\rm CR}+\delta_{\rm CR}F(z)) \ln \left(\frac{M_{500}^{\rm true}}{M_p}\right)\nonumber \\
   & & +\gamma_{\rm CR}F(z),  \label{eq:cr-m}\\
    \ln \left(\frac{\lambda}{\lambda_p}\right)&=& \alpha_\lambda +(\beta_\lambda+\delta_{\lambda }F(z))\ln \left(\frac{M_{500}^{\rm true}}{M_p}\right)\nonumber \\
    & & +\gamma_{\lambda }F(z), \label{eq:lambda-m}
\end{eqnarray} 
where $M_p$, $CR_p$, and $\lambda_p$ are a pivot mass $10^{14}h_{70}^{-1}M_\odot$, a pivot count-rate $1\,{\rm cnt\,s}^{-1}$, and a pivot richness $40$, respectively. The normalization, the mass-dependent slope, the redshift-dependent slope in the normalization, and the redshift-dependent slope in the mass-related slope are expressed by $\alpha$, $\beta$, $\gamma$, and $\delta$, respectively. 
The deviation from the self-similar model is quantified by $\gamma$ and $\delta$ \citep[e.g.,][]{2019ApJ...871...50B, 2022A&A...661A..11C,2024A&A...689A.298G}.

The intrinsic covariance \citep{Okabe10c} is described by 
\begin{eqnarray}
    C_{\rm int}=\begin{pmatrix}
\sigma_{\rm WL}^2 &  r_{{\rm CR,WL}}\sigma_{\rm CR}\sigma_{\rm WL} & r_{\lambda,{\rm WL}}\sigma_\lambda\sigma_{\rm WL}\\
 r_{{\rm CR,WL}}\sigma_{\rm CR}\sigma_{\rm WL} & \sigma_{\rm CR}^2 & r_{{\rm CR},\lambda} \sigma_{\rm CR}\sigma_\lambda \\
 r_{\lambda,{\rm WL}}\sigma_\lambda\sigma_{\rm WL}  & r_{{\rm CR},\lambda} \sigma_{\rm CR}\sigma_\lambda & \sigma_\lambda^2 \\
\end{pmatrix},
\end{eqnarray}
where $\sigma$ and $r$ are the intrinsic scatter of the left-hand-side quantities of Eqs. (\ref{eq:mwl-m})-(\ref{eq:lambda-m}) and the intrinsic coefficient between two variables, respectively. We assume that the intrinsic covariance is independent of the redshift. We use the WL mass bias parameters as priors (Appendix. \ref{app2}), taking into account their error covariance matrix.

We first performed a regression analysis for three cases:(1) $r_{{\rm CR,WL}}=r_{\lambda,{\rm WL}}=0$, (2) $\delta=0$, and (3) all the free parameters. 
The resulting parameters are shown in Table \ref{table:scaling}. 
The intrinsic coefficients, $r_{\rm CR,WL}$ and $r_{\lambda,{\rm WL}}$, are not well constrained in the three cases and are consistent with 0.
In contrast to \citet{2024A&A...689A.298G},  the errors of redshift-dependent slopes $\delta$ and $\gamma$ are large because of our small sample size. 
We then fix $r_{\rm CR,WL}= r_{\lambda,{\rm WL}}=0$ and $\delta=0$, which is referred to as (4).

We assessed the models by employing the Akaike's information criterion (AIC) and Bayesian information criterion (BIC) to determine which model aligns most closely with reality. The deviations of the AIC and BIC of (1)-(3) from (4) give larger values with $+5\sim+22$. 
When we additionally fixed $\gamma=0$, $\Delta {\rm AIC}$ and $\Delta {\rm BIC}$ worsened even  by $+9$ and $+3$, respectively.  Therefore, we chose to focus on the result of (4), as shown in Fig. \ref{fig:lambda-Mwl}. The blue and magenta solid lines represent the best-fit scaling relations concerning the WL and true masses (Fig. \ref{fig:lambda-Mwl}), respectively.  
 
 The mass-dependent slope for CR and mass scaling relation is consistent with 1. 
 The mass-dependent slope for the richness and mass scaling relation confirms that the number of cluster galaxy members is proportional to the cluster mass. 
  Although the redshift dependencies of the normalization are not well constrained, the normalization for the CR and the richness are likely to increase and decrease as the redshift increases, respectively.
   The intrinsic scatter for the two scaling relations is $\sim$30\%. We find that the intrinsic coefficient between the CR and richness is consistent with zero.

Stacked lensing results are indicated by blue diamonds, regardless of the success of individual mass measurements. The three sub-samples are divided by the richness with $\lambda\leq 40$, $40<\lambda<70$, and $70\leq \lambda$. Since the number of background galaxies and the lensing efficiency for individual clusters are different from each other, we computed the average values with a lensing weight of $w_{\rm lens,i}=\sum_{i} \Sigma_{\rm cr}^{-1}(z_{l,j},z_{s,i})/\sum_{i,j} \Sigma_{\rm cr}^{-1}(z_{l,j},z_{s,i})$, where $i$ and $j$ are the indexes for the clusters and their background galaxies, respectively.  The stacked quantities coincide with the baseline with the WL mass.

The Bayesian analysis also gives us the parent population of the true mass $M_{500}^{\rm true}$, which leads to the $M_{500}^{\rm WL}$ distribution of the parent population.  As shown in Fig. \ref{fig:z_vs_M500}, the $M_{500}^{\rm WL}$ distribution represents well the measured $M_{500}^{\rm WL}$ of the observed sample. When we change the order of the polynomial function to either $1$ or $3$, the result does not change. When we use the 2nd order polynomial function for the standard error, both the AIC and BIC get worse by $+1$ and $+4$, respectively. The current model is thus sufficient to describe the sample.

We artificially change the intrinsic scatter of the WL mass calibration to $\sigma_{\rm WL}=0.214$ calibrated with synthetic weak-lensing observations \citep{2020ApJ...890..148U} of 639 cluster halos in a dark-matter-only realization of BAHAMAS simulations \citep{2017MNRAS.465.2936M} at a single redshift $z=0.25$. 
The analysis was repeated with just $\sigma_{\rm WL}$ adjusted to this redshift-independent value, while keeping the other parameters the same. We find that $\gamma$ and $\sigma_{\rm int}$ for the two scaling relations change by $\sim+20\%$ and $\sim-10\%$, respectively.

Since our WL mass measurements do not use informative priors, it is easy to find out the outliers, such as the poor-fit clusters. When we remove the 12 poor-fit clusters, we find that the baseline parameters do not change significantly and the intrinsic scatter of the CR and the richness become $\sim 8\%$ lower, though they are consistent with each other within $1\sigma$.

\begin{table*}[!ht]
\caption{Best-fit parameters for the mass-richness-CR relation.} \label{table:scaling}
\centering   
\begin{tabular}{c|cccc}
& {\bf (4)} & (1) & (2) & (3) \\
\hline 
 $\alpha_{\rm CR}$ & $-1.722_{-0.372}^{+0.318}$ 
                     & $-1.812_{-0.426}^{+0.357}$ 
                     & $-1.400_{-0.345}^{+0.241}$ 
                     & $-1.785_{-0.437}^{+0.341}$ \\
 $\beta_{\rm CR}$  & $1.310_{-0.277}^{+0.294}$ 
                     & $1.388_{-0.302}^{+0.354}$ 
                     & $0.995_{-0.212}^{+0.278}$ 
                     & $1.361_{-0.301}^{+0.304}$ \\
 $\gamma_{\rm CR}$  & $1.984_{-1.148}^{+1.136}$ 
                      & $2.570_{-3.554}^{+2.638}$ 
                      & $3.374_{-1.116}^{+1.036}$ 
                      & $0.139_{-0.995}^{+1.145}$ \\
 $\delta_{\rm CR}$  & 0 (fixed) 
                      & $-0.268_{-2.043}^{+2.721}$ 
                      & 0 (fixed) 
                      & $1.196_{-0.894}^{+1.095}$ \\
 $\alpha_{\lambda}$ & $-1.165_{-0.244}^{+0.252}$ 
                     & $-1.283_{-0.300}^{+0.381}$ 
                     & $-1.220_{-0.237}^{+0.312}$ 
                     & $-1.176_{-0.270}^{+0.224}$ \\
 $\beta_{\lambda}$  & $1.066_{-0.216}^{+0.183}$ 
                     & $1.181_{-0.265}^{+0.220}$ 
                     & $1.107_{-0.210}^{+0.163}$ 
                     & $1.113_{-0.190}^{+0.193}$ \\
 $\gamma_{\lambda}$  & $-0.368_{-1.012}^{+0.905}$ 
                      & $-2.075_{-3.150}^{+4.637}$ 
                      & $0.074_{-1.003}^{+0.942}$ 
                      & $-0.068_{-1.185}^{+1.252}$ \\
 $\delta_{\lambda}$  & 0 (fixed) 
                      & $0.789_{-2.534}^{+2.402}$ 
                      & 0 (fixed) 
                      & $-0.610_{-0.884}^{+1.498}$ \\
 $\sigma_{\rm CR}$  & $0.353_{-0.135}^{+0.094}$ 
                      & $0.380_{-0.147}^{+0.109}$ 
                      & $0.443_{-0.090}^{+0.076}$ 
                      & $0.320_{-0.184}^{+0.109}$ \\
 $\sigma_{\lambda}$  & $0.399_{-0.128}^{+0.096}$ 
                       & $0.434_{-0.151}^{+0.155}$ 
                       & $0.249_{-0.162}^{+0.120}$ 
                       & $0.401_{-0.101}^{+0.096}$ \\
 $r_{{\rm CR,WL}}$  & $0$ (fixed)
                     & $0$ (fixed)
                     & $0.129_{-0.587}^{+0.459}$ 
                     & $0.193_{-0.639}^{+0.465}$ \\
 $r_{\lambda,{\rm WL}}$  & $0$ (fixed)
                           & $0$ (fixed)
                           & $-0.476_{-0.278}^{+0.681}$ 
                           & $0.038_{-0.576}^{+0.517}$ \\
 $r_{{\rm CR},\lambda}$  & $-0.036_{-0.501}^{+0.398}$
                           & $0.085_{-0.546}^{+0.404}$
                           & $-0.232_{-0.410}^{+0.409}$
                           & $-0.140_{-0.423}^{+0.400}$ \\
                           \hline
 $\Delta$AIC   & $0$ 
               & $+4.9$
               & $+6.1$
               & $+10.2$ \\
 $\Delta$BIC   & $0$ 
               & $+10.9$
               & $+12.0$
               & $+22.0$ \\
\end{tabular}
\end{table*}

\subsection{Mass-concentration relation}

Numerical simulations \citep[e.g.,][]{Bullock01,Bhattacharya13,Child18,2019ApJ...871..168D,2021MNRAS.506.4210I} predict that the halo concentration for the NFW mass model increases with decreasing halo mass and redshift.
Such an anti-correlation between mass and concentration can be explained by hierarchical structure formation.
The progenitors of less massive halos form first and their characteristic central mass density is reflected by a critical density at higher redshifts. More massive halos form later, with lower mass densities than the less massive halos, and grow by mass accretion and mergers of smaller objects. 
Some simulations show that the concentration turns upward for the most massive halos \citep[e.g.,][]{2012MNRAS.423.3018P,2016MNRAS.457.4340K}. This upturn feature is still controversial \citep[e.g.,][]{2012MNRAS.423.3018P,2015ApJ...799..108D,2016MNRAS.457.4340K,Child18}.
The concentration also depends on the dynamical state, that is, relaxed halos have a higher concentration than unrelaxed halos \citep[e.g.,][]{2014MNRAS.441.3359D,2016MNRAS.457.4340K,2021MNRAS.506.4210I}.

The mass and concentration give us a unique opportunity to test how structures form at cluster scales.
Thanks to X-ray emissivity, X-ray centroid determination does not suffer from projection effects as in the case of optically selected clusters \citep[e.g.,][]{Okabe16b,2016ApJ...821..116U,2019PASJ...71...79O,2020ApJ...890..148U}.

 \begin{figure*}[!ht]
   \centering
   \includegraphics[width=\linewidth]{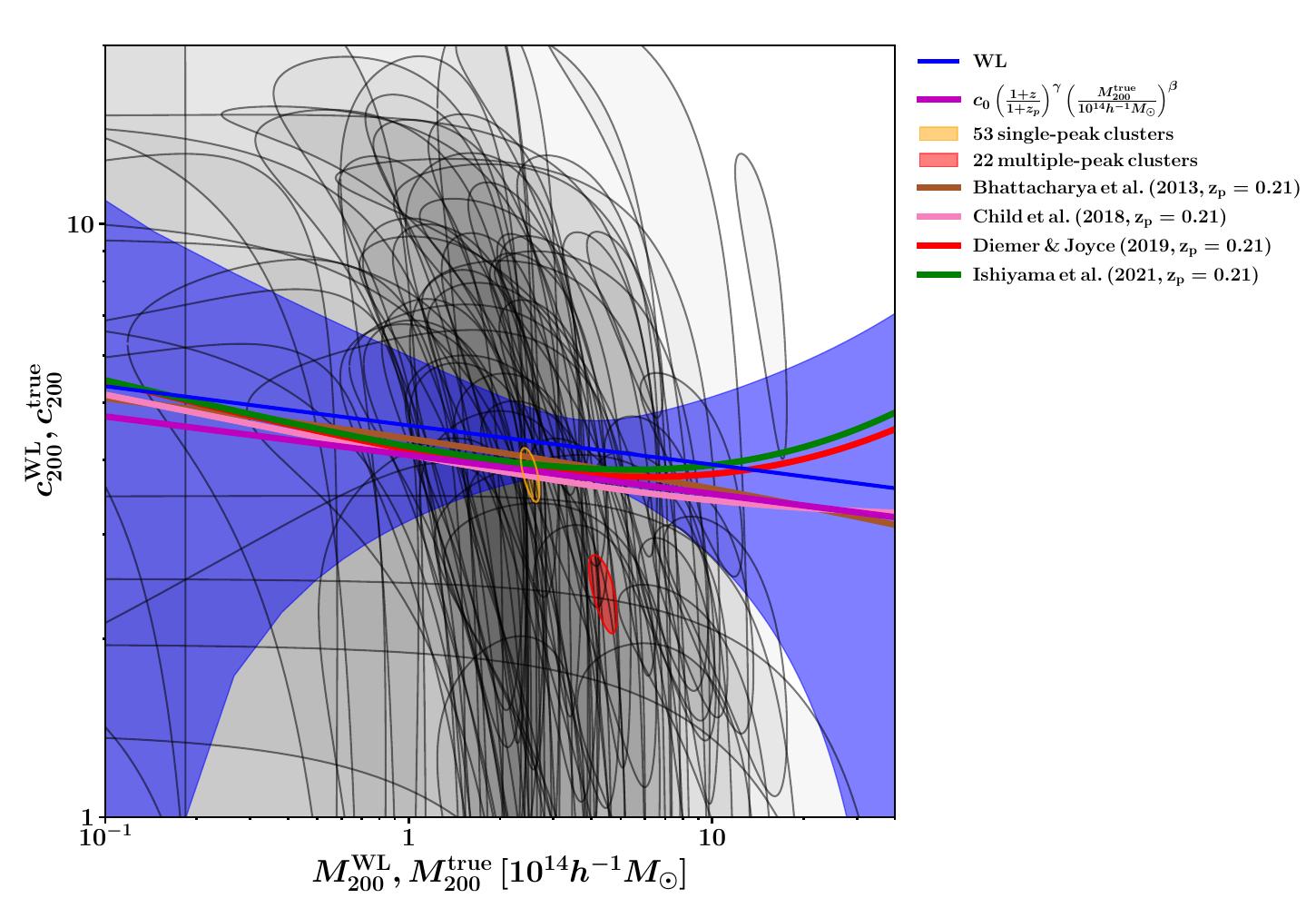}
      \caption{WL mass and concentration relation at $\Delta=200$. The transparent gray banana regions represent the $1\sigma$ constraint on mass and halo concentration for the 72 association clusters. The orange and red shaded regions represent the $1\sigma$ constraints by stacked lensing analysis for the 53 clusters with single galaxy peaks and the 22 clusters with multiple galaxy peaks, respectively. The best-fit line and its $1\sigma$ uncertainty are shown as the blue solid line and region, respectively. The magenta line is the best-fit for the true mass and the true concentration. The brown, pink, red, and green lines are the results of numerical simulations of \cite{Bhattacharya13}, \cite{Child18}, \cite{2019ApJ...871..168D}, and \cite{2021MNRAS.506.4210I}, respectively. 
              }
         \label{fig:c200_M200}
   \end{figure*}

\begin{figure}[!ht]
    \centering
    \includegraphics[width=\linewidth]{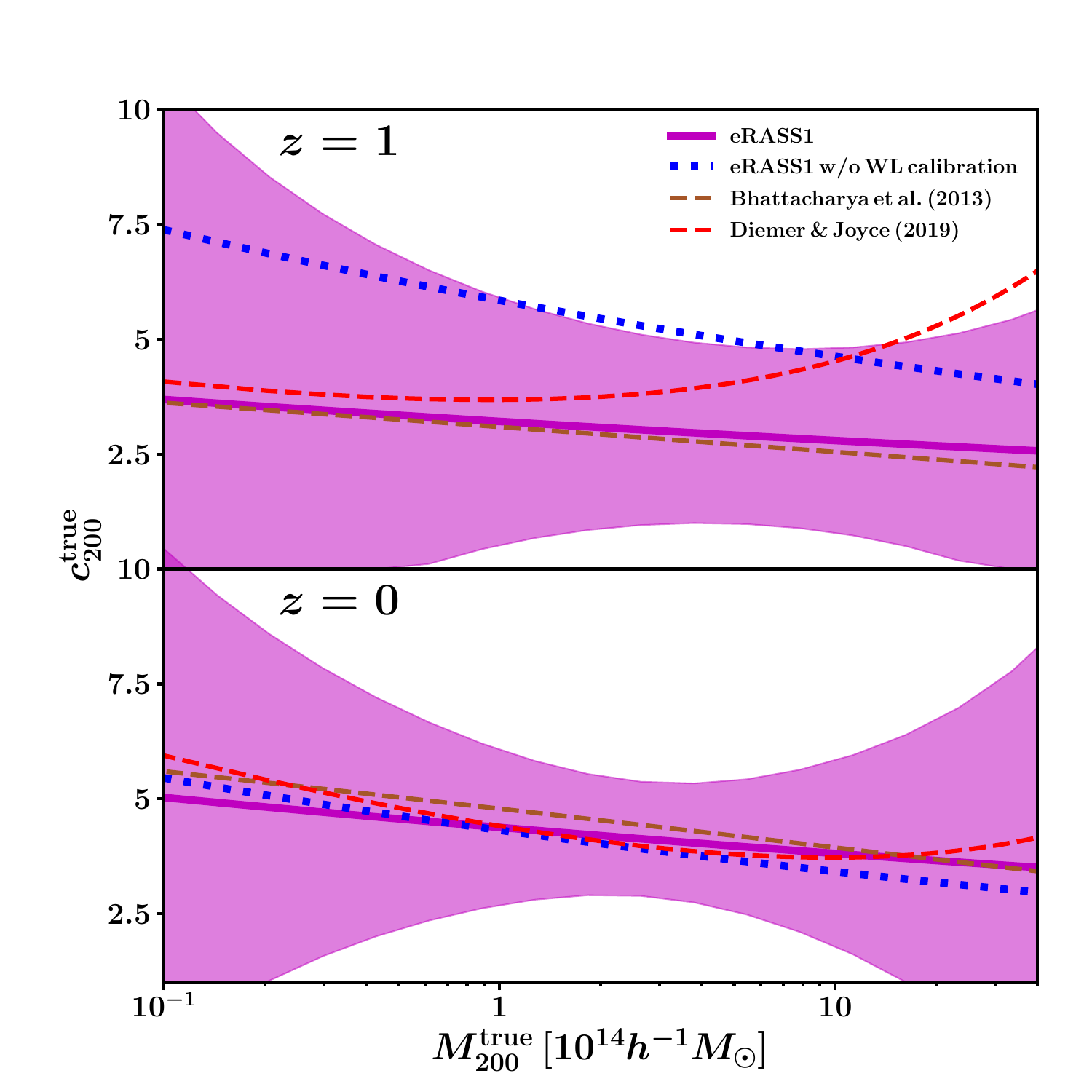}
    \caption{Redshift evolution for the mass-concentration relation. The best-fit line and its $1\sigma$ uncertainty are shown as the magenta solid line and region, respectively. The blue dotted lines are the results without the WL calibration. The brown and red dashed lines denote the normalizations of numerical simulations of \citet{Bhattacharya13} and \cite{2019ApJ...871..168D}, respectively.}
    \label{fig:M-C_norm_vs_z}
\end{figure}

The mass and concentration of the 72 association clusters for which WL masses were measured are shown in Fig. \ref{fig:c200_M200}.
Since these errors are correlated, the $1\sigma$ banana-shaped regions are shown in transparent gray.
To model the mass and concentration relation, we adopted the following relation
\begin{eqnarray}
  \ln\left( \frac{c_{200}}{c_p}\right)&=&\ln c_0  + \gamma \ln \left(\frac{1+z}{1+z_{p}}\right) \nonumber \\
  & &+\left(\beta+\delta \ln \left(\frac{1+z}{1+z_{p}}\right)\right)\ln \left(\frac{M_{200}}{10^{14}h^{-1}M_\odot}\right),
\end{eqnarray}
with $c_p=4$ and an intrinsic scatter of $\sigma_{c}$. The reference redshift is set to be the lensing weight average of $z_p=0.21$ for the 72 association clusters. 
We considered the correlation between the errors in mass and concentration, as well as the WL mass calibration, during the fitting process (see Appendix \ref{app}). Since the mass and concentration parameters were measured simultaneously, we evaluated the WL mass calibration as a mass and concentration plane with the intrinsic covariance (Appendix \ref{app2}) and used it as a prior. 
Table \ref{table:m-c} lists the best-fit parameters with and without the WL calibration, where we fixed $\delta=0$ because we could not constrain it. 
In Fig. \ref{fig:c200_M200}, the best-fit and its $1\sigma$ uncertainty for the WL mass and concentration relation are shown as the blue solid line and the shaded region. The normalization of the WL mass and concentration relation is overestimated by $\sim12\%$ compared to the true one (the magenta line). The mass-dependent slope is in good agreement with the predictions of recent numerical simulations \citep{Bhattacharya13,Child18,2019ApJ...871..168D,2021MNRAS.506.4210I}, although the uncertainties are too large to distinguish between negative and positive values. 
Figure \ref{fig:M-C_norm_vs_z} shows the redshift evolution of the normalization of the true mass and concentration relation. The error is too large to make a strong conclusion, but the normalization decreases as the redshift increases, which is consistent with the numerical simulation \citep{Bhattacharya13}.
In contrast, when we do not apply the WL mass calibration, the normalization increases as the redshift increases (blue dotted lines in Fig. \ref{fig:M-C_norm_vs_z}). 
When we remove the poor-fit clusters, we find that our results are not changed; namely, we have $\ln c_0=-0.023_{-0.388}^{+0.305}$, $\beta=-0.043_{-0.215}^{+0.272}$, and $\gamma=-0.625_{-1.244}^{+1.102}$.

\begin{table}[!ht]
\caption{Best-fit parameters for the mass and concentration relation.} \label{table:m-c}
\centering   
\begin{tabular}{c|c|c}
 & WL cal & no WL cal \\
\hline$\ln c_0$ & $0.019_{-0.343}^{+0.285}$ & $0.160_{-0.336}^{+0.281}$  \\
$\alpha$ & $-0.065_{-0.202}^{+0.254}$ & $-0.101_{-0.192}^{+0.233}$ \\
$\gamma$ & $-0.500_{-1.224}^{+1.107}$ & $0.436_{-0.974}^{+0.855}$ \\
$\sigma_{\ln c}$ & $<0.198$ & $<0.198$ \\
\hline
\end{tabular}
\end{table}

We divided the clusters into two subsamples using the number of peaks in the galaxy map. We selected clusters containing massive galaxy subhalos by applying the criterion that the ratio of the peak height of the most massive subhalo to that of the main cluster exceeds 0.5.
We refer to these as multiple-peak clusters and the remaining clusters as single-peak clusters. In the analysis, we considered 75 association clusters, including those for which the WL masses have not been successfully measured, to focus on their average characteristics. The halo concentration for the multiple-peak clusters is $2.40_{-0.35}^{+0.38}$, which is about 0.63 times that of the single-peak clusters $3.80_{-0.38}^{+0.41}$.
The halo concentration for the single-peak clusters is in good agreement with the numerical simulations, while the result for the multiple-peak clusters is $\sim 3.5\sigma$ lower than those for the given mass.
The two clusters (J141507.1-002905 and J141457.8-002050) in multiple-peak clusters are members of a three-cluster system. When we exclude the two clusters, the resulting $c_{200}^{\rm WL}=2.52_{-0.38}^{+0.43}$ does not change significantly.  This characteristic agrees with the lower concentration for merging clusters \citep[e.g.,][]{2014MNRAS.441.3359D,2016MNRAS.457.4340K,2021MNRAS.506.4210I}. The same result was reported in optically selected clusters \citep{2019PASJ...71...79O}.

\subsection{Miscentering effect from 2D WL analysis}\label{subsec:mis_center}

In the 2D analysis, the central positions are treated as a free parameter, allowing the calculation of the distance between the eRASS1 centroids and WL-determined centers, which offers crucial insights into the mis-centering effect.
Since the S/N of the WL signals is not high enough (Fig. \ref{fig:cluster_sample}), 
the centers and masses of some clusters are not well determined from the marginalized posterior distributions. 
If the WL-determined centers are associated with other clusters, we removed them from the results.   
The sample of the 2D WL analysis using the spherical NFW model is limited to 36 clusters. 
The S/N of the lensing signal of the 36 clusters is $S/N\simgt 4$, with mean and median values of $5.4$ and $5.1$, respectively, belonging to a high S/N population in the parent samples (Sec \ref{subsec:M_WL}).

\begin{figure*}[!ht]
   \centering
   \includegraphics[width=8cm]{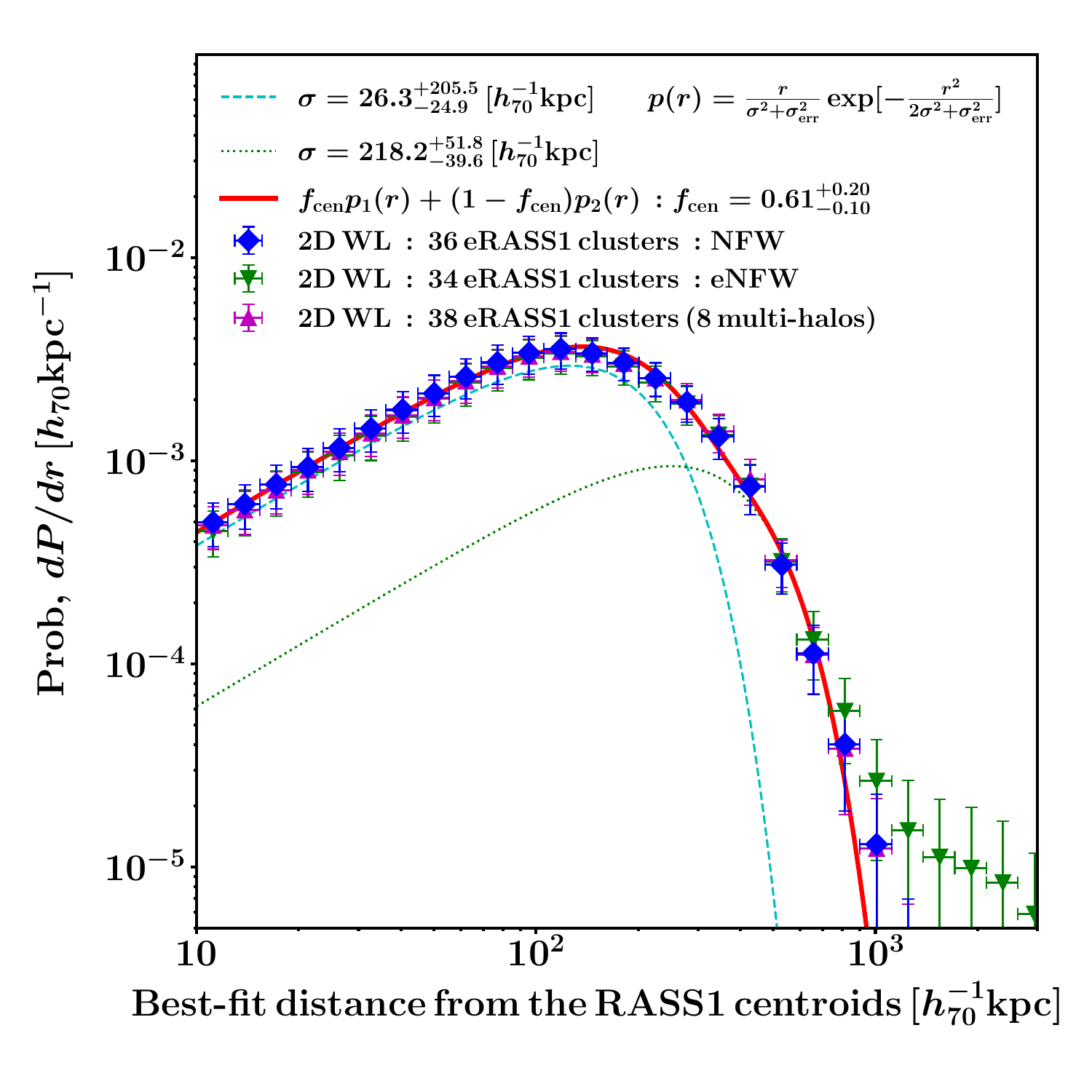}
   \includegraphics[width=8cm]{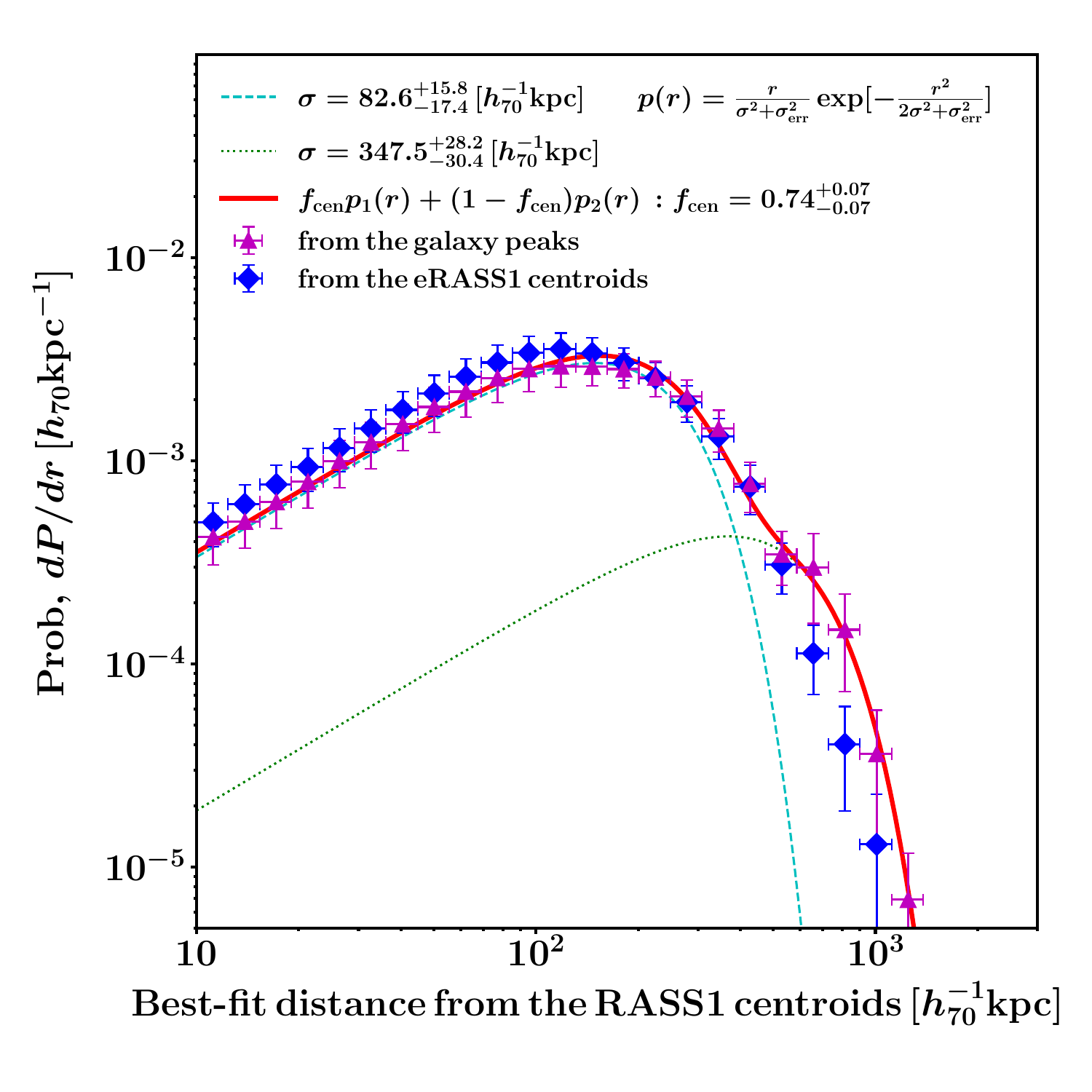}
      \caption{Left: Best-fit distance from the eRASS1 centroids. The blue diamonds, green down-triangles, and magenta up-triangles represent the results of the spherical NFW model, the elliptical NFW model, and the inclusion of the multiple spherical NFW components, respectively. The red solid line, the light-blue dashed line, and the green dotted line are the best-fit for all the components, the inner Gaussian and the outer Gaussian, respectively. Right: Magenta up-triangles represent the distance from galaxy map peaks. The blue diamonds are the same as in the left panel.
              }
         \label{fig:d_vs_n}
   \end{figure*}

Figure \ref{fig:d_vs_n} shows the probability distribution of the best-fit distance from the eRASS1 centroids. 
The errors are estimated by randomly drawing from the posterior distributions. The distribution has a peak around $\sim150\,h_{70}^{-1}{\rm kpc}$ and a tail at larger distances. To quantify the distribution, we introduced a model composed of two Gaussian components (Rayleigh distributions), as  done in  previous studies \citep{2010MNRAS.405.2215O,2018PASJ...70S..20O,2023A&A...669A.110O}:
\begin{eqnarray}
    \frac{dP}{dr}= f_{\rm cen}p_1(r)+(1-f_{\rm cen})p_2, \hspace{1em} p_i(r)=\frac{r}{\sigma_i^2}\exp\left[-\frac{r^2}{2r\sigma_i^2}\right]. \label{eq:dPdr}
\end{eqnarray}
Here, $f_{\rm cen}$ is the fraction of the inner Gaussian and $\sigma_i$ is a scale parameter (standard error). Since there are the measurement uncertainties of the WL-determined centers, we convolved the function (Eq. \ref{eq:dPdr}) with the typical measurement error of $\sigma_{\rm err}\simeq 123\,h_{70}^{-1}{\rm kpc}$. The best-fit result is shown in Table \ref{tab:mis-center}. The scale parameters of the inner and outer components are at most $\sigma_1 \sim26\,h_{70}^{-1}{\rm kpc}$ and $\sigma_2 \sim218\,h_{70}^{-1}{\rm kpc}$, respectively. 

\begin{table}[!ht]
    \centering
    \caption{Parameters of the miscentering effect.}
    \begin{tabular}{c|cc}
       & X-ray centroid & Galaxy-peak\\
       \hline
      $\sigma_1\,[h_{70}^{-1}{\rm kpc}]$ & $26.3_{-24.8}^{+205.5}$ & $82.6_{-17.4}^{+15.8}$ \\
      $\sigma_2\,[h_{70}^{-1}{\rm kpc}]$ & $218.2_{-25.7}^{+47.2}$ & $347.5_{-30.4}^{+28.2}$ \\
      $f_{\rm cen}$ & $0.61_{-0.10}^{+0.20}$ & $0.74_{-0.07}^{+0.07}$ \\
      \hline
    \end{tabular}
    \label{tab:mis-center}
\end{table}

Due to the presence of several mass structures within certain cluster fields, we additionally conducted a multi-component analysis. Compared to the result of the single NFW model, J141457.8-002050 and J141507.1-002905 (Fig. \ref{fig:massmap_J141457.8-002050}) are added to the result. The probability distribution for the multiple-halo model is almost the same as that for the single-halo model (left panel of Fig. \ref{fig:d_vs_n}). The probability distribution for the single elliptical NFW model is also similar to those used for the single and multiple NFW models. The best-fit results do not change. However, the errors for the two clusters become $10^3$ times larger and the tail over $10^3\,h_{70}^{-1}{\rm kpc}$ is found (left panel of Fig. \ref{fig:d_vs_n}). Therefore, the WL centers do not change significantly with a choice of mass models and the miscentering effect is small.

We also computed the probability distribution for the single NFW model centering the peaks in the galaxy maps, as shown in the right panel of Fig. \ref{fig:d_vs_n} and Table \ref{tab:mis-center}. The scale parameter of the inner component is about three times higher than that for the X-ray centroids.  
It shows that the X-ray centroids are a better tracer of the mass centers than the center of optical galaxies, thanks to the X-ray emissivity of $n_e^2$.

\subsection{Stacked mass map}\label{subsec:massmap_result}

\begin{figure}[!ht]
   \centering
   \includegraphics[width=0.9\hsize]{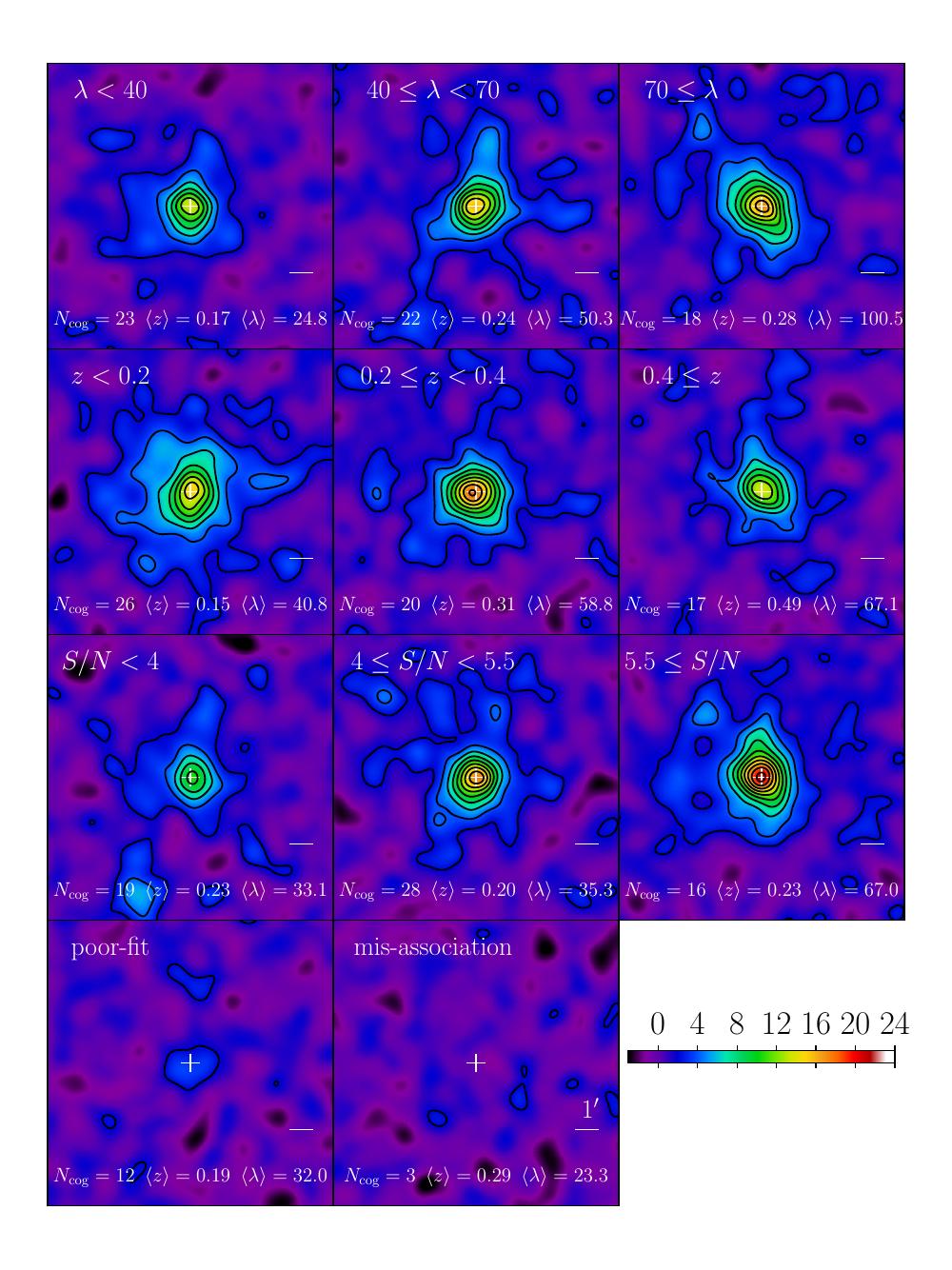}
   \caption{S/N of stacked mass maps for the subsamples ($12'\times12'$). The black contours represent the reconstructed WL mass map spaced in units of 2$\sigma$ bootstrapping error starting from $2\sigma$. The first, second, and third rows from the top show the maps for the subsamples divided by richness, redshift, and the S/N in the tangential shear, respectively. 
   These samples exclude the poor-fit and misassociation clusters. 
      The left and right panels in the bottom row show the mass maps for the poor-fit and misassociation clusters, respectively. The selection criteria are described in the upper part of each panel. The white crosses denote the eRASS1 centroids. The white horizontal line corresponds to the smoothing FWHM of $1'$. The number of clusters, the average redshift, and the average richness are described in the lower part of each panel.}
              \label{fig:stacked_massmap}
\end{figure}

Since the measurement of mis-centering effect by the 2D WL analysis is limited to clusters with high WL S/N (Sec \ref{subsec:mis_center}), the effect is unclear for clusters with low WL S/N. 
We make stacked mass maps for the subsamples (Fig. \ref{fig:stacked_massmap}), centering the eRASS1 X-ray centroids to visually investigate the mis-centering effect. The smoothing FWHM is $1$ arcmin.
The top three rows of Fig. \ref{fig:stacked_massmap} show the S/N maps for different bins of richness, redshift, and S/N in the tangential shear profile, respectively. The subsamples do not include the poor-fit or misassociation clusters.  The central region above $6\sigma$ has an almost concentric distribution. All of the peak centers coincide with the X-ray centroids within the smoothing scale.  The average redshifts are about the same at $\sim 0.2$, suggesting a similar lensing efficiency. 
The peak S/N increases with increasing richness and WL S/N, as expected from the mass-richness correlation described in Sect. \ref{subsec:scaling}. 
The peak S/N in the redshift bins is highest at $0.2\le z < 0.4$, due to good lensing efficiency and a larger number of background galaxies.
The peak SN for the poor-fit clusters has only a $\sim 3.6\sigma$ peak and no clear concentration. Even when we remove the substructure of the multiple-component cluster, J141457.8-002050 (Fig. \ref{fig:massmap_J141457.8-002050}), the result does not change significantly. The SN map for the misassociation clusters shows no lensing signal.

We compare the peak S/N ($(S/N)_{\rm peak}$) in the mass maps with the S/N in the stacked tangential shear ($(S/N)_{\langle \Delta \Sigma+\rangle }$) in the range of 100-3000 $h_{70}^{-1}{\rm kpc}$.  They are highly correlated for the S/N subsamples $\ln (S/N)_{\rm peak}=1.06+0.91\ln (S/N)_{\langle \Delta \Sigma+\rangle}$. The subsample with $S/N<4$, which could not be analyzed in the 2D WL, has a peak height similar to that expected from the S/N of the other two subsamples. Therefore, by analogy with the results of the 2D WL analysis of the other two samples, a subsample with $S/N<4$ would have a similar level of mis-centering effect as the result of the 2D WL analysis.

The peak height is also weakly correlated with the cluster richness $\langle \lambda \rangle$; $\ln (S/N)_{\rm peak}=8.40+0.15\ln\langle \lambda \rangle$. 
All the peaks expected from the richness are consistent with each other.

In contrast, the peak height of the poor-fit clusters is only 60 and 30 percent 
of those expected from $(S/N)_{\Delta \Sigma+}\simeq 4.1$ and $\langle \lambda \rangle\simeq 30$, respectively. As a result, the mass distribution of clusters with poor fits significantly differs from that of the remaining 63 clusters.

\subsection{Poor-fit clusters}\label{subsec:poorfit}

\subsubsection{WL mass measurements}

Both the stacked tangential shear profile (Fig. \ref{fig:g+_stack_log}) and the stacked mass map (Fig. \ref{fig:stacked_massmap}) for the 12 poor-fit clusters have different characteristics from the other 63 association clusters. Here, the 12 poor-fit clusters are composed of 5 clusters including J141457.8-002050 (Fig. \ref{fig:massmap_J141457.8-002050}) at $z<0.2$, 5 clusters at $0.2<z<0.5$, and 2 clusters at $0.5<z$, respectively. Since the lensing-weighted redshift is $\langle z\rangle=0.19$, these average lensing properties are affected by the clusters at $z<0.2$ because the number of background galaxies at $z>0.4$ is much smaller than that at $z<0.2$. Since the mass parameters of the subsamples split by redshifts are poorly constrained and their difference is not statistically significant, we examined the average characteristics for the poor-fit clusters.

\begin{figure}[!ht]
    \centering
    \includegraphics[width=\hsize]{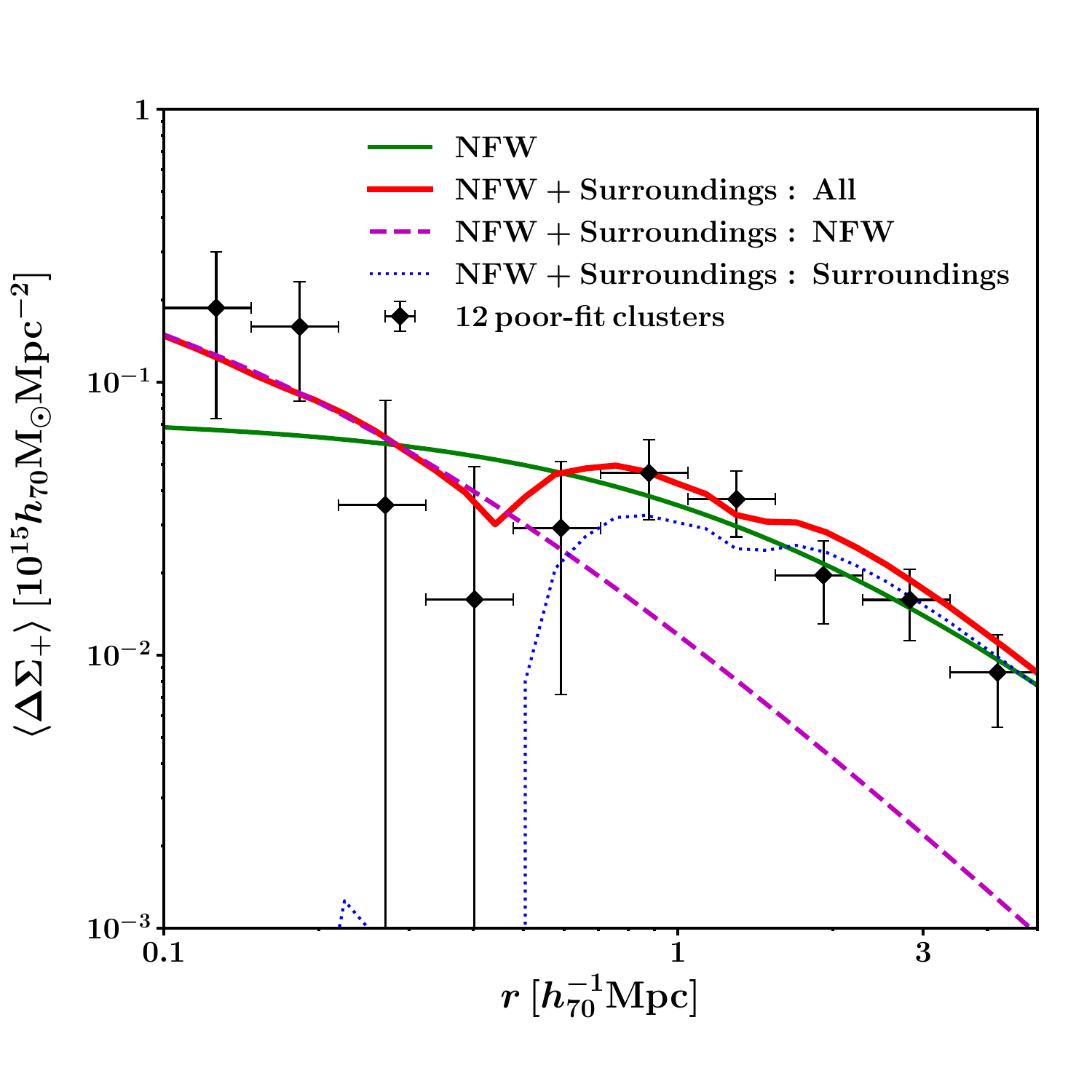}
    \caption{Stacked tangential shear profile for the 12 poor-fit clusters (same as Fig. \ref{fig:g+_stack_log}). The green solid line denotes the best-fit model for a single NFW model, but showing $c_{200}=1.16_{-0.48}^{+0.68}$. The red solid denotes the total lensing profile composed of the main NFW model and the surrounding halos. The
    magenta dashed and blue dotted lines are the tangential shear profiles of the NFW component and the surrounding halos, respectively.}
    \label{fig:g+_poor}
\end{figure}

When we fit the stacked tangential shear profile with a single NFW model, we obtain the extremely low concentration parameter of $c_{200}=1.15_{-0.47}^{+0.66}$ (Figs. \ref{fig:g+_poor} and \ref{fig:poorfit_property}). Indeed, the best-fit line is lower than the lensing signal observed in the central region of $r\simlt 200\,h_{70}^{-1}{\rm kpc}$. To understand the discrepancy, we first add the point source at the center to the lensing model, but the best-fit point mass is too massive $\sim3\times10^{13}h_{70}^{-1}M_\odot$ and the concentration parameter remains below 1. Since the lensing signals in the intermediate radius are low, the lensing profiles might be affected by the surrounding mass structures. Since the lensing signals for the poor-fit clusters are relatively low, we cannot identify the mass structure in the individual mass maps or find the eRASS1 clusters around the poor-fit clusters, except for J141457.8-002050 (Fig. \ref{fig:massmap_J141457.8-002050}). Therefore, we instead used the galaxy maps as a proxy to compute the lensing signal from the surrounding halos. We fixed the multi-halo positions and the best-fit WL masses for J141457.8-002050 obtained by 2D WL analysis and the offset positions of the galaxy peaks.  The surrounding halos are distributed over $0.2-1.9h_{70}^{-1}\,{\rm Mpc}$ for six clusters. The lensing-weighted average of the distance is $1.1h_{70}^{-1}\,{\rm Mpc}$. We parameterize the average mass associated with the galaxy peaks. Here, we assume that all the masses for the galaxy peaks are the same. Taking into account lensing weight in the stacked tangential shear profile, we computed the lensing signal from the surrounding halos as a function of $M_{200}$, where $c_{200}=4$ or $c_{500}=2.6$ is fixed. The modeled tangential shear profiles are computed by synthesizing the NFW mass model of the main cluster and the surrounding halos.

The constraints on the NFW parameters are weak, but not in contradiction with the baseline (left panel of Fig. \ref{fig:poorfit_property}). 
The halo masses for the main cluster and its surroundings are $M_{200}=0.31_{-0.12}^{+0.24}\times10^{14}h_{70}^{-1}M_{\odot}$ and $M_{200}^{\rm sur}=4.07_{-1.30}^{+2.12}\times10^{14}h_{70}^{-1}M_{\odot}$, respectively. 
Given the masses, we obtain $r_{200}\simeq 0.6\,h_{70}^{-1}{\rm Mpc}$ and $r_{200}^{\rm sur}\simeq 1.4\,h_{70}^{-1}{\rm Mpc}$, respectively.
Since the lensing-weighted average of the offset distances of the surrounding structures is $\sim 1.1h_{70}^{-1}\, {\rm Mpc}$, $r_{200}$ slightly overlaps with each other. 
We rerun fitting for the 5 multi-halo clusters (excluding J141457.8-002050), and obtain $M_{200}=0.19_{-0.06}^{+0.16}\times10^{14}h_{70}^{-1}M_{\odot}$ and $M_{200}^{\rm sur}=2.68_{-0.43}^{+0.83}\times10^{14}h_{70}^{-1}M_{\odot}$, respectively. The results do not change significantly.
Since the surrounding halos are more massive than the eRASS1 clusters, the eRASS1 clusters are likely to be subhalos or halos accompanied by the surrounding halos. 
We note that the surrounding structures are not listed by the main eRASS1 X-ray catalog \citep{2024A&A...682A..34M} except for J141457.8-002050. As for the single-peak clusters, we obtained a slightly higher mass of $M_{200}^{\rm single}=0.53_{-0.26}^{+0.33}\times10^{14}h_{70}^{-1}M_\odot$, along with $c_{200}^{\rm single}=5.47_{-5.46}^{+16.20}$.

As mentioned above, the poor-fit clusters are less massive objects at ${\mathcal O}(10^{13})\,h_{70}^{-1}M_\odot$. They are categorized into two categories: one includes structures with surrounding mass structures, while the other comprises those without. As for the clusters without the surrounding mass structures, the S/N in the stacked lensing profile is only $3.4$. The WL mass measurements for these clusters are difficult simply because of the less massive objects. As for the clusters with the surrounding mass structures, the S/N in the stacked lensing profile is $9.4$ but the lensing contamination from the surrounding structures cannot be ignored.

\subsubsection{Scaling relations}

\begin{figure*}[!ht]
    \centering
    \includegraphics[width=0.33\hsize]{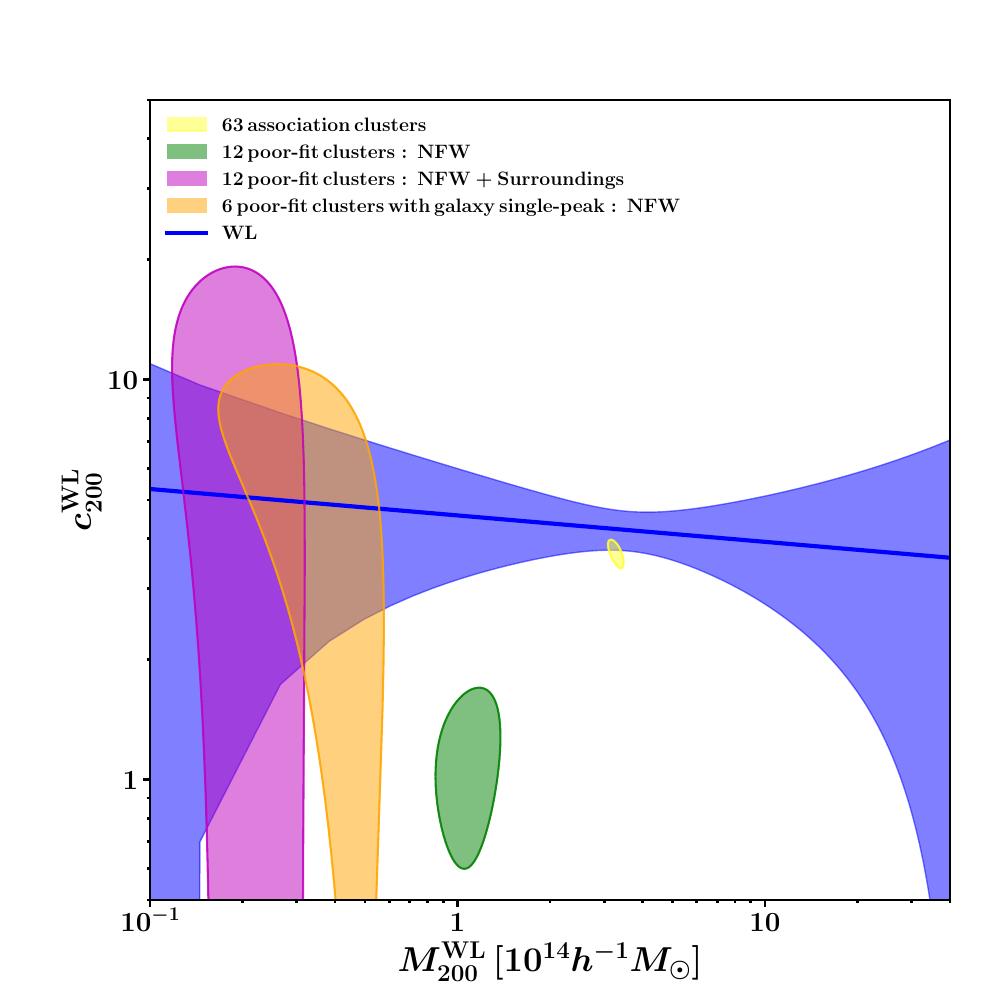}   
    \includegraphics[width=0.33\hsize]{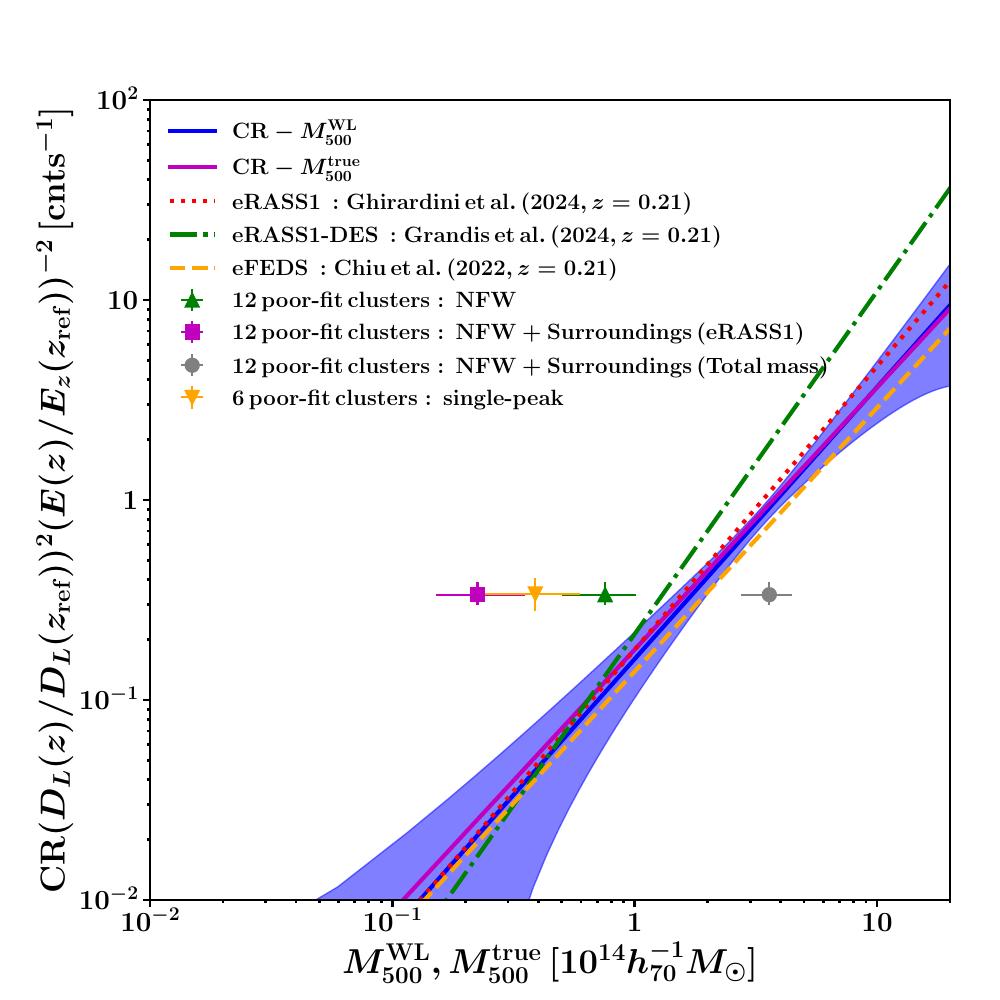}
    \includegraphics[width=0.33\hsize]{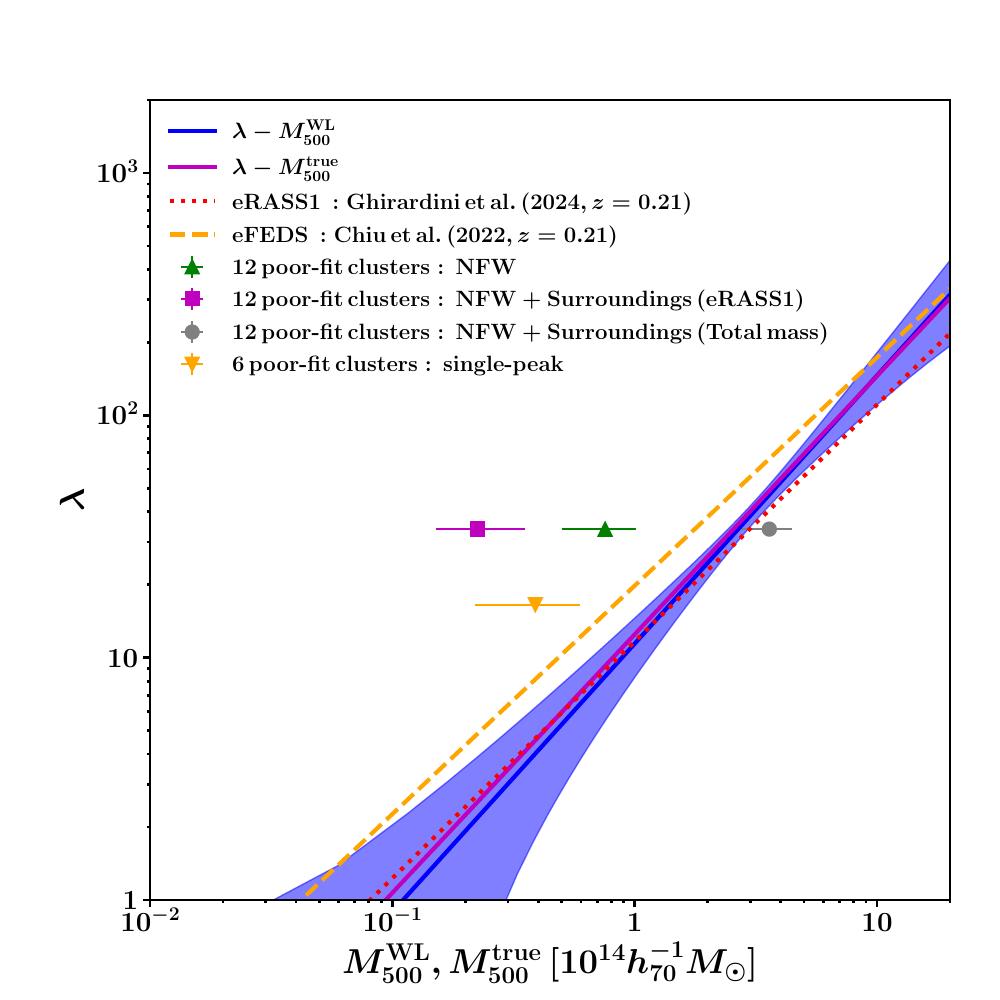}
    \caption{Comparison of the baselines and the results of the poor-fit clusters. The green region and up-triangles are the results of the single NFW model for the 12 poor-fit clusters. The magenta region and squares are the results of the eRASS1 clusters considering the surrounding mass halos in the model fitting. The gray circles are the results of the total components, including the surrounding mass.
    The orange region and down-triangles are the results of the single NFW model for the six poor-fit clusters with a single galaxy peak. 
    Left: Mass-concentration relation. Middle: Mass and the corrected CR relation. Right: Mass and richness relation. The lensing contribution from the surrounding halos cannot be ignored for the poor-fit clusters. Once we consider the surrounding halos, the concentration agrees with the baseline, but the masses expected from the count-rate or the richness are significantly overestimated.}
    \label{fig:poorfit_property}
\end{figure*}

We computed the stacked $\lambda$ and the corrected CR for the 12 poor-fit clusters. The WL masses taken with and without the surrounding halos are shown by magenta squares and green triangles in Fig. \ref{fig:poorfit_property}. In the mass-CR relation, the WL masses estimated with and without the surroundings are $\sim 8 \sigma_{\rm int}$ and $\sim 4 \sigma_{\rm int}$ lower than those expected from the CR, respectively. Here, $\sigma_{\rm int}=\sigma_{\rm CR}/\beta_{\rm CR}$ is the intrinsic scatter for the mass. Similarly, in the mass-richness relation, the WL masses with and without the surroundings are $\sim 7\sigma_{\rm int}$ and $\sim 3\sigma_{\rm int}$ lower, where $\sigma_{\rm int}=\sigma_{\lambda}/\beta_\lambda$. These results are similar to the results of the stacked mass maps showing the different mass distribution (Sect. \ref{subsec:massmap_result}).
When we consider the total mass derived by modeling the eRASS1 clusters along with the surrounding mass structures, we find that the mass is marginally above the expectation from the mass-CR baseline, yet it aligns well with the expectation from the mass-$\lambda$ baseline. This suggests that although galaxies within the surrounding structures are included in the richness calculation, the centrally concentrated X-ray count-rate is excluded from the surrounding mass structure.

This result highlights the importance of taking the surrounding mass structure into account when dealing with less massive clusters, for WL mass measurements, richness estimations, and X-ray count-rate analyses.

\subsection{2D halo ellipticity} \label{sec:2DhaloE}

\begin{figure}[!ht]
   \centering
   \includegraphics[width=8cm]{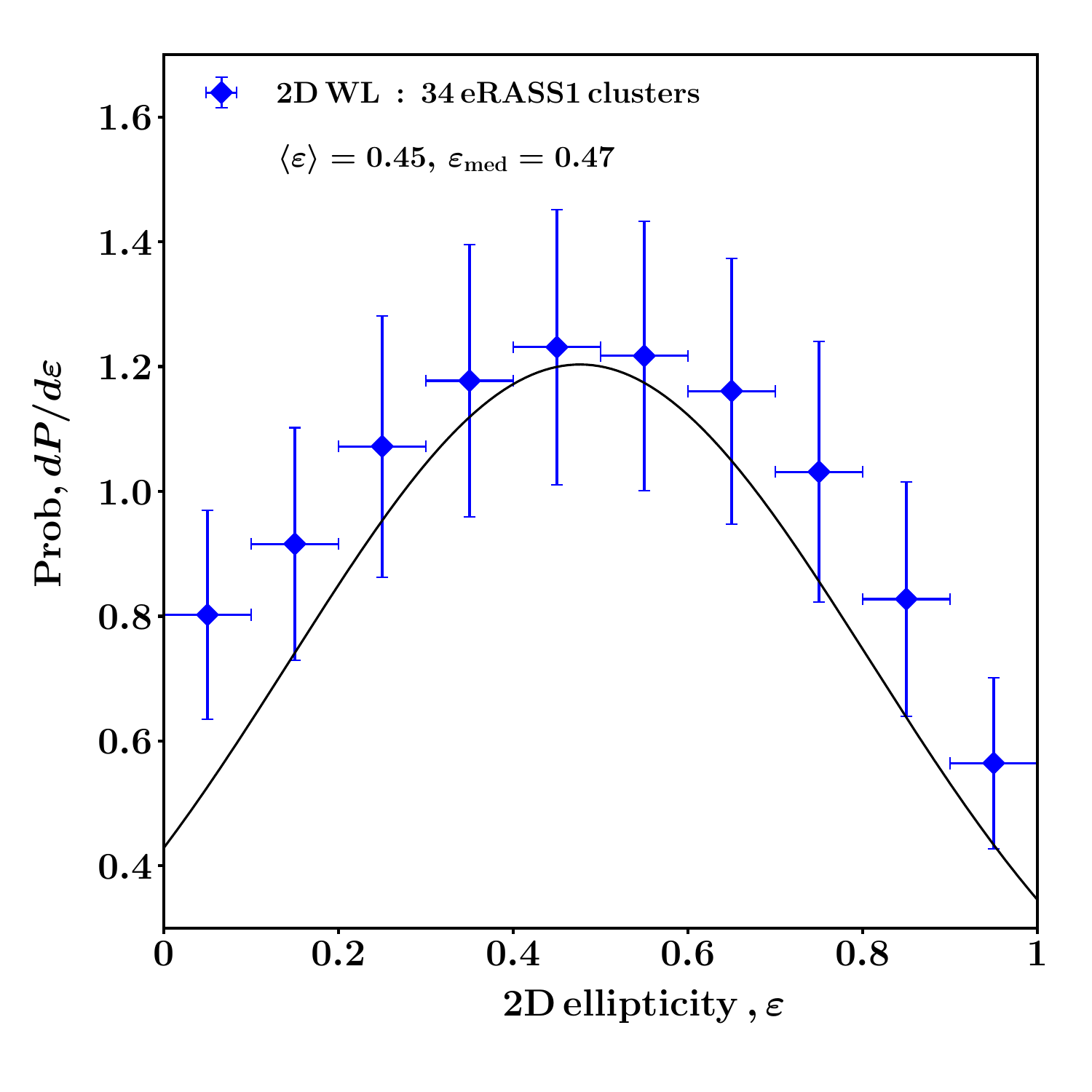}
      \caption{2D halo ellipticity distribution. The solid line is the best-fit Gaussian distribution convolved with the average measurement error. 
              }
         \label{fig:ell}
   \end{figure}

The probability distribution of the 2D halo ellipticity ($\varepsilon$) measured by the 2D WL analysis is shown in Fig. \ref{fig:ell}. The number of clusters is 34. The errors are estimated by 50,000 Monte Carlo redistribution of the ellipticity parameter of each cluster. The average and median ellipticities are $0.45$ and $0.47$, respectively. The average measurement error for each cluster is $0.29$.  The probability distribution is slightly skewed; the probability at the lowest ellipticity bin is 1.4 times higher than that at the highest ellipticity bin. This is because we considered avoiding the zero bound by treating the absolute value but do not the bound at $\varepsilon=1$. It is, therefore, difficult to conclude whether the skewed distribution is intrinsic or is due to the measurement technique. We fit the Gaussian distribution by convolving the average measurement error and obtain the mean ellipticity, $0.47$, and the standard error $0.16$.

\begin{figure}[!ht]
    \centering
    \includegraphics[width=8cm]{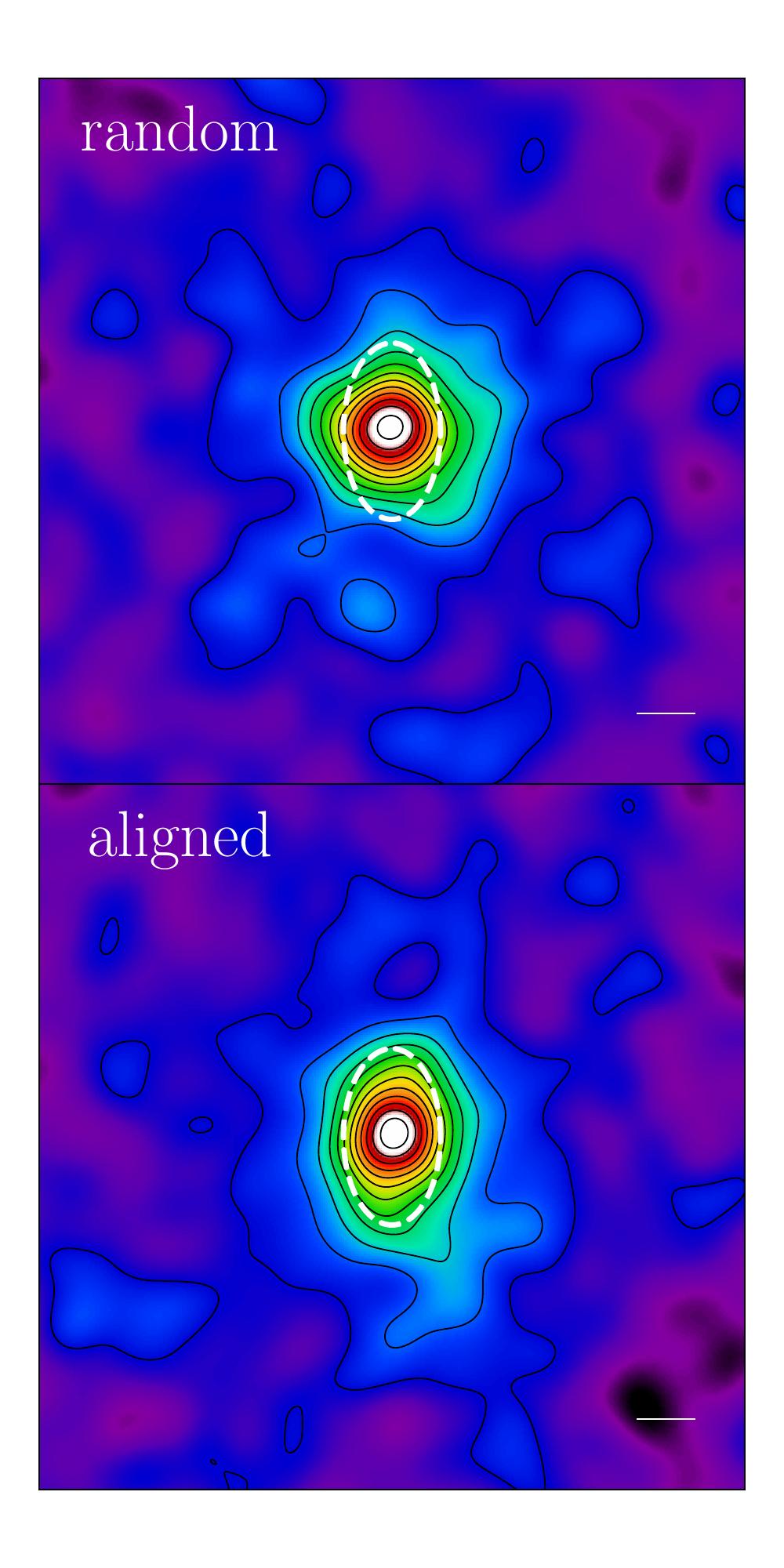}
    \caption{Stacked mass maps for the 34 clusters analyzed with the elliptical NFW model. Top: Random orientations.  Bottom: Aligned the major axis with the $y$-axis. The white dashed lines are auxiliary lines reflecting the median ellipticity. The white solid lines at the bottom right represent the 1 arcmin smoothing scale.}
    \label{fig:stacked_masmap_eNFW}
\end{figure}

We simultaneously obtained the orientation angle with  measurement error of a few percent. The small errors enable us to make a stacked mass map aligning the major axis with the $y$-axis (Fig. \ref{fig:stacked_masmap_eNFW}), which gives a sanity check to our measurement. We adopted X-ray centroids as the center. Since the miscentering effect is small, the choice of centers does not change the result. The 2D mass distribution is elongated along the $y$-axis, as expected by the 2D WL analysis. The degree and orientation angle of the elongation in the model-independent mass map are consistent with the result derived through the elliptical NFW model using discrete shear data.
As a control sample, we made a stacked mass map with random orientations as shown in the top panel of Fig. \ref{fig:stacked_masmap_eNFW}. The mass distribution is almost concentric, in contrast to the aligned mass map.

\begin{figure}[!ht]
    \centering
    \includegraphics[width=8cm]{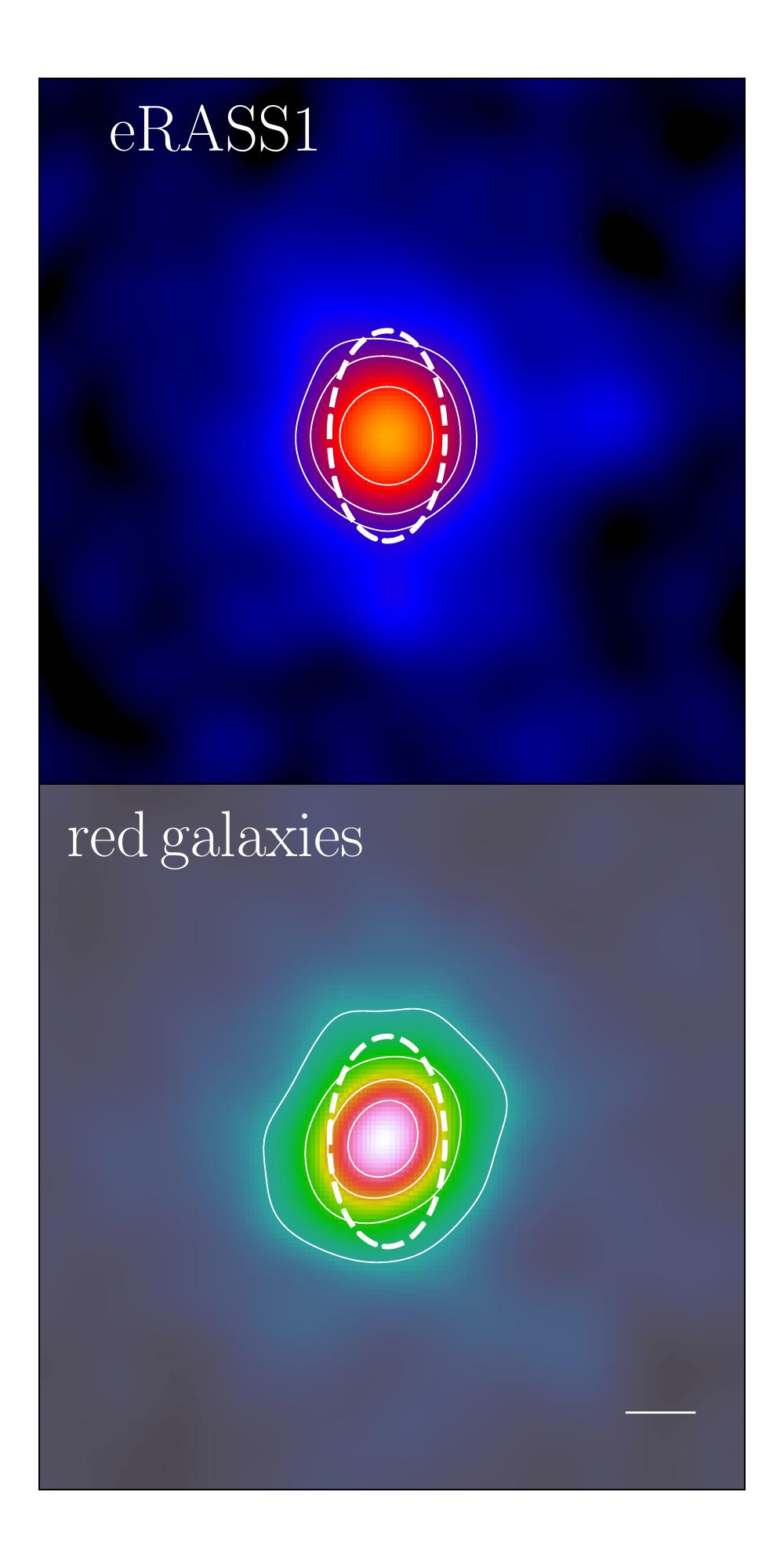}
    \caption{Stacked eRASS1 (top) and galaxy (bottom) maps for the 34 clusters analyzed with the elliptical NFW model. The alignment is the same as Fig. \ref{fig:stacked_masmap_eNFW}. The white dashed lines are auxiliary lines reflecting the median ellipticity. The white solid lines at the bottom right represent the $1$ arcmin smoothing scale.}
    \label{fig:stacked_masmap_eNFW2}
\end{figure}

We stacked eRASS1 soft-band (0.6-2.3 keV) images aligned with the major axis obtained by the 2D NFW fitting. 
We made X-ray images for individual clusters by subtracting the corresponding backgrounds and dividing them by the exposure maps. Here, we did not consider Galactic absorption at each cluster field because the clusters are located in a low Galactic column density region. We excluded X-ray images within 30 arcsec from X-ray point-sources in the main eRASS1 X-ray catalog \citep{2024A&A...682A..34M}. The stacking weight, expressed in Eq. \ref{eq:cr_wgt}, was used to standardize the flux to its expectation at the average redshifts ($z_{\rm p}=0.19$) of the 34 clusters. The resulting map is shown in the top panel of Fig. \ref{fig:stacked_masmap_eNFW2}. We fit it with an elliptical $\beta$ model; $S_X=S_0(1+(r/r_c)^2)^{-2\beta+0.5}$ where $r$ is an iso-contour computed with the ellipticity (eq. \ref{eq:elliptical}) and obtain the ellipticity $\varepsilon_X\simeq 0.1$. As expected, the best-fit major axis is aligned along the $y$ axis.
We stack the red galaxy maps, as shown in the bottom panel of Fig. \ref{fig:stacked_masmap_eNFW2}. We fit it with the elliptical $\beta$ model added to a background component and obtain ellipticity $\varepsilon_G\simeq 0.1$. When we use the galaxy map within $1$ arcmin, the ellipticity becomes slightly higher $\varepsilon_G\simeq 0.2$. The best-fit major axis is aligned with the $y$-axis.
Thus, the ellipticities of the hot and cold baryonic components are smaller than that of the mass (mainly dark matter) component.

\section{Discussion} \label{sec:dis}

\subsection{Mass-richness-CR relation comparison}

Figure \ref{fig:lambda-Mwl} shows the baselines of  eRASS1 \citep[red dotted line;][]{2024A&A...689A.298G}, eRASS-DES \citep[green dot-dashed line;][]{2024A&A...687A.178G} and eFEDS \citep[orange dashed line;][]{2022A&A...661A..11C} clusters. The mass-dependent slopes for the CR and the richness, $\beta_{\rm CR}=1.310_{-0.277}^{+0.294}$ and $\beta_{\lambda}=1.066_{-0.216}^{+0.183}$, are consistent with the slope in the soft-band X-ray luminosity and mass relation ($L_{\rm soft}^X\propto M$) and the idea that the number of cluster members is proportional to the cluster masses. In contrast, our constraints on $\delta$ in the mass-dependent slope are poor due to the small number of our sample, while the literature \citep{2022A&A...661A..11C, 2024A&A...689A.298G,2024A&A...687A.178G} constrains it well.
Therefore, we computed the mass-dependent slopes in the literature \citep{2022A&A...661A..11C, 2024A&A...689A.298G,2024A&A...687A.178G} at our lensing-weighted average redshift ($z_p=0.21$) for the following comparison. As the error covariance matrix for regression parameters is not fully detailed in the literature, we opted for the best-fit baseline. This choice does not impact the following discussion because our measurement errors are larger than those reported in the literature.

The best-fit slope of the mass-CR scaling relation is $1\sim 2\sigma$ lower than 
 $\beta_{\rm CR}=1.69$ \citep{2022A&A...661A..11C}, $\beta_{\rm CR}=1.42$ \citep{2024A&A...689A.298G}
 and $\beta_{\rm CR}=1.71$ \citep{2024A&A...687A.178G}. 
  The intrinsic scatter of the count rate, $\sigma_{\rm CR}= 0.344^{+0.097}_{-0.146}$, is comparable to the eFEDS-HSC result of $\sigma_{\rm CR}=0.301^{+0.089}_{-0.078}$ \citep{2022A&A...661A..11C}, while $\sigma_{\rm CR}=0.98_{-0.04}^{+0.05}$ \citep{2024A&A...689A.298G} and $\sigma_{\rm CR}=0.61\pm0.19$  \citep{2024A&A...687A.178G} are more than twice as large as ours. \citet{2024A&A...689A.298G} discussed the very large intrinsic scatter by introducing the mass-dependent slope in the intrinsic scatter and contamination of the active galactic nuclei and random sources. We concluded that a further study is needed to understand the causes of very large intrinsic scatter. Ultimately, we did not find such a large intrinsic scatter in the eFEDS-HSC and eRASS1-HSC analyses.

Our result of the mass and richness scaling relation agrees with the results of eRASS1 \citep{2024A&A...689A.298G} and eFEDS \citep{2022A&A...661A..11C}, although the richness measurement of \citep{2024A&A...689A.298G} is different. Our mass-dependent slope agrees well with $\beta_{\lambda}=0.98$ \citep{2024A&A...689A.298G} and $\beta_\lambda=0.94$ \citet{2022A&A...661A..11C}.
The intrinsic scatter of the richness, $\sigma_{\rm \lambda}=0.350_{-0.118}^{+0.099}$,  agrees well with $\sigma_{\rm \lambda}=0.274^{+0.078}_{-0.055}$ \citet{2022A&A...661A..11C} and $\sigma_{\rm \lambda}=0.32_{-0.03}^{+0.04}$ \citep{2024A&A...689A.298G}.

\subsection{Miscentering effect} \label{subsec:mis_center_dis}

The results of the 2D WL analysis (Sect. \ref{subsec:mis_center}) show that the probability distribution of the offset distance (Fig. \ref{fig:d_vs_n}) is composed of a double peak structure. The standard error (scale parameter) of the inner component of the X-ray centroids is at most $\sim26\,h_{70}^{-1}{\rm kpc}$, which indicates that the X-ray centroids are a good tracer of the bottom of the gravitational potential projected on to the sky.

To confirm this result independently, we also fit the stacked lensing profile for the 36 clusters used in the 2D WL analysis and the 75 association clusters with the mis-centering NFW model. We chose X-ray centroids as the centers. Here we assume that the concentration parameter follows \cite{2019ApJ...871..168D} and that the offset distances, $d_{\rm off}$, are the same for all the clusters. In other words, the fitting procedure determines $d_{\rm off}$ to represent the concentration parameter \citep{2019ApJ...871..168D}. We used the flat prior, $[\ln 0.02, \ln 0.5]\,h_{70}^{-1}{\rm Mpc}$, for $\ln d_{\rm off}$. We obtained $d_{\rm off}=46.2_{-16.8}^{+20.7}\,h_{70}^{-1}{\rm kpc}$ and $38.0_{-11.5}^{+18.4}\,h_{70}^{-1}{\rm kpc}$ for the 36 and 75 association clusters, respectively. They are consistent with the result obtained in the 2D WL analysis. Therefore, the miscentering effect with the X-ray centroids is, on average, negligible when describing the lensing properties.

The scale parameter of the inner component using the galaxy peaks, $\sim 83h_{70}^{-1}{\rm kpc}$, is about three times larger than that using the X-ray centroids. In the stacked lensing analysis, we chose the innermost radius as $100\,h_{70}^{-1}{\rm kpc}$, the miscentering effect would appear in the concentration parameter derived by the stacked lensing analysis. 
We repeated the stacked lensing analysis for the 36 clusters used in the 2D WL analysis using the galaxy peaks and obtained $c_{200}^{\rm WL}=3.02_{-0.28}^{+0.31}$, which is about $\sim 12.7\%$ lower than that using the eRASS1 centroids. The discrepancy is at the $1\sigma$ uncertainty level. 
Similarly, we computed the stacked lensing profiles for the 27 non-poor-fit clusters that were not suitable for study using the 2D WL analysis. The resulting concentration parameters using the galaxy peak centers and the eRASS1 centroids agree within the error. Hence, it is difficult to identify that the galaxy peaks are statistically more misaligned than the eRASS1 centroids for the less massive clusters.

\begin{figure}[!ht]
    \centering
    \includegraphics[width=\linewidth]{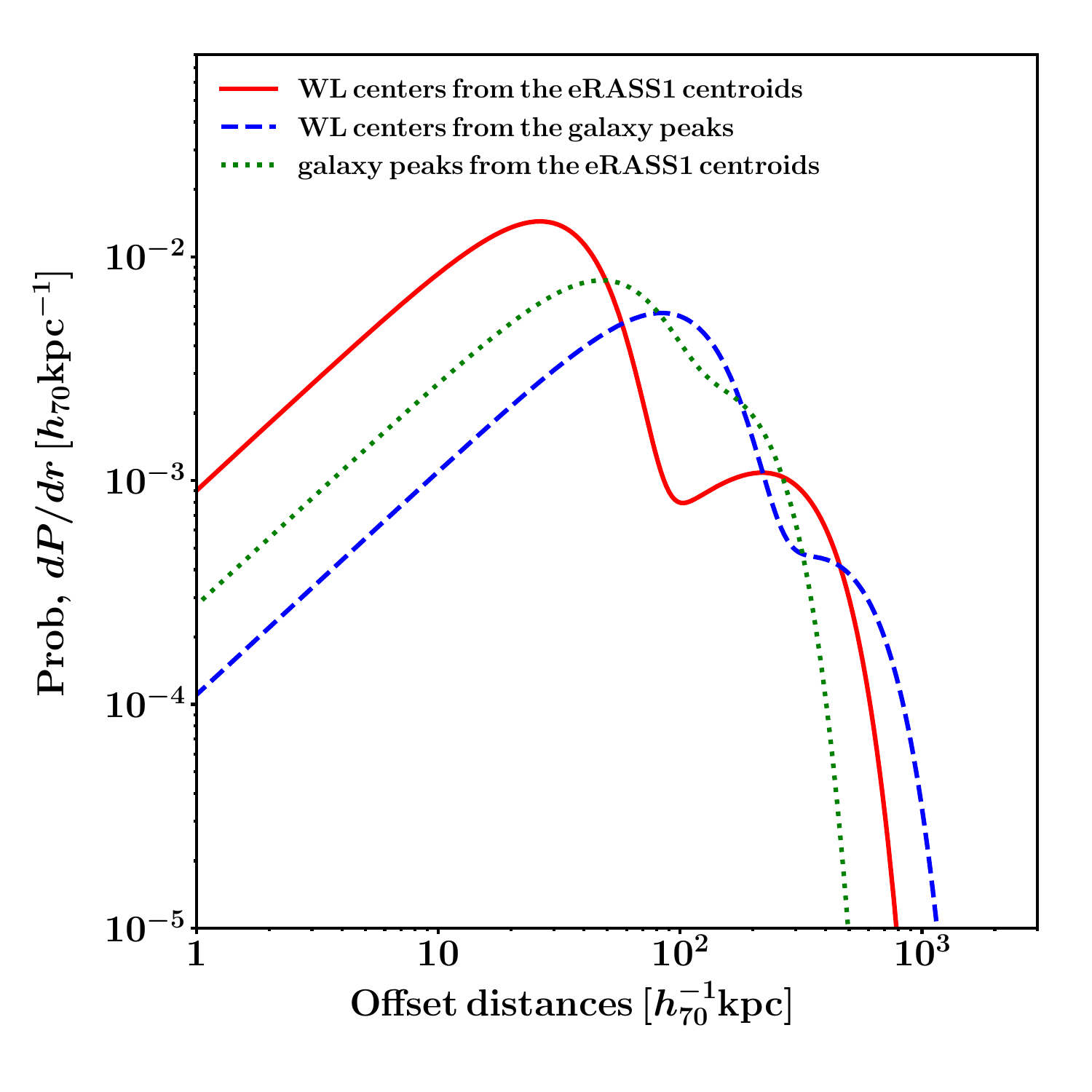}
    \caption{Offset distances between the WL centers and the eRASS1 centroids (red solid line), between the WL centers and the galaxy peaks (blue dashed line), and between the galaxy peaks and the eRASS1 centroids (green dotted line). It indicates that the offset directions of the galaxy peaks are different from those of the WL centers. }
    \label{fig:offset_all}
\end{figure}

We calculated the distance between galaxy peaks and the eRASS1 X-ray centroids. The resulting parameters of the 75 association clusters are $f_{\rm cen}=0.46_{-0.10}^{+0.10}$ $\sigma_1=43.4_{-5.6}^{+6.2}\,h_{70}^{-1}{\rm kpc}$, and $\sigma_2=128.9_{-12.5}^{+17.7}\,h_{70}^{-1}{\rm kpc}$. The resulting parameters of the 36 clusters used in the 2D WL analysis are similar to those for the 75 association clusters. Interestingly, the scale parameter of the inner component is smaller than that of the inner one between the WL-determined centers and galaxy peaks.
Figure \ref{fig:offset_all} presents a comparison of the deconvolved separations among three distinct components. The X-ray centroids are the best tracer of the WL centers. The directions in which the galaxy peaks are offset from the eRASS1 centroids differ from the offset directions of the WL centers. As a result, the offset distance between the WL centers and the galaxy peaks is the largest.

Among the 12 poor-fit clusters, 6 clusters have surrounding halos. The average mass for the surrounding halos is more massive than that for the eRASS1 clusters (Sec. \ref{subsec:poorfit}). Since these surrounding structures, except for J141457.8-002050, are not identified by X-ray, it causes a possible source of mis-centering effect. We computed the probability distribution of the offset distance (Fig. \ref{fig:d_vs_n}) including the five poor-fit clusters accompanied by the undetected surrounding structures, assuming that the true centers are in the surrounding halos and their uncertainty of the center determination is the mean value obtained by the 2D WL analysis.
The scale parameters, $\sigma_1$ and $\sigma_2$ (eq. \ref{eq:dPdr}), change by only $\sim +5$ and $\sim+7\%$, respectively. Therefore, the mis-centering effect is statistically small in the whole sample.

\cite{2022A&A...661A..11C} have computed the offset distribution between X-ray centroids and optical BCGs for the 313 eFEDS clusters and characterize it by a combination of a Rayleigh distribution for the outer component and a modified Rayleigh distribution for the inner component. The best-fit scale parameters for the inner and outer components are $\sigma_1=0.17\pm0.01 h_{70}^{-1}\,{\rm Mpc}$ and $\sigma_2=0.61\pm0.03 h_{70}^{-1}\,{\rm Mpc}$, and the fraction for the inner component is $f_{\rm cen}=1-0.54\pm0.02=0.46\pm0.02$. These parameters for the outer component are used as Gaussian priors for a simultaneous WL fit. The scale parameters are larger than our results between the galaxy peaks and the eRASS1 centroids. It might be due to the following feature that the galaxy peaks are insensitive to bright galaxies but sensitive to the number of member galaxies.

\cite{2024A&A...687A.178G} have studied an intrinsic miscentering using the hydrodynamical cosmological simulations. With 100 random viewing angles for the simulated 191 clusters (116 clusters at $z=0.252$ and 75 clusters at $z=0.518$), the offset between the peak of X-ray surface brightness distribution in 0.5-2 keV and the center-of-mass on the sky is well described by a single Rayleigh distribution with mean $\langle r_{\rm mis}^{\rm int}\rangle=\sigma_{\rm int}r_{500}$ and $\sigma_{\rm int}=0.104\pm0.016$. Given the mean value, we obtain the scale parameter, $\sigma_1=\langle r_{\rm mis}^{\rm int}\rangle(\pi/2)^{-1/2}\simeq 0.082 r_{500}$ and  $\sigma_{1}\simeq80\,h_{70}^{-1}{\rm kpc}$ when we insert $r_{500}$ for our sample. They also studied the observational miscentering effect using the offset between input and output position of clusters in the "eROSITA all-sky survey twin"  \cite{2022A&A...665A..78S}. Taking into account a realistic active galactic nucleus and cluster properties under the eRASS1 observational conditions, they obtained a typical miscentering $\sim 11$ arcsec \citep[see Sec. 4.2.2 in][]{2024A&A...687A.178G}. They synthesized the two miscentering effects in the WL mass measurement. Their miscentering effect corresponds to our inner component.

\cite{2023A&A...669A.110O}  studied the offset distribution between X-ray centroids and NFW-determined centers by 2D WL analysis for the 23 HSC-CAMIRA clusters \citep{2018PASJ...70S..20O} of which richness is higher than $N>40$ in the eFEDS field. They obtained $\sigma_1=174\,h_{70}^{-1}{\rm kpc}$, $\sigma_2=1339\,h_{70}^{-1}{\rm kpc}$ and $f_{\rm cen}=0.56$ for X-ray centroids of the CAMIRA clusters, where they did not consider the typical measurement errors for the WL centers. We refit it with the double-peak structure convolved with the typical measurement error of $\sigma_{\rm err}=170\,h_{70}^{-1}\,{\rm kpc}$ and obtain $f_{\rm cen}=0.56_{-0.05}^{+0.05}$ $\sigma_1=37.2_{-30.4}^{+46.7}\,h_{70}^{-1}{\rm kpc}$ and $\sigma_2=1328.5_{-111.8}^{+111.3}\,h_{70}^{-1}{\rm kpc}$. The scale parameter of the inner component is similar to ours. However, the scale parameter of the outer component is seven times larger than ours. 
It might be caused by the fact that they did not check whether the WL-determined centers belong to other clusters.
The scale parameters of the inner component from the CAMIRA centers and the galaxy peaks are $149.9_{-32.0}^{+30.0}\,h_{70}^{-1}{\rm kpc}$ and $124.7_{-26.5}^{+25.3}\,h_{70}^{-1}{\rm kpc}$, respectively. 
Similarly, the central positions identified by galaxies are further away from the WL centers.

\cite{2017MNRAS.469.4899M} carried out the WL mass calibration of \texttt{redMaPPer} clusters, using the DES Science Verification field. They compared the centers of the \texttt{redMaPPer} clusters with X-ray data and SZ data from the South Pole Telescope. They considered the uncertainty of the X-ray and SZ centers \citep{2016ApJS..224....1R} and found that the fraction of the central component is $f_{\rm cen}=0.76$ and the scale parameter is about 35\% of the cluster radius determined by richness. Their results are similar to what we found in our study.

\cite{2024PhRvD.110h3509B} evaluated the miscentering effect by combining the optical data of the DES and the Hubble Space Telescope and the SZ data from the South Pole Telescope. The central fractions for the SZ and optical distributions are $0.88$ and $0.89$. The scale parameters for the optical distribution are $\simeq 10\,h_{70}^{-1}{\rm kpc}$ and $\simeq 260\,h_{70}^{-1}{\rm kpc}$. The SZ distribution has similar properties. 
They also investigate the mis-centering effect using 70 clusters from Chandra X-ray data and found $f_{\rm cen}=0.8$, $\sigma_1\simeq25\,h_{70}^{-1}{\rm kpc}$ and $\sigma_2\simeq80\,h_{70}^{-1}{\rm kpc}$. Their results are similar to our findings from the 2D WL analysis.

\subsection{Background selection}

The main systematics of cluster mass measurements is a background selection. If cluster members are included in the shape catalog, the concentration parameter, $c_{200}$, or the masses, $M_\Delta$ ($\Delta \simgt 500$), within inner radii, are underestimated because the fraction of member galaxies to background galaxies increases with decreasing radius.  The effect is known as a dilution effect \citep[e.g.,][]{Broadhurst05,2014ApJ...795..163U,Okabe16b,2016MNRAS.463.4004Z, Medezinski18}.  Historically, the number of available imaging bands was limited, and background galaxies were selected in the color-magnitude plane \citep{Okabe16b} or the color-color plane \citep{2014ApJ...795..163U}, which is referred to as the color-color (CC) selection. The advantage of the CC selection is independent of the photometric redshifts. In the multi-band survey era, background galaxies can be selected using photometric redshifts, which is referred to as the ${\rm P_z}$ selection.

\begin{figure}[!ht]
    \centering
    \includegraphics[width=8cm]{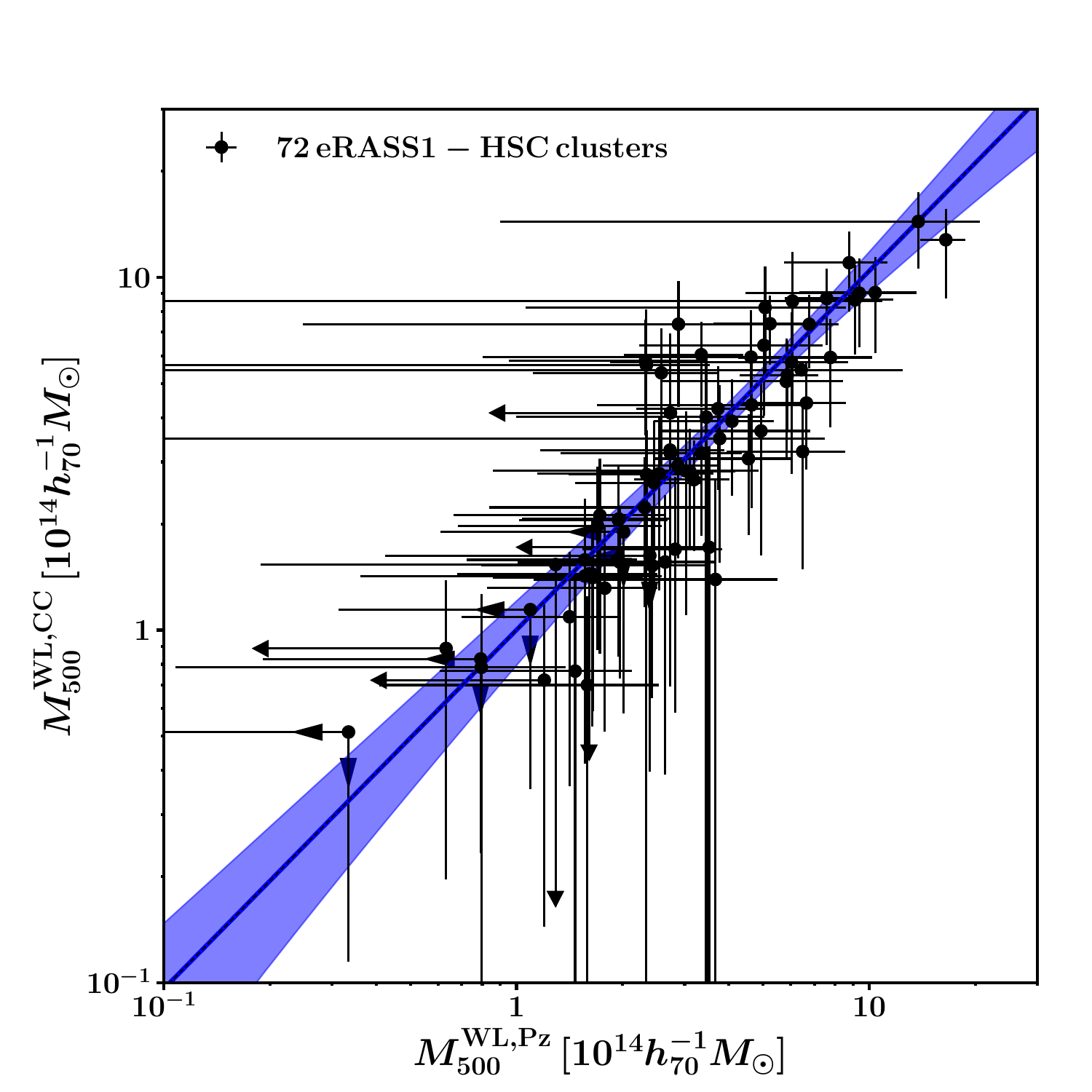}
    \caption{Mass comparison with the Pz and CC selections. The blue solid line and the 
blue region are the best-fit and the $1\sigma$ uncertainty. }
    \label{fig:Pz_vs_CC}
\end{figure}

\cite{Medezinski18} have developed a new CC selection using the HSC five-band imaging through a monitor of colors and lensing signals. They also have found that the masses estimated with the CC and ${\rm P_z}$ selections are in good agreement. We repeat the same analysis using the CC selection as the ${\rm P_z}$ selection. A mass comparison is shown in Fig. \ref{fig:Pz_vs_CC}. 
We perform a regression analysis with $\ln M_{500}^{\rm WL,CC}=\alpha_{\rm CC,Pz}+\beta_{\rm CC,Pz}\ln M_{500}^{\rm WL,Pz}$ and obtain $\alpha_{\rm CC,Pz}=0.001_{-0.228}^{+0.197}$ and $\beta_{\rm CC,Pz}=1.018_{-0.121}^{+0.141}$. The intrinsic scatter is $<0.138$. When we fix $\beta_{\rm CC,Pz}=1$ and no intrinsic scatter, we obtain $\alpha_{\rm CC,Pz}=0.030_{-0.076}^{+0.077}$, that is, the weighted geometric mean of the mass ratio of $1.030_{-0.076}^{+0.081}$. The masses of the two background selections are therefore in excellent agreement.

\subsection{Mass-concentration relation}

As discussed in Sect. \ref{subsec:mis_center_dis}, the miscentering effect is statistically negligible and thus does not affect the estimation of the halo concentration.
The mass and concentration relation for the eRASS1 clusters agrees well with the numerical simulations as shown in Fig. \ref{fig:c200_M200}, which supports the assumptions of the mass-concentration relation adopted in the simultaneous WL mass measurements for the eRASS1 clusters \citep{2022A&A...661A..11C,2024A&A...687A.178G,2025A&A...695A.216K}.

\begin{figure}[!ht]
   \centering
   \includegraphics[width=8cm]{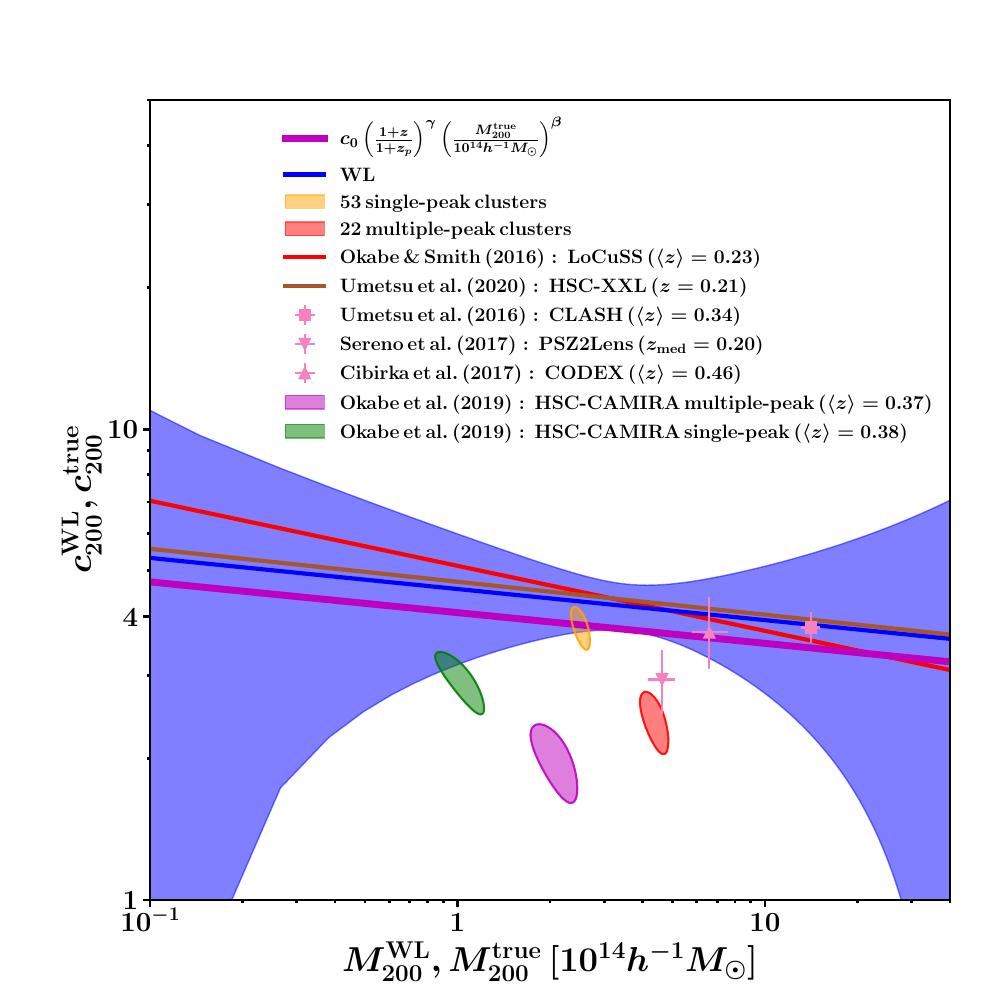}
      \caption{Comparison with the literature. The blue solid line and the 
blue region are the best-fit and the $1\sigma$ uncertainty (Fig. \ref{fig:c200_M200}), respectively. 
The magenta line is the true mass and concentration relation.
The red and brown lines are the best-fit lines of \citet{Okabe16b} and \citet{2020ApJ...890..148U}, respectively, where the redshift $z_{\rm ref}=0.3$ of \citet{2020ApJ...890..148U} is converted to $z=0.21$. 
The orange and red shaded regions represent the $1\sigma$ constraints by stacked lensing analysis for the 53 clusters with single galaxy peak, and the 22 clusters with multiple galaxy peaks, respectively.
The green and pink shaded regions represent the $1\sigma$ constraints by stacked lensing analysis for the HSC-CAMIRA clusters \citep{2019PASJ...71...79O} with single galaxy peak and multiple galaxy peaks, respectively. The pink square, up-triangle and down-triangle are the results of stacked lensing analyses of \citet{2016ApJ...821..116U}, \citet{Cibirka17} and \citet{2017MNRAS.472.1946S}, respectively.  }
         \label{fig:c200_M200_prev}
   \end{figure}

Since the mass and concentration relation provides us with an opportunity to test hierarchical structure formation on a scale of Mpc, it has been relatively well investigated in previous studies. \citep[Fig. \ref{fig:c200_M200_prev}; e.g.,][]{2014ApJ...795..163U,Okabe16b, 2016ApJ...821..116U,Cibirka17,2017MNRAS.472.1946S,2019PASJ...71...79O,2020ApJ...890..148U}. The cluster catalog of the ROSAT All Sky Survey \citep[RASS; e.g.,][]{1999A&A...349..389V,Bohringer04} were used for the WL analyses \citep{2014ApJ...795..163U,2016ApJ...821..116U,Okabe16b,Cibirka17}.
The Subaru/Suprime-cam WL analyses of CLASH \citep{2014ApJ...795..163U,2016ApJ...821..116U} and  LoCuSS \citep{Okabe16b} selected massive ROSAT clusters $M_{200}\simgt 5\times 10^{14}\,h_{70}^{-1}M_\odot$ at a redshift slice $z\sim0.2$. Since the number densities of background galaxies are more than twice that of our paper, they did not apply WL calibration. They found a negative mass-dependent slope but could not constrain the redshift evolution because of narrow redshift ranges. \cite{Cibirka17} measured stacked mass and concentration for 27 massive clusters $M_{200}\sim 7\times 10^{14}\,h_{70}^{-1}M_\odot$  at a redshift slice of $z\sim0.5$ using the Canada-France-Hawaii Telescope (CFHT) data.
Using the HSC-SSP shape catalog, the XXL survey covers less massive clusters and groups, whose masses cover a range down to $M_{200}\simgt  10^{13}\,h_{70}^{-1}M_\odot$ \citep{2020ApJ...890..148U} in the wide redshift range of $0.031\leq z \leq 1.033$.
Our result agrees well with the mass-concentration relation for the RASS clusters \citep{2014ApJ...795..163U,2016ApJ...821..116U,Okabe16b,Cibirka17} and the XXL clusters \citep{2020ApJ...890..148U}. The concentration for the PSZ2 clusters from the Planck mission is somewhat lower than our concentration \citep{2017MNRAS.472.1946S}.

As for optically selected clusters, \citep{2019PASJ...71...79O} performed a stacked lensing analysis for the HSC-CAMIRA clusters \citep{2018PASJ...70S..20O} to compare the concentration parameter for the single and multiple peaks in the galaxy maps. They did not consider the miscentering effect in the mass modeling. The concentration parameter for the single-peak clusters is about $60\%$ of the baseline. Since the mass measurement techniques are the same as our analysis, the miscentering effect for the optically selected clusters cannot be ignored, as found in \citet{2023A&A...669A.110O} and Sect. \ref{subsec:mis_center_dis}. in \citet{2019PASJ...71...79O} have also shown that the halo concentration for the multiple-peak clusters is $\sim33\%$ lower than that for the single-peak clusters. Similar results are found in our analysis; the ratio between the concentration parameters for the eRASS1 clusters with multiple-peak and single-peak is $\sim0.63$.

The cluster catalog covering a wide redshift range enables us to measure the redshift evolution of normalization. \citep{2020ApJ...890..148U} have shown the slope $\gamma=-0.03\pm0.47$. Our result gives a similar result, namely, $\gamma=-0.449^{+1.152}_{-1.245}$. Since both measurement errors are still large, we need a larger sample of clusters over wide redshift ranges.

As shown in Fig. \ref{fig:c200_M200}, the resulting baseline is in good agreement with dark matter-only numerical simulations \citep{Bhattacharya13,Child18,2019ApJ...871..168D,2021MNRAS.506.4210I}. 
Similar results have been reported in the literature \citep{2014ApJ...795..163U,2016ApJ...821..116U,Okabe16b,Cibirka17}. For instance,
\cite{Child18} have shown that the concentration parameter for the unrelaxed clusters is $69\%$ of that for the relaxed clusters at $z=0$. The feature is similar to our results. 
 The redshift evolution is similar to the numerical simulations (Fig. \ref{fig:M-C_norm_vs_z}). However, the concentration parameters of different numerical simulations at $z=1$ vary by up to $\simgt 50$\% at the high mass end. To further constrain the redshift evolution in future studies, we need a large number of clusters at high redshifts and carefully treat the WL mass bias caused by a small number of background galaxies.

\subsection{2D halo ellipticity}

We do not adopt any WL mass calibration in the halo ellipticity measurement. We leave this as next task using numerical simulations.

Our result agrees well with $\langle \varepsilon \rangle =0.46\pm0.04$ of the pilot 2D WL study \citep{2010MNRAS.405.2215O}. \citet{2018ApJ...860..104U} measured 2D halo ellipticity for 20 high-mass galaxy clusters selected from the CLASH survey and found $\langle \varepsilon \rangle =0.33\pm0.07$ at $M_{\rm vir}\simeq 10^{15}M_{\odot}$. They discussed the CLASH selection function based on X-ray morphological regularity, with its average value being slightly lower than theoretical predictions. \citet{2018ApJ...860..126C} constrained the minor-to-major axis ratio to a mass scaling relation for the 20 CLASH clusters by a decomposition method using a strong and weak lensing dataset. Their result using the informative prior from dark matter simulations \citep{2015MNRAS.449.3171B} shows $\langle \varepsilon^{\rm 3D} \rangle\simeq 0.50$ at a stacked WL mass $M_{200}^{\rm WL}=4.09_{-0.27}^{+0.29}\times10^{14}h^{-1}M_\odot$ for the 34 clusters measured with the elliptical NFW model. Their result is similar to ours. 
The findings impose constraints on the halo ellipticity within a mass range of about a half or less compared to previous studies; for example, $M_{200}^{\rm WL} \sim 7 \times 10^{14} \, h^{-1} M_\odot$ \citep{2010MNRAS.405.2215O} and $M_{200}^{\rm WL} \sim 10^{15} \, h^{-1} M_\odot$ \citep{2018ApJ...860..104U,2018ApJ...860..126C}.
Although it is difficult to make a direct comparison with the 3D halo ellipticity, our results are similar to $\varepsilon\sim 0.45-0.55$ from numerical simulations \citep{2014MNRAS.443.3208D,2015MNRAS.449.3171B,2007MNRAS.376..215B}.

\citet{2018MNRAS.478.1141O}  studied the projected ellipticity and orientation distributions for gas, stars, and dark matter using Horizon-AGN Simulation \citep{2014MNRAS.444.1453D}. They found the mean projected ellipticities measured within 1 Mpc radius are 0.33 for the total mass, 0.48 for the galaxies, and 0.17 for the gas, respectively. Their findings are consistent with ours, except for the galaxies. Given that we stacked galaxies based on WL-determined orientations, any mis-centering between galaxies and mass could obscure the ellipticities of the stacked galaxies map. When we choose optical centers, the project ellipticity of the galaxies becomes $\varepsilon_G\sim0.35$, which marginally agrees with \citet{2018MNRAS.478.1141O}. Furthermore, if the orientation angles of different components are misaligned, our measurements of gas and galaxy ellipticities would be underestimated. Individual analysis for massive nearby clusters would be suitable for more detailed studies because of good statistics of photon and member counts.

\subsection{eRASS1 mass comparison}

We compared the individual WL masses with the eRASS1 masses \citep{2024A&A...685A.106B}. 
The eRASS1 masses have been computed with the posterior distributions obtained by 
a simultaneous fit for the target cluster with priors of count-rate and mis-centering effect under an assumption of the mass and concentration relation \citep{2022A&A...661A..11C,2024A&A...687A.178G,2025A&A...695A.216K}.
It is important to compare masses in a complementary way to our analysis.

\begin{figure}[!ht]
   \centering
   \includegraphics[width=8cm]{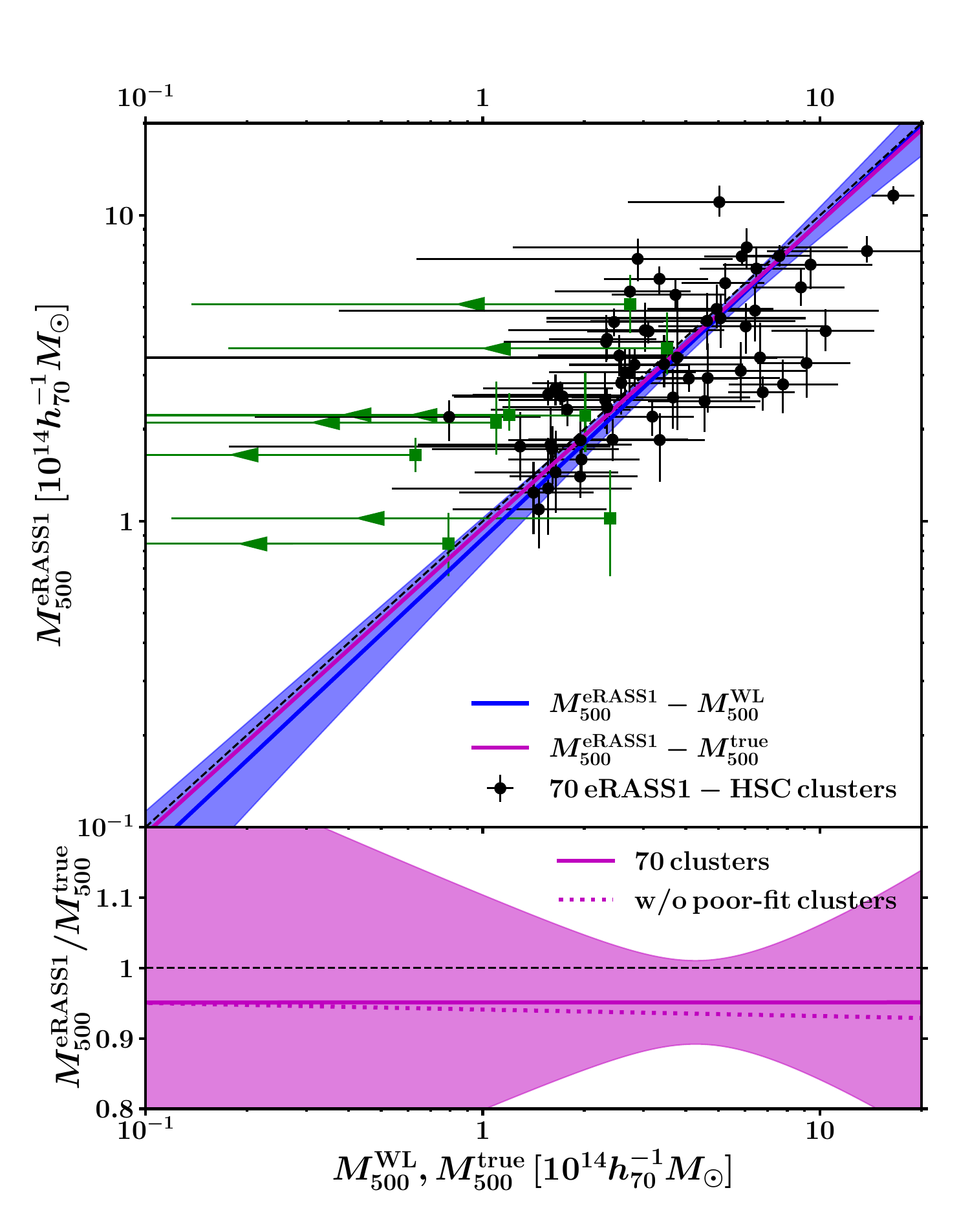}
      \caption{Top: Mass comparison between $M_{500}^{\rm WL}$ in this study and the eRASS1 mass in the primary sample \citep{2024A&A...685A.106B}. Black circles and green squares denote the association clusters and the poor-fit clusters, respectively. The magenta and blue lines denote the best-fit $M_{500}^{\rm eRASS1}-M_{500}^{\rm true}$ and $M_{500}^{\rm eRASS1}-M_{500}^{\rm WL}$ relations, respectively. The blue region is the $1\sigma$ uncertainty of the baseline. Bottom: Ratio of $M_{500}^{\rm eRASS1}$ to $M_{500}^{\rm WL}$. The magenta solid and dotted lines are the baseline for all the clusters and the 62 clusters excluding the poor-fit clusters, respectively. The magenta region is the $1\sigma$ uncertainty for all the clusters.
              }
         \label{fig:mass_erass1}
   \end{figure}

The number of clusters for the comparison is 76 in total because the two misassociation clusters, J114647.4-012428 and J124503.8-002823, are listed in the cosmology sample \citep{2024A&A...689A.298G}, but not in the primary sample \citep{2024A&A...685A.106B}.  Among them, 
the masses of four clusters, including a misassociation cluster and two poor-fit clusters, are not listed in the primary sample, and our analysis cannot measure the WL masses for the two clusters. Thus, the number of samples is 70 clusters.
Figure \ref{fig:mass_erass1} shows a mass comparison, showing a good agreement. We perform the regression analysis with no redshift evolution of the normalization.
The mass ratio between $M_{500}^{\rm eRASS1}$ and $M_{500}^{\rm true}$ is shown in the bottom panel of Fig. \ref{fig:mass_erass1}. The baseline is consistent with unity within the 1$\sigma$ level, though the best-fit value is $-5\%$ at $M_{500}^{\rm true}=10^{15}h_{70}^{-1}M_\odot$ and $-5\%$ at $10^{13}h_{70}^{-1}M_\odot$.  As mentioned in Sec. \ref{subsec:poorfit}, a treatment of surrounding halos at $0.2-1.9h_{70}^{-1}{\rm Mpc}$ from the poor-fit clusters of which masses are $\mathcal{O}(10^{13}h_{70}^{-1}M_\odot)$ is important. 
Even when we exclude the poor-fit clusters, the results do not change significantly.
When we set the mass-dependent slope to unity, we obtain the ratio $0.937_{-0.057}^{+0.067}$ and $0.949_{-0.053}^{+0.059}$ for all the clusters and the sample excluding the poor-fit clusters, respectively.

\section{Conclusions} \label{sec:con}

In this work, we  present the individual WL mass measurements for 78 eRASS1 clusters in the HSC-SSP footprint. 
In our analysis, we did not adopt any priors for the eRASS1 X-ray quantities and the richness and the assumption of the mass and concentration relation, which is complementary to other WL mass measurements employed for the eFEDS or eRASS1 clusters \citep{2022A&A...661A..11C,2024A&A...687A.178G,2025A&A...695A.216K}. 

The cluster sample was selected from the 103 eRASS1 clusters within the HSC-SSP footprint in order to ensure a sufficient number of background galaxies were available. Specifically, clusters lacking a uniform azimuthal distribution of background galaxies were excluded. By visual inspection of the HSC-SSP optical images, we  found that 3 out of the 78 eRASS1 clusters do not exhibit a match with the optical counterparts (Fig. \ref{fig:junk}).  This finding was further supported by the null-lensing signal in its stacked lensing profile (Fig. \ref{fig:g+_stack}).

Taking into account the WL mass calibration, the scaling relations between the true mass and cluster richness and  X-ray count-rate are in good agreement  with the results of the eRASS1 Western Galactic Hemisphere region based on count-rate-inferred masses, which have been calibrated with the HSC-SSP, DES, and KiDS surveys. 

We developed a Bayesian framework to measure the true mass and concentration relation, considering the WL mass calibration in the mass and concentration plane. The redshift-dependent mass and concentration measurements are in excellent agreement with numerical simulations and previous studies.

Based on the 2D WL analysis, the offsets between the WL-determined centers and X-ray centroids for the 36 eRASS1 clusters with high WL S/N are described by two Gaussian components. 
The miscentering effect with the X-ray centroids is small, as also assumed in the previous studies \citep{2022A&A...661A..11C,2024A&A...687A.178G}. The stacked mass maps support it even for less massive clusters. In contrast, when we use the galaxy peaks, the miscentering effect becomes slightly larger. 

The projected halo ellipticity driven by the 2D WL analysis shows $\langle \varepsilon \rangle=0.45$, which agrees with the results of numerical simulations and previous studies. 
The stacked mass map, aligned with the major axes, represents the elongated structure obtained from the 2D WL analysis. The optical and X-ray ellipticities are $\varepsilon\simeq 0.1-0.2$ and smaller than the result from the dark matter distribution.

The average mass for the 12 poor-fit clusters changes from $\sim 10^{14}h_{70}^{-1}M_\odot$ to $\sim 2\times 10^{13}h_{70}^{-1}M_\odot$ when lensing contamination from surrounding mass structures is taken into account. In that case, the concentration parameter is significantly improved from $\sim 1$ by including lensing contamination. An accurate WL mass measurement would require the modeling procedure to include the surrounding halos or a choice of the maximum radius in the tangential shear profile. 

\bibliographystyle{aa} 
\bibliography{newbib} 

\begin{appendix}

\section{Bayesian analysis for the mass-concentration relation} \label{app}

Taking into account the WL calibration on mass and concentration, a Bayesian framework for the mass and concentration analysis is slightly different from that for the multivariate scaling relations. To specify it, we summarize the formulation of the Bayesian analysis for the mass-concentration relation. 

We define the terms for the logarithms pertaining to the true and actual WL mass, as well as the true and actual WL concentration, as follows: $X=\ln M_{200}^{\rm true}$, $\tilde{x}=\ln \tilde{M}_{200}^{\rm WL}$, $Y=\ln c_{200}^{\rm true}$, $\tilde{y}=\ln \tilde{M}_{200}^{\rm WL}$. 
We assume a linear regression between the true mass and the true concentration,
\begin{eqnarray}
    Y&=&\alpha+(\beta + \delta F(z))X+\gamma F(z), \\
   F(z)&=&\ln \left(\frac{1+z}{1+z_p}\right), 
\end{eqnarray}
where $\alpha$, $\beta$, $\gamma$, and $\delta$ are the normalization, the mass-dependent slope, the redshift-dependent slope in the normalization, and the redshift-dependent slope in the mass-dependent slope, respectively. 
The true concentration, $Y_n$, of the $n$-th cluster is randomly varied with the intrinsic scatter, $\sigma_{Y}(z_n)=\exp((\ln \sigma_Y)(1+\gamma_{\sigma_Y} F(z_n)))$, from the baseline $Y$. Here, since the intrinsic scatter is a positive quantity, $\ln \sigma_Y$ is utilized as a parameter, and its redshift dependence is proportional to $\ln \sigma_Y$ to avoid negative values in $\sigma_Y$.
The probability of $Y_n$ given by $X_n$ is written by
\begin{eqnarray}
    p(Y_n|X_n,{\bm \theta})=\mathcal{N}\left(\alpha+(\beta + \delta F(z_n))X_n+\gamma F(z_n),\sigma_Y(z_n)\right),
\end{eqnarray}
where $\mathcal{N}(\mu,\sigma)$ is a Gaussian distribution with a mean, $\mu$, and a standard error, $\sigma$, and $\bm{\theta}$ is parameters in the mass and concentration relation composed of $\{\alpha, \beta, \gamma, \delta, \ln \sigma_Y,\gamma_{\sigma_Y}\}$, respectively. 
The parent population of $X$ assumes a Gaussian distribution, $\mathcal{N}(\mu_X(z),\sigma_X(z))$, with a redshift-dependent hyper-parameters, $\bm{\psi}=(\mu_X(z),\sigma_X(z))$. Here, $\bm{\psi}(z)=\bm{\alpha}_{\psi} + \sum_k \bm{\gamma}_{\psi,k} F(z)^k$. 
The WL calibration between the true values and the actual WL values is described by
\begin{eqnarray}
    p(\tilde{\bm{x}}_n|\bm{X}_n)&=& \mathcal{N}\left(\bm{\alpha}_{\rm WL}'(z_n)+\bm{\beta}_{\rm WL}'(z_n)\bm{X}_n, \bm{\Sigma}_{\rm WL}(z_n)\right),\nonumber \\
    \bm{\alpha}_{\rm WL}'(z) &=& \bm{\alpha}_{\rm WL}+\bm{\gamma}_{\rm WL}F(z), \nonumber \\
     \bm{\beta}_{\rm WL}'(z) &=& \bm{\beta}_{\rm WL}+\bm{\delta}_{\rm WL}F(z), \nonumber \\
     \bm{\Sigma}_{\rm WL}(z)&=&\Bigl(\begin{smallmatrix}
       \sigma_{{\rm WL},x}^2(z) &  r_{{\rm WL},xy}(z)\sigma_{{\rm WL},x}(z)\sigma_{{\rm WL},y}(z)  \\
       r_{{\rm WL},xy}(z)\sigma_{{\rm WL},x}(z)\sigma_{{\rm WL},y}(z) & \sigma_{{\rm WL},y}^2(z) 
    \end{smallmatrix}\Bigr), \nonumber \\
    \sigma_{\rm WL}(z)&=&\exp((\ln \sigma_{\rm WL})(1+\gamma_{\sigma_{\rm WL}}F(z))), \nonumber \\
    r_{{\rm WL},xy}(z)&=&r_{\rm WL,xy}(1+\gamma_{\rm r_{\rm WL}} F(z)), \nonumber 
\end{eqnarray}
where $\tilde{\bm{x}}_n=(\tilde{x}_n,\tilde{y}_n)$, $\bm{X}_n=(X_n,Y_n)$, the WL mass calibrations ($\bm{\alpha}_{\rm WL}$, $\bm{\beta}_{\rm WL}$, $\bm{\delta}_{\rm WL}$, and $\bm{\gamma}_{\rm WL}$), and the intrinsic WL covariance ${\bm \Sigma}_{\rm WL}$, respectively. The observed WL mass and concentration, $\tilde{\bm{x}}=\left(\ln M_{200}^{\rm WL},\ln c_{200}^{\rm WL}\right)$, are related to the actual WL values with the error covariance matrix ${\bm \Sigma}_{\rm err}$, specified by
\begin{eqnarray}
    p(\bm{x}_n|\tilde{\bm{x}}_n)=\mathcal{N}\left(\tilde{\bm{x}}_n,{\bm \Sigma}_{\rm err}\right).
\end{eqnarray}
The likelihood function is obtained by the Bayesian chain rule, as follows:
\begin{eqnarray}
  && \hspace{-2em}  p(\bm{x}|\bm{\theta},\bm{\psi}),  \\
    & \propto&\prod_n^N \int d^2\tilde{\bm{x}}_n \int d^2\bm{X}_n p(\bm{x}_n|\tilde{\bm{x}}_n)p(\tilde{\bm{x}}_n|\bm{X}_n)p(Y_n|X_n,\bm{\theta})p(X_n|\bm{\psi}), \nonumber \\
    & \propto&\prod_n^N \frac{1}{2\pi {\rm det}(\Sigma_{{\rm tot},n})^{1/2}}\exp\left[-\frac{1}{2}\bm{v}_n^\top \Sigma_{{\rm tot},n}^{-1}\bm{v}_n\right],
\end{eqnarray}
Here, the vector, $\bm{v}$, and the total covariance matrix, $\Sigma_{\rm tot}$, are given by
\begin{eqnarray}
    {\bm v}&=&\begin{pmatrix}
      x- \left(\alpha_{{\rm WL},x}' + \beta_{{\rm WL},x}'\mu_X\right) \\
      y- \left(\alpha_{{\rm WL},y}' + \beta_{{\rm WL},y}' (\alpha'+\beta' \mu_X) \right)
    \end{pmatrix}, \nonumber 
    \end{eqnarray}
    and
    \begin{eqnarray}
     \alpha'(z)&=&\alpha+\gamma F(z), \\
    \beta'(z)&=&\beta+\delta F(z), \\
   {\bm \Sigma}_{\rm tot}(z)&=&{\bm \Sigma}_{\rm err}+{\bm \Sigma}_{\rm int}(z)+{\bm \Sigma}_{\rm WL}(z)+ {\bm \Sigma}_{\rm X}(z),\\
   {\bm \Sigma}_{\rm int}&=&\begin{pmatrix}
        0  & 0  \\
      0 & \beta_{{\rm WL},y}'^2 \sigma_{Y}^2 
    \end{pmatrix}, \nonumber  \\
  {\bm   \Sigma}_{\rm X}&=&\begin{pmatrix}
        \beta_{{\rm WL},x}'^2 \sigma_X^2 & \beta'\beta_{{\rm WL},x}'\beta_{{\rm WL},y}'\sigma_X^2 \\
        \beta'\beta_{{\rm WL},x}'\beta_{{\rm WL},y}'\sigma_X^2 &  \beta'^2\beta_{{\rm WL},y}'^2 \sigma_X^2
    \end{pmatrix}, \nonumber
\end{eqnarray}
respectively.

Then, the posterior distribution is given by
\begin{eqnarray}
  p(\bm{\theta},\bm{\psi}|\bm{x}) \propto  p(\bm{x}|\bm{\theta},\bm{\psi})p(\bm{\theta})p(\bm{\psi}).
\end{eqnarray}
Here, $p(\bm{\theta})$ and $p(\bm{\psi})$ are priors. We adopted a flat prior $[-10^4,10^4]$ to the normalization and 
a student’s t1 distribution with one degree of freedom on the slopes so that the slope angles become uniformly distributed. 
As for a noninformative prior distribution on the normalization of the variance of the parent distribution, a scaled inverse $\chi^2$
distribution as a conjugate prior, satisfying that posterior distributions have the same probability distribution family as the prior distribution.

\section{WL mass calibrations} \label{app2}

\begin{figure}[!ht]
    \centering
    \includegraphics[width=\linewidth]{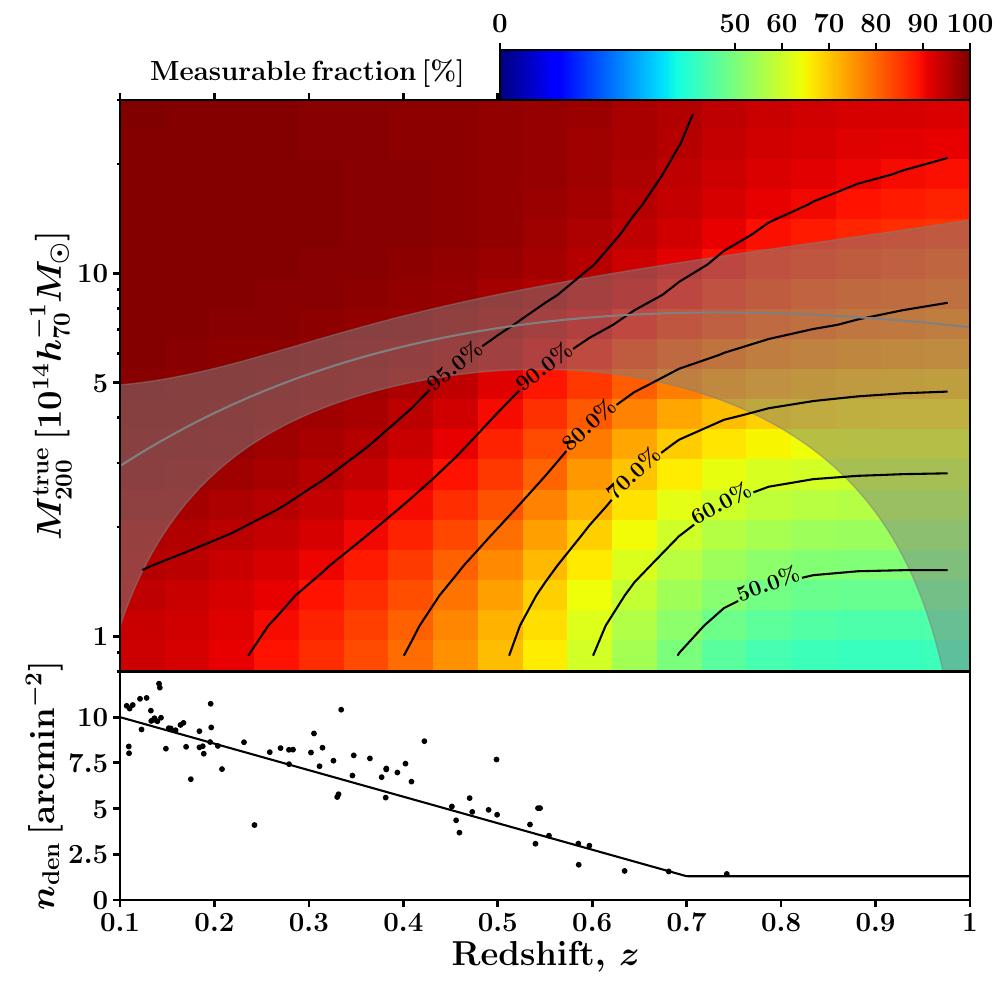}
    \caption{Top:  Fraction of 9000 synthetic clusters for which WL masses can be measured in the $M_{500}^{\rm true}$ and $z$ plane. The background colors are smoothed with a Gaussian kernel of $dz=0.1$ and $d\ln M_{500}^{\rm true}=0.35$ for a visual purpose. The gray shadow region represents the true mass distribution obtained by a scaling relation analysis (Sec \ref{subsec:scaling}). Bottom: Number densities as a function of cluster redshifts. The black solid line represents the model distribution used in the mock analysis.}
    \label{fig:z_vs_M_WLcal}
\end{figure}

Cluster WL analysis extracts masses by an ensemble average of galaxy ellipticity which is composed of
coherent WL signals synthesized to the intrinsic ellipticity. Thus, measurement accuracy depends on the number of background galaxies. 
For instance, there is no mass bias if an infinite number of background galaxies is available. In contrast,  
if the number of background galaxies was not sufficient, given the WL signal, each measured mass would be accidentally overestimated or underestimated. 
With the secure background selection using $P(z)$ in the HSC-SSP Survey,  the number density of background galaxies depends strongly on the cluster redshift (Figs. \ref{fig:z_vs_M500} and \ref{fig:z_vs_M_WLcal}), varying from $\sim 10$ to $1$ [arcmin$^{-2}$] by changing the cluster redshift from $0.1$ to $0.8$.  Even though we measure individual cluster masses, there might be bias and scatter in the measured WL masses.

We evaluated such systematics using a mock shape catalog, realizing our observing conditions. We assume a spherical NFW model and no lensing effect from the large-scale structure. Other effects (e.g., miscentering effect and halo triaxiality) as appeared in numerical simulations are not included. WL calibrations using numerical simulations are reported in the literature \citep[e.g.,][]{2010A&A...519A..90M,2011ApJ...740...25B,2012MNRAS.426.1558G,2020ApJ...890..148U,2024A&A...681A..67E}. 
The quantitative details are not comparable due to differences in the redshift range, the mass range, the overdensities, the fitting method, the number of samples, and simulation types. However, the overall trend of the bias in the relationship between WL mass and true mass is negative by several per cent and approaches zero as the true mass increases. This could be explained mainly by the fact that the probability of the major axes of halo mass distributions aligning with the sky plane is higher than in the line-of-sight direction.
Although our tests include minimal effects, we placed more emphasis on a catalog that represents the observational conditions and our analysis.  To asses the bias and scatter in the WL mass measurement, the sufficient number of clusters in each mass and redshift bin is required, otherwise, an accidental over- or underestimation cannot be ruled out. The mock clusters are uniformly distributed in the logarithmic space of mass range of $[0.8,30]h_{70}^{-1}10^{14}M_\odot$ and in the linear space of the redshift range of $[0.1,1]$. The concentration parameter uses \cite{Bhattacharya13}. The adopted number density of the background galaxies follows our result (bottom panel of Fig. \ref{fig:z_vs_M_WLcal}). We randomly pick up galaxies from the whole HSC catalog \citep{HSC-3Y-Shape, 2020arXiv200301511N}, used in our analysis by satisfying eq. (\ref{eq:Pz_sel}) as background galaxies. The catalog contains the photometric redshift, its probability, and ellipticity. 
 The spatial distributions of background galaxies are uniformly distributed, where we do not consider star mask effects. The intrinsic shape ellipticity is calculated by randomly rotating the catalog ellipticity. We then synthesize the lensing signal to the intrinsic ellipticity. We make 9000 mock clusters and then repeat the same tangential shear analysis.
 
The upper panel of Fig. \ref{fig:z_vs_M_WLcal} shows a measurable fraction in the $z$ and $M_{200}$ plane. For a visual purpose, we smoothed it with a Gaussian kernel with $dz=0.1$ and $\ln M_{200}^{\rm true} = 0.35$. We successfully measure cluster WL masses at $z\simlt 0.4$ and $M_{200}\simgt 5\times 10^{14}h_{70}^{-1}M_\odot$ with more than $95\%$. In contrast,  the higher the redshift and the lower the mass, the more difficult it is to measure WL masses. Therefore, our measured clusters selectively have a positive bias in the WL mass. 
The gray shadow region represents the $z$ and $M_{200}^{\rm true}$ distribution for our sample of the eRASS1 clusters, obtained by the scaling relation analysis (Sec. \ref{subsec:scaling}). The sample is distributed around $\sim95\%$ measurable fraction. Given the 75 association clusters, the number of the measurable clusters is $\sim 0.95 \times 75\sim 71$, which agrees with our case of the $72$ clusters.

There are two proposed WL mass calibrations. First is to calibrate between $M_{500}^{\rm WL}$ and $M_{500}^{\rm true}$ used in the scaling relations. The second is to calibrate the mass and concentration relation. 

\begin{figure}[!ht]
    \centering
    \includegraphics[width=\linewidth]{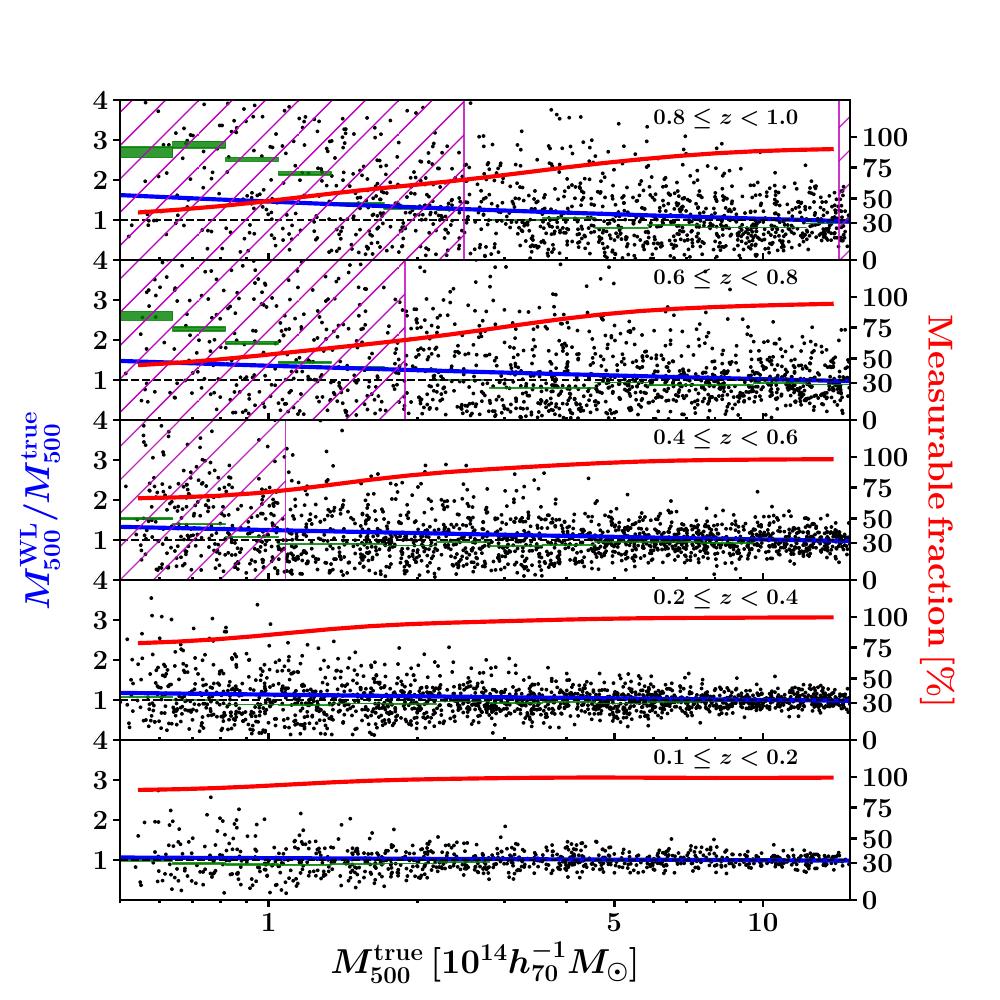}
    \caption{Mass bias, $M^{\rm WL}_{500}/M_{500}^{\rm true}$, in five redshift bins. The black points are the mass bias for individual mock clusters. We ignore their errors for a visual purpose. The green boxes are the average WL bias in each mass bin. The red solid lines are the measurable fraction. The higher the redshift and the lower the mass, the more difficult it is to measure WL masses and the lower the measurable fraction. This leads to a selection mass bias in the measured quantities. The magenta hatched boxes are excluded regions in the mass bias fit, where the probability of the target clusters being present is less than $10^{-6}$. The blue solid lines are the best-fit scaling relations for the mass bias.}
    \label{fig:mbias_vs_M_WLcal}
\end{figure}

For the WL mass calibration of $M_{500}$, we use the $x$ parameters only. 
Figure \ref{fig:mbias_vs_M_WLcal} shows the mass bias (black circles), $M_{500}^{\rm WL}/M_{\rm 500}$, for the mock clusters without the measurement errors and its average values (green boxes) in each mass and redshift bin. 
The measurable fraction (red solid lines) decreases as the mass is lower and the redshift is higher. 
In the region around the true mass where the measurable fraction is generally lower than $50\%$, the black points show a selective deviation from unity, and the average values exhibit positive biases. 
In contrast, the mass bias where the measurable fraction is more than $\sim 75\%$ is close to unity. 
The mass dependence of the mass bias is likely to be non-linear. 
Thus, it is essential to reflect our mass distribution so that a linear equation can describe the WL mass calibration. We first performed the scaling relation analysis without the WL mass calibration and measured the parent population. We then excluded the mass range where the probability of the target clusters being present is less than $10^{-6}$. The magenta hatched boxes in the Fig. \ref{fig:mbias_vs_M_WLcal} illustrate the excluded ranges for mass and redshift where the mass bias changes non-linearly. The resulting baseline parameters ( blue solid lines in Fig. \ref{fig:mbias_vs_M_WLcal} ) are listed in Table \ref{table:WLcalib}. There is no significant alteration in the results when we either apply the minimum threshold of $10^{-5}$ or choose a random $90\%$ sample of the clusters.

\begin{figure}[!ht]
    \centering
    \includegraphics[width=\linewidth]{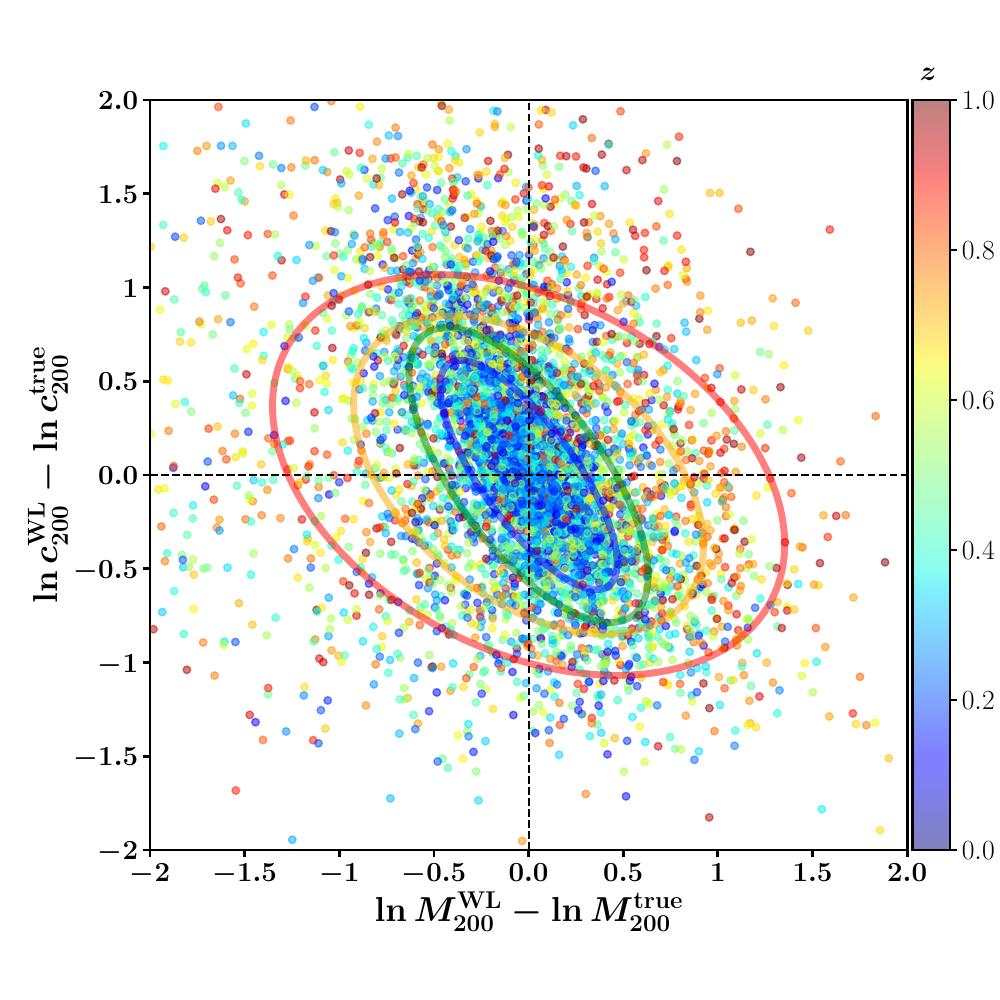}
    \caption{The points represent the logarithmic WL biases ($\ln M_{200}^{\rm WL}-\ln M_{200}^{\rm true}$ versus $\ln c_{200}^{\rm WL}-\ln c_{200}^{\rm true}$) for the sample with the presence probability that is higher than $10^{-6}$. 
    The colors denote redshifts. The elliptical lines in blue, green, orange, and red represent the covariance distribution composed of the intrinsic covariance and typical measurement errors at $z=0.2$, $0.4$, $0.6$, and $0.8$, respectively. }
    \label{fig:mbias_vs_cbias}
\end{figure}

For $M_{200}^{\rm WL}$ and $c_{200}^{\rm WL}$, we select the clusters with the condition that the probability of the target clusters being present is larger than $10^{-6}$. 
The WL biases in $M_{200}$ and $c_{200}$ are correlated, as shown in Fig. \ref{fig:mbias_vs_cbias}. We evaluated the WL mass calibration parameters by simultaneously fitting all the measurements with the covariance error matrix. 
The result is shown in Table \ref{table:WLcalib}. Since the intrinsic WL scatter in the mass and concentration is highly coupled, the intrinsic WL scatter becomes large. In the regression analysis, we used the parameters of the WL mass calibration as priors, taking into account its error covariance matrix.

\begin{table}[!ht]
\centering   
\caption{WL mass calibrations for $M_{200}^{\rm WL}$, $c_{200}^{\rm WL}$ and $M_{500}^{\rm WL}$, where $\sigma_{\rm WL}=\exp(\ln \sigma_{\rm WL})$.} \label{table:WLcalib}
\begin{tabular}{c|cc}
& $M_{200}^{\rm WL}$/$c_{200}^{\rm WL}$ & $M_{500}^{\rm WL}$ \\ 
&  $[10^{14}h^{-1}M_\odot$] &  [$10^{14}h_{70}^{-1}M_\odot$]\\
\hline
$M_p$ & $10^{14}h^{-1}M_\odot$ &  $10^{14}h_{70}^{-1}M_\odot$ \\
$c_p$ & $4$ & - \\

 $\alpha_{{\rm WL},x}$ & $0.110_{-0.021}^{+0.021}$ &  $0.079_{-0.024}^{+0.024}$\\
 $\beta_{{\rm WL},x}$ & $0.933_{-0.009}^{+0.009}$  & $0.965_{-0.011}^{+0.011}$\\
 $\gamma_{{\rm WL},x}$ & $1.609_{-0.174}^{+0.173}$ &  $0.665_{-0.198}^{+0.196}$\\
 $\delta_{{\rm WL},x}$ & $-0.375_{-0.042}^{+0.041}$ &   $-0.258_{-0.090}^{+0.090}$ \\
 $\alpha_{{\rm WL},y}$ & $0.108_{-0.017}^{+0.017}$  &-  \\
 $\beta_{{\rm WL},y}$ & $0.947_{-0.029}^{+0.028}$   &- \\
 $\gamma_{{\rm WL},y}$ & $0.613_{-0.141}^{+0.139}$   &- \\ 
 $\delta_{{\rm WL},y}$ & $-0.586_{-0.120}^{+0.120}$  &- \\
 $\sigma_{{\rm WL},x}$ & $0.182_{-0.010}^{+0.010}$  & $0.016_{-0.003}^{+0.003}$ \\
 $\gamma_{\sigma_{{\rm WL},x}}$ & $-0.619_{-0.179}^{+0.200}$  & $-2.031_{-0.025}^{+0.026}$  \\
 $\sigma_{{\rm WL},y}$ & $0.319_{-0.010}^{+0.011}$  &-  \\
 $\gamma_{\sigma_{{\rm WL},y}}$ & $-1.323_{-0.146}^{+0.150}$  \\
 $r_{{\rm WL}}$ & $-0.999_{-0.000}^{+0.001}$  &- \\
 $\gamma_{r_{{\rm WL}}}$ & $0\,({\rm fixed})$  &-  
\end{tabular}
\end{table}

 \section{Mass-richness-CR relation with $E(z)$ }

Since the redshift dependence of the overdensity masses follows $M_\Delta \propto \rho_{\rm cr}(z) r_\Delta^3 \propto E(z)^{-1}$ and the soft-band luminosity follows $L_X\propto  \rho_{\rm cr}(z)^2r_\Delta^3\propto E(z)$, we here replace $M_{500}$, ${\rm CR}$, and $\lambda$ by $M_{500}E(z)$, ${\rm CR}E(z)^{-1}$, and $\lambda E(z)$ in the scaling relation analysis, respectively. The result changes only the normalization in response to the replacement (the first column in Table \ref{tab:scaling_relation_MEz}).

\begin{table}[!ht]
    \centering
       \caption{Scaling relation parameters using $M_{500}E(z)$, ${\rm CR}E(z)^{-1}$, and $\lambda E(z)$ or CAMIRA richness, $N$. }
    \label{tab:scaling_relation_MEz}
    \begin{tabular}{c|cc}
     & $E(z)$ & CAMIRA \\
\hline   
 $\alpha_{\rm CR}$ & $-1.786_{-0.575}^{+0.339}$ 
                     & $-1.350_{-0.389}^{+0.336}$ \\
 $\beta_{\rm CR}$  & $1.306_{-0.292}^{+0.506}$ 
                     & $1.012_{-0.265}^{+0.308}$ \\
 $\gamma_{\rm CR}$  & $1.544_{-2.817}^{+1.482}$ 
                      & $2.508_{-1.539}^{+1.143}$ \\
 $\delta_{\rm CR}$  & 0 (fixed) 
                      & 0 (fixed) \\
 $\alpha_{\lambda}$ & $-1.167_{-0.362}^{+0.256}$ 
                     & $-1.109_{-0.241}^{+0.207}$ \\
 $\beta_{\lambda}$  & $1.108_{-0.184}^{+0.244}$ 
                     & $1.053_{-0.161}^{+0.168}$ \\
 $\gamma_{\lambda}$  & $-0.753_{-1.056}^{+0.967}$ 
                      & $-2.168_{-0.916}^{+0.808}$ \\
 $\delta_{\lambda}$  & 0 (fixed) 
                      & 0 (fixed) \\
 $\sigma_{\rm CR}$  & $0.226_{-0.164}^{+0.180}$ 
                      & $0.463_{-0.134}^{+0.136}$ \\
 $\sigma_{\lambda}$  & $0.368_{-0.140}^{+0.106}$ 
                       & $0.357_{-0.196}^{+0.115}$\\ 
 $r_{{\rm CR,WL}}$  & $0$ (fixed)
                     & $0$ (fixed)\\
 $r_{\lambda,{\rm WL}}$  & $0$ (fixed)
                           & $0$ (fixed)\\
 $r_{{\rm CR},\lambda}$  & $-0.299_{-0.420}^{+0.518}$
                           & $-0.419_{-0.344}^{+0.369}$\\
    \end{tabular}
\tablefoot{The fixed parameters are the same as (4) in Table \ref{table:scaling}.} 
\end{table}

\section{Mass-richness-CR relation with the CAMIRA richness }

We cross-match the HSC CAMIRA clusters \citep[][Oguri  et al. in prep.]{2018PASJ...70S..20O} with the eRASS1 clusters. We used the internal CAMIRA catalog with the S21A version. The tolerances for the spatial and redshift differences are 4 arcmin and 0.1 in the crossmatching. The 68 clusters out of the 78 eRASS1 clusters are matched. In the scaling relation analysis, we replace the richness, $\lambda$, by the CAMIRA richness $N_{\rm camira}$. Since the masses of two clusters are not measurable, the number of clusters is 66 in the scaling relation analysis. The resulting parameters are described in the second column in Table \ref{tab:scaling_relation_MEz}. The result changes little except for the redshift-dependent slope in the richness. 

\begin{figure}[!ht]
    \centering
    \includegraphics[width=\linewidth]{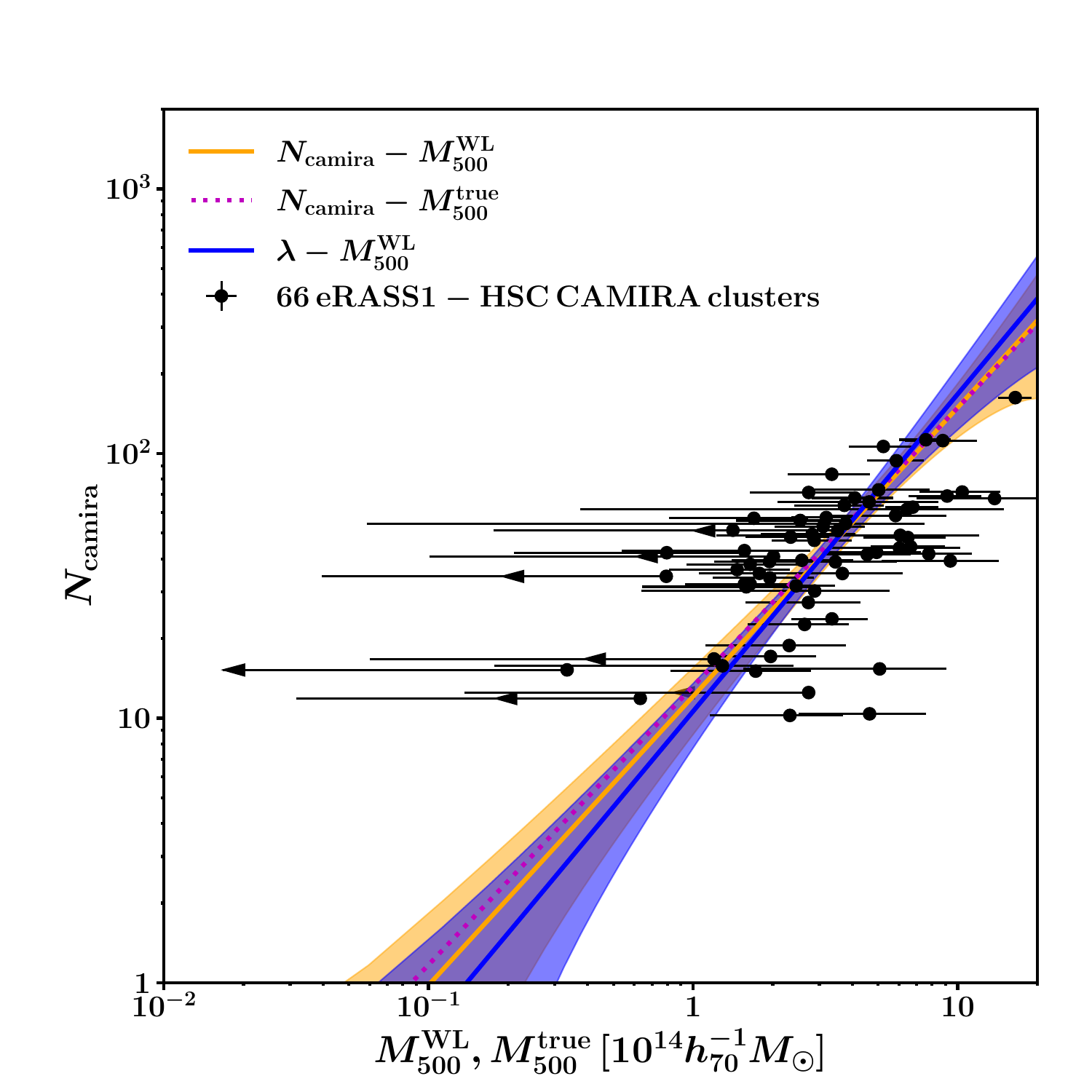}
    \caption{The CAMIRA richness and mass scaling relation.  The black circles denote the overlapped 66 clusters.  The orange solid and magenta dotted lines are the best-fit scaling relations concerning the WL and true masses, respectively. The orange region is the $1\sigma$ uncertainty of the scaling relation with the WL masses. The blue solid line and region are the best-fit and $1\sigma$ uncertainty with the eRASS1 richness (Fig. \ref{fig:lambda-Mwl}), respectively.}
    \label{fig:camira richness}
\end{figure}

\section{WL Mass table}

Table \ref{table:cog1} presents the WL masses and concentrations of the eRASS1 clusters.

\clearpage
\onecolumn
\begin{landscape}
\begin{longtable}{c|cccccc}
\caption{\label{table:cog1} Cluster properties and their NFW weak-lensing parameters of $M_{500}^{\rm WL}$, $M_{200}^{\rm WL}$, and $c_{200}^{\rm WL}$.} \\
\hline
Name\tablefootmark{a} & alt-name\tablefootmark{b} & z\tablefootmark{c} & Richness\tablefootmark{d}, $\lambda$ & $M_{500}^{\rm WL}$\tablefootmark{e}, & $M_{200}^{\rm WL}$\tablefootmark{f}, & $c_{200}^{\rm WL}$\tablefootmark{g}, \\
      &         &   &  & [$10^{14}\hubbleMsol$] &  [$10^{14}\hubbleMsol$] & \\
\hline
\endfirsthead
\caption{continued.}\\
\hline
\endhead
\hline
\endfoot
\hline
\endlastfoot
J083652.6+030000  &  {\small WHL J083651.5+030001}
                  & $0.189$ 
                  & $41.68\pm3.39$ 
                  & $4.55_{-1.45}^{+2.10}$ 
                  & $6.54_{-2.72}^{+3.69}$ 
                  & $4.08_{-4.08}^{+18.85}$ \\
J084306.7+002834  &  {\small WHL J084305.1+002847}
                  & $0.270$ 
                  & $30.15\pm1.98$ 
                  & $1.64_{-0.70}^{+0.87}$ 
                  & $2.22_{-1.03}^{+1.40}$ 
                  & $5.87_{-2.99}^{+9.23}$ \\
J084342.8+040323  &  {\small WHL J084343.1+040335}
                  & $0.208$ 
                  & $20.42\pm2.65$ 
                  & $1.72_{-0.90}^{+1.06}$ 
                  & $2.25_{-1.74}^{+1.74}$ 
                  & $7.34_{-7.34}^{+11.80}$ \\
J084527.7+032736  &  {\small ZwCl 0842.8+0336 / WHL J084528.3+032638}
                  & $0.326$ 
                  & $109.32\pm5.40$ 
                  & $5.24_{-1.35}^{+1.62}$ 
                  & $6.99_{-2.10}^{+2.78}$ 
                  & $6.22_{-2.58}^{+5.38}$ \\
J085029.4+001453  &  {\small WHL J085023.9+001536}
                  & $0.196$ 
                  & $63.28\pm4.34$ 
                  & $2.82_{-1.02}^{+1.28}$ 
                  & $4.81_{-2.07}^{+3.45}$ 
                  & $2.08_{-1.93}^{+3.19}$ \\
J085217.5-010126\tablefootmark{$\natural$}  &  {\small WHL J085216.7-010136}
                  & $0.459$ 
                  & $89.51\pm5.09$ 
                  & $1.20_{-1.20}^{+1.53}$ 
                  & $1.61_{-1.61}^{+2.27}$ 
                  & $6.20_{-6.20}^{+30.66}$ \\
J085230.8+002509  &  {\small WHL J085230.3+002511}
                  & $0.283$ 
                  & $41.38\pm3.06$ 
                  & $2.87_{-0.89}^{+1.11}$ 
                  & $3.72_{-1.28}^{+1.73}$ 
                  & $7.62_{-3.08}^{+6.44}$ \\
J085435.8+003858  &  {\small WHL J085436.6+003833 / HSCS J085437+003946}
                  & $0.110$ 
                  & $22.37\pm2.57$ 
                  & $1.95_{-0.76}^{+0.91}$ 
                  & $2.74_{-1.13}^{+1.51}$ 
                  & $4.62_{-2.68}^{+6.65}$ \\
J085751.0+031014  &  {\small ABELL 0732 / WHL J085754.0+031035}
                  & $0.204$ 
                  & $89.35\pm4.60$ 
                  & $3.34_{-1.06}^{+1.32}$ 
                  & $4.51_{-1.60}^{+2.23}$ 
                  & $5.79_{-2.45}^{+4.57}$ \\
J085932.4+030832  &  {\small WHL J085932.3+030841}
                  & $0.197$ 
                  & $45.27\pm3.62$ 
                  & $3.35_{-1.00}^{+1.21}$ 
                  & $4.67_{-1.59}^{+2.11}$ 
                  & $4.85_{-2.03}^{+3.69}$ \\
J090131.4+030055  &  {\small WHL J090121.9+030208}
                  & $0.196$ 
                  & $60.05\pm5.23$ 
                  & $2.45_{-0.91}^{+0.99}$ 
                  & $3.13_{-1.19}^{+1.55}$ 
                  & $8.61_{-8.61}^{+14.50}$ \\
J091414.7+001922  &  {\small WHL J091422.8+001658}
                  & $0.167$ 
                  & $19.35\pm5.64$ 
                  & $1.29_{-1.11}^{+1.10}$ 
                  & $2.48_{-1.75}^{+2.66}$ 
                  & $1.47_{-1.47}^{+40.37}$ \\
J091453.6+041611  &  {\small NSC J091454+041559}
                  & $0.143$ 
                  & $27.60\pm2.84$ 
                  & $2.43_{-1.06}^{+1.63}$ 
                  & $3.48_{-2.50}^{+2.50}$ 
                  & $4.12_{-4.12}^{+17.53}$ \\
J091608.3-002355  &  {\small ABELL 0776 / WHL J091605.7-002324 / HSCS J091606-002338}
                  & $0.330$ 
                  & $105.57\pm5.16$ 
                  & $8.78_{-2.46}^{+3.04}$ 
                  & $14.15_{-4.57}^{+6.63}$ 
                  & $2.54_{-1.01}^{+1.37}$ \\
J092050.7+024512  &  {\small WHL J092049.7+024514 / HSCS J092050+024606}
                  & $0.279$ 
                  & $65.36\pm10.62$ 
                  & $2.54_{-1.08}^{+1.40}$ 
                  & $4.06_{-2.15}^{+3.48}$ 
                  & $2.62_{-2.62}^{+11.72}$ \\
J092121.0+031735  &  {\small WHL J092121.1+031713}
                  & $0.347$ 
                  & $106.39\pm3.94$ 
                  & $5.86_{-1.32}^{+1.56}$ 
                  & $7.68_{-1.95}^{+2.40}$ 
                  & $7.21_{-2.04}^{+3.26}$ \\
J092210.0+034626  &  {\small WHL J092207.6+034558}
                  & $0.279$ 
                  & $91.24\pm4.55$ 
                  & $3.73_{-1.31}^{+1.54}$ 
                  & $6.64_{-2.51}^{+3.61}$ 
                  & $1.82_{-1.07}^{+1.65}$ \\
J093025.3+021707  &  {\small WHL J093024.2+021728}
                  & $0.540$ 
                  & $75.50\pm4.36$ 
                  & $4.62_{-2.54}^{+3.82}$ 
                  & $6.24_{-4.97}^{+7.01}$ 
                  & $5.85_{-5.85}^{+32.71}$ \\
J093150.9-002220  &  {\small WHL J093151.0-002213}
                  & $0.346$ 
                  & $23.91\pm4.05$ 
                  & $2.73_{-1.15}^{+1.56}$ 
                  & $3.85_{-1.78}^{+2.87}$ 
                  & $4.51_{-2.56}^{+5.85}$ \\
J093459.9+005438  &  {\small WHL J093500.7+005414}
                  & $0.377$ 
                  & $47.25\pm3.00$ 
                  & $3.02_{-1.83}^{+2.17}$ 
                  & $4.90_{-2.75}^{+3.98}$ 
                  & $2.49_{-2.49}^{+43.08}$ \\
J093512.7+004735  &  {\small WHL J093513.0+004846}
                  & $0.365$ 
                  & $128.92\pm5.33$ 
                  & $13.78_{-6.82}^{+12.88}$ 
                  & $29.15_{-29.15}^{+43.73}$ 
                  & $1.17_{-1.17}^{+1.65}$ \\
J093521.3+023222  &  {\small WHL J093522.5+023324}
                  & $0.499$ 
                  & $136.83\pm5.88$ 
                  & $5.03_{-2.34}^{+2.80}$ 
                  & $6.36_{-3.37}^{+4.32}$ 
                  & $9.45_{-9.45}^{+108.49}$ \\
J094023.3+022824  &  {\small ABELL 0847 / WHL J094024.6+022840}
                  & $0.154$ 
                  & $53.12\pm3.11$ 
                  & $6.77_{-1.43}^{+1.67}$ 
                  & $11.43_{-2.86}^{+3.83}$ 
                  & $2.15_{-0.72}^{+0.92}$ \\
J094611.9+022201  &  {\small ABELL 0869 / WHL J094612.0+022208}
                  & $0.123$ 
                  & $20.81\pm1.60$ 
                  & $1.94_{-0.74}^{+0.93}$ 
                  & $2.95_{-1.27}^{+1.86}$ 
                  & $3.20_{-1.92}^{+4.57}$ \\
J094844.3+020019  &  {\small }
                  & $0.490$ 
                  & $20.93\pm6.53$ 
                  & $4.64_{-2.14}^{+2.95}$ 
                  & $7.03_{-5.45}^{+5.45}$ 
                  & $3.24_{-3.24}^{+28.70}$ \\
J095341.6+014200  &  {\small SDSS-C4-DR3 1212 / RXC J0953.6+0142}
                  & $0.109$ 
                  & $10.82\pm1.78$ 
                  & $1.96_{-0.77}^{+0.95}$ 
                  & $2.67_{-1.18}^{+1.66}$ 
                  & $5.59_{-3.12}^{+7.56}$ \\
J095736.6+023430  &  {\small WHL J095737.0+023426}
                  & $0.382$ 
                  & $41.68\pm3.71$ 
                  & $2.32_{-1.17}^{+1.37}$ 
                  & $3.77_{-1.88}^{+2.64}$ 
                  & $2.48_{-1.77}^{+3.96}$ \\
J095858.4-001323\tablefootmark{$\natural$}  &  {\small SDSS CE J149.746765-00.198446}
                  & $0.175$ 
                  & $48.87\pm3.17$ 
                  & $0.55_{-0.55}^{+0.64}$ 
                  & $0.72_{-0.72}^{+1.01}$ 
                  & $7.43_{-7.43}^{+21.34}$ \\
J101851.0-010105\tablefootmark{$\natural$}  &  {\small WHL J101852.4-005857}
                  & $0.114$ 
                  & $9.20\pm1.32$ 
                  & $0.26_{-0.26}^{+0.38}$ 
                  & $0.37_{-0.37}^{+0.60}$ 
                  & $4.13_{-4.13}^{+9.01}$ \\
J102250.3-000309  &  {\small SDSS CE J155.708923-00.050793}
                  & $0.302$ 
                  & $29.02\pm2.08$ 
                  & $1.56_{-1.02}^{+1.20}$ 
                  & $2.05_{-1.60}^{+2.17}$ 
                  & $7.06_{-7.06}^{+33.71}$ \\
J105039.5+001707  &  {\small SDSS J1050+0017 Cluster}
                  & $0.597$ 
                  & $64.21\pm3.44$ 
                  & $9.38_{-4.22}^{+4.92}$ 
                  & $11.83_{-7.89}^{+7.89}$ 
                  & $9.47_{-9.47}^{+13.43}$ \\
J111111.5+004454  &  {\small ABELL 1191 / WHL J111111.2+004508}
                  & $0.188$ 
                  & $44.23\pm4.49$ 
                  & $2.34_{-0.76}^{+0.93}$ 
                  & $3.14_{-1.14}^{+1.51}$ 
                  & $6.00_{-2.70}^{+5.84}$ \\
J112626.5+003625\tablefootmark{$\natural$}  &  {\small }
                  & $0.305$ 
                  & $10.84\pm1.55$ 
                  & $0.45_{-0.45}^{+0.65}$ 
                  & $0.70_{-0.70}^{+1.08}$ 
                  & $2.86_{-2.86}^{+4.90}$ \\
J112818.3-005859  &  {\small SDSS CE J172.069611-00.983567}
                  & $0.456$ 
                  & $74.51\pm3.60$ 
                  & $10.40_{-3.22}^{+4.07}$ 
                  & $17.56_{-6.31}^{+9.81}$ 
                  & $2.15_{-0.93}^{+1.28}$ \\
J113655.7+000612  &  {\small WHL J113658.7+000602}
                  & $0.585$ 
                  & $124.71\pm10.25$ 
                  & $5.07_{-3.52}^{+4.01}$ 
                  & $9.32_{-7.59}^{+13.31}$ 
                  & $1.66_{-1.66}^{+5.07}$ \\
J113843.1+031510\tablefootmark{$\natural$}  &  {\small BLOX J1138.7+0315.6}
                  & $0.137$ 
                  & $8.29\pm1.34$ 
                  & - 
                  & - 
                  & - \\
J114441.9+004414  &  {\small SDSS CE J176.174454+00.738345 / HSCS J114441+004304}
                  & $0.334$ 
                  & $19.87\pm1.96$ 
                  & $2.31_{-1.20}^{+1.47}$ 
                  & $3.02_{-1.48}^{+2.79}$ 
                  & $7.11_{-7.11}^{+26.43}$ \\
J114647.4-012428\tablefootmark{*}  &  {\small WHL J114641.6-012324}
                  & $0.331$ 
                  & $5.31\pm1.23$ 
                  & - 
                  & - 
                  & -\\
      J115019.2-003637  &  {\small ABELL 1392$^\dagger$ / WHL J115018.7-003631 / HSCS J115028-003723}
                  & $0.142$ 
                  & $32.67\pm2.68$ 
                  & $3.19_{-0.83}^{+1.03}$ 
                  & $4.21_{-1.23}^{+1.65}$ 
                  & $6.73_{-2.31}^{+3.92}$ \\
J115208.5-004726\tablefootmark{$\natural$}  &  {\small WHL J115205.2-004616 / HSCS J115212-004511}
                  & $0.259$ 
                  & $4.84\pm1.34$ 
                  & $0.02_{-0.02}^{+0.31}$ 
                  & $0.03_{-0.03}^{+0.44}$ 
                  & $14.18_{-14.18}^{+266.93}$ \\
J115214.5+003057  &  {\small WHL J115214.2+003126 / HSCS J115216+003055}
                  & $0.470$ 
                  & $85.97\pm3.88$ 
                  & $9.12_{-2.60}^{+3.16}$ 
                  & $13.13_{-4.22}^{+5.65}$ 
                  & $4.08_{-1.33}^{+1.89}$ \\
J115235.7+035642\tablefootmark{$\natural$}  &  {\small WHL J115237.0+035434 / ACT-CL J1152.5+0356}
                  & $0.681$ 
                  & $67.55\pm7.79$ 
                  & $1.43_{-1.43}^{+2.08}$ 
                  & $1.74_{-1.74}^{+2.91}$ 
                  & $14.07_{-14.07}^{+43.30}$ \\
J115417.0+022123  &  {\small ACT-CL J1154.2+0221}
                  & $0.743$ 
                  & $42.78\pm3.56$ 
                  & $6.06_{-4.84}^{+6.01}$ 
                  & $7.97_{-5.35}^{+9.62}$ 
                  & $6.96_{-3.76}^{+18.28}$ \\
J115620.0-001220  &  {\small ABELL 1419 / WHL J115620.1-001220 / HSCS J115612-001237}
                  & $0.110$ 
                  & $22.66\pm2.45$ 
                  & $1.59_{-0.95}^{+1.18}$ 
                  & $2.40_{-2.40}^{+2.71}$ 
                  & $3.28_{-3.28}^{+6.19}$ \\
J120024.4+032112  &  {\small ABELL 1437 / WHL J120025.3+032049}
                  & $0.136$ 
                  & $101.18\pm4.36$ 
                  & $7.58_{-1.56}^{+1.81}$ 
                  & $12.57_{-3.07}^{+4.03}$ 
                  & $2.29_{-0.73}^{+0.93}$ \\
J120143.7-001110  &  {\small ABELL 1445 / WHL J120143.7-001104 / HSCS J120141-000953}
                  & $0.170$ 
                  & $40.73\pm2.91$ 
                  & $3.10_{-1.06}^{+1.36}$ 
                  & $4.77_{-1.88}^{+2.97}$ 
                  & $3.03_{-1.64}^{+2.62}$ \\
J120417.3+025402  &  {\small }
                  & $0.149$ 
                  & $7.91\pm1.81$ 
                  & $1.61_{-0.90}^{+0.93}$ 
                  & $2.21_{-1.11}^{+1.36}$ 
                  & $5.20_{-5.20}^{+13.98}$ \\
J120954.5-003328  &  {\small SDSS CE J182.479843-00.560958 / HSCS J120946-003512}
                  & $0.184$ 
                  & $26.87\pm2.24$ 
                  & $1.78_{-0.73}^{+0.96}$ 
                  & $2.27_{-1.03}^{+1.51}$ 
                  & $8.93_{-7.60}^{+19.24}$ \\
J121016.3+022338  &  {\small WHL J121017.7+022343}
                  & $0.381$ 
                  & $54.40\pm3.42$ 
                  & $4.94_{-1.86}^{+2.34}$ 
                  & $6.43_{-2.70}^{+3.82}$ 
                  & $7.39_{-4.01}^{+10.53}$ \\
J121336.5+025331  &  {\small WHL J121334.5+025355}
                  & $0.394$ 
                  & $87.86\pm15.88$ 
                  & $2.57_{-1.17}^{+1.46}$ 
                  & $3.48_{-1.75}^{+2.51}$ 
                  & $5.80_{-5.80}^{+33.75}$ \\
J122420.6+021206  &  {\small WHL J122422.0+021211}
                  & $0.451$ 
                  & $46.70\pm3.38$ 
                  & $6.48_{-2.08}^{+2.54}$ 
                  & $10.18_{-3.71}^{+5.17}$ 
                  & $2.79_{-1.16}^{+1.64}$ \\
J122528.9+004237\tablefootmark{$\natural$}  &  {\small WHL J122529.3+004334 / XMMXCS J122528.0+004229.2}
                  & $0.242$ 
                  & $17.91\pm4.67$ 
                  & - 
                  & - 
                  & - \\
J122644.5-003724  &  {\small NSC J122644-003738}
                  & $0.159$ 
                  & $36.01\pm2.30$ 
                  & $0.80_{-0.58}^{+0.69}$ 
                  & $1.18_{-0.77}^{+1.01}$ 
                  & $3.50_{-3.50}^{+18.06}$ \\
J123108.4+003653\tablefootmark{*}  &  {\small }
                  & $0.473$ 
                  & $9.60\pm9.27$ 
                  & - 
                  & - 
                  & - \\
J123755.2-001611  &  {\small ABELL 1577}
                  & $0.141$ 
                  & $37.91\pm2.41$ 
                  & $1.41_{-0.56}^{+0.72}$ 
                  & $1.88_{-1.04}^{+1.27}$ 
                  & $6.48_{-6.48}^{+18.95}$ \\
J124503.8-002823\tablefootmark{*}  &  {\small }
                  & $0.231$ 
                  & $34.71\pm20.69$ 
                  & - 
                  & - 
                  & - \\
J125035.8+003646  &  {\small ACT-CL J1250.6+0036}
                  & $0.634$ 
                  & $38.93\pm3.29$ 
                  & $3.77_{-3.71}^{+3.71}$ 
                  & $4.95_{-4.95}^{+5.61}$ 
                  & $6.65_{-6.65}^{+8.63}$ \\
J131129.4-012026  &  {\small ABELL 1689 / WHL J131132.1-011946}
                  & $0.184$ 
                  & $183.14\pm5.32$ 
                  & $16.50_{-2.28}^{+2.55}$ 
                  & $21.08_{-3.72}^{+4.52}$ 
                  & $8.58_{-3.30}^{+7.24}$ \\
J134840.1+003907\tablefootmark{$\natural$}  &  {\small SDSS CE J207.185104+00.662313}
                  & $0.422$ 
                  & $33.69\pm2.49$ 
                  & $0.87_{-0.87}^{+1.14}$ 
                  & $1.70_{-1.67}^{+2.02}$ 
                  & $1.41_{-1.41}^{+14.08}$ \\
J135326.5+000255  &  {\small WHL J135326.3+000248}
                  & $0.107$ 
                  & $13.57\pm1.95$ 
                  & $1.64_{-0.64}^{+0.79}$ 
                  & $2.33_{-1.01}^{+1.36}$ 
                  & $4.39_{-2.29}^{+4.98}$ \\
J135424.2-010250  &  {\small ZwCl 1351.8-0048 / WHL J135421.8-010000}
                  & $0.152$ 
                  & $52.01\pm3.51$ 
                  & $4.08_{-1.28}^{+1.64}$ 
                  & $6.45_{-2.38}^{+3.78}$ 
                  & $2.74_{-1.35}^{+2.09}$ \\
J140707.1-001450  &  {\small SDSS CE J211.765610-00.254689 / HSCS J140709-001436}
                  & $0.554$ 
                  & $74.25\pm3.73$ 
                  & $2.88_{-2.24}^{+2.63}$ 
                  & $4.51_{-3.12}^{+5.49}$ 
                  & $2.83_{-2.83}^{+10.42}$ \\
J141452.4+001316\tablefootmark{$\natural$}  &  {\small SDSS CE J213.734085+00.213623}
                  & $0.121$ 
                  & $24.87\pm2.31$ 
                  & $0.27_{-0.27}^{+0.52}$ 
                  & $0.49_{-0.49}^{+1.12}$ 
                  & $1.80_{-1.80}^{+4.48}$ \\
J141457.8-002050\tablefootmark{$\natural$}  &  {\small SDSS-C4 1031}
                  & $0.133$ 
                  & $99.48\pm4.26$ 
                  & $0.61_{-0.61}^{+1.78}$ 
                  & $0.84_{-0.84}^{+2.95}$ 
                  & $4.44_{-4.44}^{+16.28}$ \\
J141507.1-002905  &  {\small ABELL 1882 / WHL J141508.4-002935}
                  & $0.133$ 
                  & $70.83\pm4.07$ 
                  & $2.34_{-1.19}^{+4.10}$ 
                  & $3.01_{-2.66}^{+6.84}$ 
                  & $8.19_{-7.59}^{+16.50}$ \\
J141547.1+001530  &  {\small WHL J141548.8+001537 / HSCS J141547+001709}
                  & $0.128$ 
                  & $29.03\pm3.25$ 
                  & $1.56_{-0.63}^{+0.84}$ 
                  & $2.28_{-1.03}^{+1.61}$ 
                  & $3.85_{-2.16}^{+5.55}$ \\
J142016.3+005716  &  {\small WHL J142016.6+005719}
                  & $0.499$ 
                  & $60.52\pm2.59$ 
                  & $6.64_{-1.94}^{+2.32}$ 
                  & $10.10_{-3.31}^{+4.46}$ 
                  & $3.19_{-1.31}^{+1.97}$ \\
J142724.6-001509  &  {\small SDSS CE J216.852905-00.249845 / HSCS J142723-001638}
                  & $0.164$ 
                  & $25.85\pm7.55$ 
                  & $1.47_{-0.65}^{+0.86}$ 
                  & $2.17_{-1.07}^{+1.69}$ 
                  & $3.58_{-2.28}^{+5.31}$ \\
J143044.6+004815  &  {\small WHL J143042.1+004902}
                  & $0.311$ 
                  & $36.78\pm2.61$ 
                  & $7.77_{-2.40}^{+3.53}$ 
                  & $14.82_{-5.75}^{+13.79}$ 
                  & $1.50_{-0.90}^{+1.11}$ \\
J143121.5-005336  &  {\small SDSS CE J217.834290-00.898368}
                  & $0.402$ 
                  & $70.96\pm3.67$ 
                  & $2.73_{-1.09}^{+1.40}$ 
                  & $3.49_{-1.53}^{+2.24}$ 
                  & $8.40_{-4.72}^{+14.27}$ \\
J143736.1-001740  &  {\small ABELL 1938 / WHL J143743.2-001512 / HSCS J143741-001819}
                  & $0.140$ 
                  & $44.59\pm3.10$ 
                  & $1.70_{-0.89}^{+1.02}$ 
                  & $3.59_{-1.58}^{+2.13}$ 
                  & $1.14_{-0.73}^{+1.13}$ \\
J144050.2+004103\tablefootmark{$\natural$}  &  {\small WHL J144048.9+004020}
                  & $0.545$ 
                  & $30.01\pm4.06$ 
                  & - 
                  & - 
                  & - \\
J144133.6-005418  &  {\small SDSS CE J220.382751-00.899307 / ACT-CL J1441.5-0053}
                  & $0.543$ 
                  & $51.99\pm3.73$ 
                  & $5.82_{-2.63}^{+3.25}$ 
                  & $12.66_{-5.91}^{+10.81}$ 
                  & $1.06_{-0.79}^{+1.14}$ \\
J144309.6+010214  &  {\small WHL J144309.3+010211}
                  & $0.534$ 
                  & $60.35\pm3.95$ 
                  & $6.02_{-2.71}^{+4.18}$ 
                  & $8.38_{-4.27}^{+8.15}$ 
                  & $4.83_{-4.17}^{+9.34}$ \\
J144950.6+005558  &  {\small WHL J144945.7+005705 / HSCS J144954+005717}
                  & $0.409$ 
                  & $47.60\pm3.44$ 
                  & $3.67_{-1.83}^{+2.55}$ 
                  & $6.70_{-3.59}^{+8.59}$ 
                  & $1.69_{-1.69}^{+2.56}$ \\
J145004.7+004931  &  {\small SDSS CE J222.476242+00.844170 / ACT-CL J1450.0+0049}
                  & $0.382$ 
                  & $61.27\pm5.61$ 
                  & $3.45_{-1.59}^{+2.45}$ 
                  & $4.62_{-2.37}^{+4.32}$ 
                  & $6.12_{-4.91}^{+10.83}$ \\
J145218.8+001002  &  {\small WHL J145218.9+001027 / HSCS J145218+000942}
                  & $0.586$ 
                  & $90.04\pm5.31$ 
                  & $6.42_{-6.04}^{+8.52}$ 
                  & $13.29_{-13.29}^{+17.24}$ 
                  & $1.20_{-1.20}^{+6.43}$ \\
J145459.3-011455  &  {\small SDSS CE J223.754150-01.250201}
                  & $0.314$ 
                  & $26.06\pm2.48$ 
                  & $2.64_{-1.03}^{+1.25}$ 
                  & $4.05_{-1.74}^{+2.40}$ 
                  & $3.08_{-1.80}^{+4.43}$             
\end{longtable}
\tablefoot{
\tablefoottext{a}{The eRASS1 cluster name.}
\tablefoottext{b}{The alt-name from literature by matching within $3$ arcmin from the centroid of the eRASS1 cluster, though the cluster denoted by $^\dagger$ has 3.5 arcmin offset distance.}
\tablefoottext{c}{The cluster redshift.}
\tablefoottext{d}{The cluster richness.}
\tablefoottext{e}{WL mass at $\Delta=500$ obtained by tangential shear fitting.}
\tablefoottext{f}{WL mass at $\Delta=200$.}
\tablefoottext{g}{WL concentration parameter at $\Delta=200$.}
\tablefoottext{*}{The misassociation clusters by visual inspection. }
\tablefoottext{$\natural$}{The poor-fit clusters.}
}
\end{landscape}
\clearpage
\twocolumn
\end{appendix}

\begin{acknowledgements}

The Hyper Suprime-Cam (HSC) collaboration includes the astronomical communities of Japan and Taiwan, and Princeton University.  The HSC instrumentation and software were developed by the National Astronomical Observatory of Japan (NAOJ), the Kavli Institute for the Physics and Mathematics of the Universe (Kavli IPMU), the University of Tokyo, the High Energy Accelerator Research Organization (KEK), the Academia Sinica Institute for Astronomy and Astrophysics in Taiwan (ASIAA), and Princeton University.  Funding was contributed by the FIRST program from the Japanese Cabinet Office, the Ministry of Education, Culture, Sports, Science and Technology (MEXT), the Japan Society for the Promotion of Science (JSPS), Japan Science and Technology Agency  (JST), the Toray Science  Foundation, NAOJ, Kavli IPMU, KEK, ASIAA, and Princeton University.

This paper makes use of software developed for the Large Synoptic Survey Telescope. We thank the LSST Project for making their code available as free software at  http://dm.lsst.org

This paper is based [in part] on data collected at the Subaru Telescope and retrieved from the HSC data archive system, which is operated by Subaru Telescope and Astronomy Data Center (ADC) at NAOJ. Data analysis was in part carried out with the cooperation of Center for Computational Astrophysics (CfCA), NAOJ. We are honored and grateful for the opportunity of observing the Universe from Maunakea, which has the cultural, historical and natural
significance in Hawaii.

The Pan-STARRS1 Surveys (PS1) and the PS1 public science archive have been made possible through contributions by the Institute for Astronomy, the University of Hawaii, the Pan-STARRS Project Office, the Max Planck Society and its participating institutes, the Max Planck Institute for Astronomy, Heidelberg, and the Max Planck Institute for Extraterrestrial Physics, Garching, The Johns Hopkins University, Durham University, the University of Edinburgh, the Queen's University Belfast, the Harvard-Smithsonian Center for Astrophysics, the Las Cumbres Observatory Global Telescope Network Incorporated, the National Central University of Taiwan, the Space Telescope Science Institute, the National Aeronautics and Space Administration under grant No. NNX08AR22G issued through the Planetary Science Division of the NASA Science Mission Directorate, the National Science Foundation grant No. AST-1238877, the University of Maryland, Eotvos Lorand University (ELTE), the Los Alamos National Laboratory, and the Gordon and Betty Moore Foundation.

This work is based on data from eROSITA, the soft X-ray instrument aboard SRG, a joint Russian-German science mission supported by the Russian Space Agency (Roskosmos), in the interests of the Russian Academy of Sciences represented by its Space Research Institute (IKI), and the Deutsches Zentrum f\"{u}r Luft- und Raumfahrt (DLR). The SRG spacecraft was built by Lavochkin Association (NPOL) and its subcontractors, and is operated by NPOL with support from the Max Planck Institute for Extraterrestrial Physics (MPE). The development and construction of the eROSITA X-ray instrument was led by MPE, with contributions from the Dr. Karl Remeis Observatory Bamberg \& ECAP (FAU Erlangen-Nuernberg), the University of Hamburg Observatory, the Leibniz Institute for Astrophysics Potsdam (AIP), and the Institute for Astronomy and Astrophysics of the University of T\"ubingen, with the support of DLR and the Max Planck Society. The Argelander Institute for Astronomy of the University of Bonn and the Ludwig Maximilians Universit\"{a}t Munich also participated in the science preparation for eROSITA.

N. Okabe and M. Oguri acknowledge JSPS KAKENHI Grant Number JP19KK0076. K.U. acknowledges support from the National Science and Technology Council of Taiwan (grant NSTC 112-2112-M-001-027-MY3) and the Academia Sinica Investigator Award (grant AS-IA-112-M04). E. Bulbul, A. Liu, V. Ghirardini, and X. Zhang acknowledge financial support from the European Research Council (ERC) Consolidator Grant under the European Union’s Horizon 2020 research and innovation program (grant agreement CoG DarkQuest No 101002585). VG acknowledges the financial contribution from the contracts Prin-MUR 2022 supported by Next Generation EU (M4.C2.1.1, n.20227RNLY3 The concordance cosmological model: stress-tests with galaxy clusters).

This paper is dedicated to our late friend, Dr. Yuying Zhang, who had hoped to pursue research using eROSITA and to contribute to the eROSITA–HSC collaboration, but sadly passed away before this could be realized; we will always cherish our shared aspirations.

\end{acknowledgements}

\section*{Affiliations}
\begin{enumerate}
\item Physics Program, Graduate School of Advanced Science and Engineering, Hiroshima University, 1-3-1 Kagamiyama, Higashi-Hiroshima, Hiroshima 739-8526, Japan 
\item Hiroshima Astrophysical Science Center, Hiroshima University, 1-3-1 Kagamiyama, Higashi-Hiroshima, Hiroshima 739-8526, Japan
\item Core Research for Energetic Universe, Hiroshima University, 1-3-1, Kagamiyama, Higashi-Hiroshima, Hiroshima 739-8526, Japan
\item Argelander-Institut f\"{u}r Astronomie (AIfA), Universit\"{a}t Bonn, Auf dem H\"{u}gel 71, 53121 Bonn, Germany
\item Universit\"{a}t Innsbruck, Institut f\"{u}r Astro- und Teilchenphysik, Technikerstr. 25/8, 6020 Innsbruck, Austria
\item Center for Frontier Science, Chiba University, 1-33 Yayoi-cho, Inage-ku, Chiba 263-8522, Japan 
\item Department of Physics, Graduate School of Science, Chiba University, 1-33 Yayoi-Cho, Inage-Ku, Chiba 263-8522, Japan
\item Institute of Astronomy and Astrophysics, Academia Sinica, P.O. Box 23-141, Taipei 10617, Taiwan
\item Max Planck Institute for Extraterrestrial Physics, Giessenbachstrasse 1, 85748 Garching, Germany
\item IRAP, Universit\'{e} de Toulouse, CNRS, UPS, CNES, Toulouse, France
\item INAF, Osservatorio di Astrofisica e Scienza dello Spazio, via Piero Gobetti 93/3, I-40129 Bologna, Italy
\item Institute for Frontiers in Astronomy and Astrophysics, Beijing Normal University, Beijing 102206, China
\end{enumerate}

\end{document}